\documentclass[12pt]{article}
\usepackage[utf8]{inputenc}
\usepackage{cite}
\usepackage{comment}
\usepackage{xcolor}
\usepackage{graphicx}
\usepackage{float}
\usepackage{amssymb,amsmath,amsfonts,palatino,amsthm}
\usepackage{epstopdf}
\usepackage{slashed} 
\usepackage{authblk}
\usepackage{xcolor}
\usepackage{hyperref}
\usepackage{tikz}
\usetikzlibrary{arrows.meta}
\usepackage{ulem}
\usepackage[nottoc]{tocbibind}
\newcommand{\f}{\begin{equation}}
\newcommand{\ff}{\end{equation}}

\DeclareMathOperator{\Tr}{Tr}

\begin{document}

 
\title{The Autodidactic Universe}

\author[1,2]{Stephon Alexander}
\author[3,4]{William J.\ Cunningham}
\author[5]{Jaron Lanier}
\author[3]{\mbox{Lee Smolin}}
\author[5,6]{Stefan Stanojevic}
\author[1,5]{Michael W.\ Toomey}
\author[5]{Dave Wecker}

\affil[1]{Brown Theoretical Physics Center and Department of Physics, Brown University, Providence, RI 02906, USA}
\affil[2]{Center for Computational Astrophysics, CCA, Flatiron Institute, New York, NY, 10010,  USA}
\affil[3]{Perimeter Institute for Theoretical Physics, \protect\\31 Caroline Street North, Waterloo, ON N2J2Y5, Canada}
\affil[4]{Agnostiq Inc., 180 Dundas St. W., Toronto, ON M5G1Z8, Canada}
\affil[5]{Microsoft Research, Redmond, WA 98052, USA}
\affil[6]{University of Michigan, Ann Arbor, MI 48109, USA}

\date{\today}
\maketitle

\begin{abstract}
We present an  approach to cosmology in which the Universe learns its own physical laws.  It does so by exploring a landscape of possible laws, which we express as a certain class of  matrix models.
We discover maps that put each of these matrix models in correspondence with both a gauge/gravity theory and a mathematical model of a learning machine, such as a deep recurrent, cyclic neural network.  This establishes a correspondence between each solution of the physical theory and a run of a  neural network.  

This correspondence is not an equivalence, partly because gauge theories emerge from $N \rightarrow \infty $ limits of the matrix models, whereas the same limits of the neural networks used here are
not well-defined.

We discuss in detail what it means to say that learning takes place in autodidactic systems, where there is no supervision.  We propose that if the neural network model 
can be said to learn without supervision,  the same can be said for the corresponding physical theory.  

We consider other protocols  for autodidactic physical
systems, such as optimization of graph variety, subset-replication using self-attention and look-ahead, geometrogenesis guided by reinforcement learning, structural learning using renormalization group techniques, and extensions. These protocols together provide a number of directions in which to explore the origin of physical laws based on 
putting machine learning architectures in correspondence with physical theories.

\end{abstract}
\newpage
\tableofcontents

\newpage

\section{Introduction}

We present ideas at the intersection of theoretical physics, computer science, and philosophy of science with a discussion from all three perspectives.  We begin with a philosophical introduction, proceed with a technical argument that combines physics and computer science, and conclude with further non-technical discussion.

Until recently, most of the research done by theoretical physicists has had the aim of discovering what the laws of physics are.  While we haven't finished that task, we seem to know enough to take a few steps towards answering a deeper question: why are these -- and not others that seem equally consistent mathematically -- the actual laws~\cite{CSP,Rees,BarrowandTipler,LOTC,Susskind-anthropic}?

It used to be thought that our standard model - including general relativity - is the low energy description of a unique 
consistent theory satisfying a short list of principles.   But research in string theory, loop quantum gravity and other approaches to quantum gravity point to the opposite conclusion:  there is a vast landscape of equally consistent theories
~\cite{LOTC,Susskind-anthropic,bousso2004string,taylor2015f}. 

What is then called for is a very different approach to the ``Why these laws?"   question:  Let us seek a dynamical mechanism by which laws can evolve and change, and search within that setting for a reason why the present laws are more likely than another set.   For example, the coupling constants might turn out to be dynamical variables.   This opens the door to several kinds of explanation, new to fundamental physics, but well understood in other parts of science.  These range from deterministic evolution in a space of couplings to stochastic forms of evolution on a 
  landscape of theories. 
The possibility of a useful explanation for some properties of accepted laws, analogous to the explanatory power of natural selection in biology, becomes imaginable, and has been explored~\cite{evolve,LOTC,cns-review}.

In biology there used to be a ``Why these species?" problem:  explain why dogs and cats exist while unicorns and werewolves do not.  With the rise of the modern Darwinian perspective, it became clear that knowing the general principles, which apply to all biology, is a necessary prelude to understanding the detailed - often highly contingent and complex - stories that explain why a particular species likely emerged.

It turns out that there are a number of puzzles concerning the values of the parameters of the standard model, which in one way or another indicate that their present values are special in that they lead to a universe far more complex than would be obtained with typical values~\cite{Rees,BarrowandTipler}.  These suggest that explanations of the sort which we find in biology might be useful~\cite{evolve,LOTC}.  

The application of natural selection to cosmology was first proposed by the American philosopher, Charles Sanders Pierce, in 1893~\cite{CSP}.  
 
The idea has been independently discovered by others since, and has been studied in some detail, to the point that several testable predictions have been recognized\cite{evolve,LOTC,cns-review}.

There is more that could be done in that direction, but that is not our purpose here.

In this paper we want to go even further and suggest that the universe has the capacity to {\it learn its laws.}   Learning is, as we will explain, a much more general notion than evolving by natural selection, it is also a more complex and demanding idea. But we hope to help the reader to see it as within the realm of the possible.

How are these kinds of explanations possible?   Why would they be helpful?    What could be the benefit of taking on such radical ideas?

What do we mean by the laws of nature learning?  The answer is simply that we can construct a correspondence between
possible laws of nature described by gauge fields, gravity and matter fields, and some of the states of a machine learning model such as the restricted Boltzmann machine.  In the simplest language, the choice of a vacuum of a $QFT$ is mapped to a process of pattern  recognition.  

The structure of this paper is as follows.  

The correspondence between a class of gauge field theories and a class of neural network models of machine learning is developed in the next two chapters.   This is based in each of them being describable in the language of matrix models~\cite{matrix-general}.
The field theories involved include topological field theories such as
Chern-Simons~\cite{CST} theory and $BF$ theory\cite{CITE-BF}, as well as the closely related Plebanski~\cite{plebanski1977separation} formulations of general relativity, by itself and coupled to Yang-Mills theory\cite{CITE1,CITE2}.   This provides an example of how a system whose degrees of freedom encode possible laws-such as spacetime dimension, gauge group and matter representations can be mapped onto formal descriptions of neural network systems that we are comfortable describing as being capable of learning.  
We note that as we are talking about the universe, there is no supervision; we call such systems autodidactic.

Note that at this point these are correspondences, not equivalences.

These results came out of a more general search for physical systems that could learn without supervision. 
In a search of physical systems that could qualify for autodidactic interpretation, we studied a variety of proposals and models.  A sample
of these are presented in Chapter 4,  and employ the renormalization group,
the proposal that quantum systems have no laws except to follow precedence, self-sampling methods, systems that maximize variety and geometrical self-assembly.   The discussions here are brief, as papers are in preparation to detail each of these.

Chapter 5 is a collection of responses to philosophical questions relevant to the interpretation of earlier sections.  

\subsection{What is learning?}

When we refer to laws of physics changing, or being different in different regions of spacetime, we are speaking in the language of effective field theory, which is to say we assume that the laws relevant for low energy physics can involve collective coordinates, emergent, renormalized and averaged fields, and also can be dependent on effective coupling constants whose values depend on the values of fields, as well as on temperature, density etc.   More generally, we adopt the useful idea that the world can be analyzed in terms of a hierarchy of levels; the degrees of freedom and the regularities described at each level appear to observers there to be stable and unchanging or slowly changing.    At the same time, when expressed in terms of the degrees of freedom and laws of a lower level, they may seem emergent and variable.  We can then be agnostics as to whether there are fundamental, timeless laws down at the smallest scale.

To proceed to define {\it learning} we need to specify some terms:

We speak of “subsystems” and “processes”.   A system is an identifiable part of the universe that contains a number of processes and persists over a time scale that is long compared to the timescales of its constituent processes - and is often characterized by a boundary which isolates it from processes not contained in it.    What is not in the system is called the system's environment.  

A system has a set of activities carried out in time, by means of which it does work: it catalyzes or alters itself and other things in the environment.    Each of these is called a process, and each has a typical time scale.  Each process is closed, which means that it operates in cycles that bring it back to its initial state, after having done work.   The thing that a process does that furthers the whole of which it is a part, is called its function.  

We require that a system must be decomposable into a network of such processes which are interdependent on each other, in the sense that the continuation, and growth of the whole is dependent on and furthered by that of the processes.  This is sometimes called a "Kantian whole"~\cite{kaufmann1,kaufmann2,kaufmann3}.  

 Typical systems of this kind are found far from thermal equilibrium, where some of their processes channel and control the rates of energy and materials through them.   They develop  from the need of the whole to manage flows of energy and materials through it.

We can say then that such a system learns when it is able to alter its internal processes and actions in the world to better capture and exploit the flows of energy through or near it to further a goal, which is typically continued existence, but might become more than that.

The notion of learning encompasses a variety of circumstances, including biological natural selection and what we call learning in individual organisms such as a person.   Many things can learn and there are many ways to learn.

Biological evolution is a vast-scale learning system, but individual members of species have also typically inherited the capacity to learn.  This does not require that the “information” learned is stored symbolically - or stored in any abstract way - a moth does not know its wings are black or that the reason is its tree is black.

Darwin suggested that an esthetic dimension of mate choice, sexual selection, had become a central driver of evolution.  In human intellectual behavior, we sometimes observe learning that is not motivated by interaction with a proximate threat or survivability criteria, but instead by such things as curiosity, a quest for elegance or beauty, or other hard-to-define targets.  

We will term these cases, where a learning system constructs its own criteria, “autodidactic systems”.  Autodidactic criteria will often, or perhaps always, be coupled to more fundamental criteria, particularly survival, but indirectly.

The term “learning” has been commonly used in computer science to refer to a class of programs, such as “deep learning networks”.  These are typically driven by either imposed criteria for the feedback that drives learning, even when an algorithm is unsupervised, or by a simulation of evolution, in the case of “artificial life” experiments.

We have described a variety of learning processes: learning might occur through survival when populations of organism compete in a lethal environment and survivors pass on genes; learning might occur in an individual through life experience, and the criteria for what is worth learning might be autodidactic; learning might be engineered into a software model by an engineer, using arbitrary criteria.  

This list is not exhaustive.  For instance, we could speculate about a form of alien life which does not have a separate level of genes, but nonetheless undergoes variations randomly – and in which interpenetrating swarms of structures destabilize each other, causing a gradual emergence of more robust structures.  This might have been what happened before the DNA system emerged on Earth.  We should not assume that we are aware of all forms of learning.

In one sense, learning is nothing special; it is a causal process, conveyed by physical interactions.  And yet we need to consider learning as special to explain events that transpire because of learning.  It is clarifying to introduce language to speak of the physics of learning systems.  It helps to have a word to refer to the things the sub-processes do for the whole that contribute to the whole’s learning.

We will call these functions {\footnote{The brief answer to the common objection that functional explanations conflict with reductionist explanations is that a complete explanation requires both.}  {\it consequence accumulators}
or {\it consequencers}, for short.  A consequencer accumulates information from the past that is more influential to the future than is typical for other contents of the system.  It is the negative feedback loop in your home’s thermostat, or a star, or a naturally occurring reactor. It is a bit in a computer, or a gene.

Other naturally occurring consequencers we can keep in mind are limit cycles, which occur necessarily, and seemingly randomly  in finite state deterministic dynamical systems [FSDS].  They gain robustness from the way small random excursions from the cycles are captured and returned by basins of attraction.  These occur randomly, without supervision or design, but are first steps in the natural self-organization of life, ecosystems, economic and food webs etc.  It its perfectly plausible that the random occurrence of limit cycles could arise in a network of tunnelling channels among a landscape of vacua in a quantum field theory. 

We also note that the term "consequencer" encompasses the idea of a knowledge base, or knowledge graph, in the field of artificial intelligence~\cite{russell2021artificial}.
One of the simplest representations of ``learned knowledge" uses propositional logic or, in other words, a set (or graph) of Boolean satisfiability constraints. The choice of knowledge representation also defines the ontology of the learning system.

How might cosmological-scale learning occur?  There have been proposals of applications of natural 
selection to cosmology~\cite{evolve,LOTC}.  If one focuses only on a single universe, however, then there is no population of universes in which selection could occur; this paper focuses on a single universe, so does not dwell on cosmological natural selection.

Instead, we consider autodidactic learning in physics.
We shall consider cosmological consequencers first, and then cosmological autodidactic processes.

\subsection{How can physical law be understood as learning?}

In the following sections, notably 3.3, we will  establish the following maps between a class of physical theories and a class of neural network-based models of learning:

\begin{center}
 \begin{tikzpicture}[scale=1]
 \node (ls) at (0,0) {Matrix models};
 \node (dt) at (-3.5,-3) {Quantum gauge theories};
 \node (gft) at (3.5,-3) {Learning machines};
 \draw[latex-latex,ultra thick] (ls) -- (dt);
 \draw[latex-latex,ultra thick] (ls) -- (gft);
 \draw[latex-latex,ultra thick] (gft) -- (dt);
 \end{tikzpicture}
 \end{center}

This will be accomplished using correspondences to matrix models: 

 \begin{center}
 \begin{tikzpicture}[scale=1]
 \node (vs) at (0,0) {Vectors, spinors};
 \node[align=center] (qmf) at (-3.5,-3) {Quantum\\ matter fields};
 \node[align=center] (l) at (3.5,-3) {Machine learning\\layers};
 
 \node (m) at (0,-4) {Matrices};
 \node[align=center] (qggf) at (-3.5,-7) {Quantum gauge,\\gravity fields};
 \node[align=center] (mlw) at (3.5,-7) {Machine learning\\weights};
 
 \node[align=center] (tq) at (0,-8) {Thermalization,\\quenches};
 \node[align=center] (ce) at (-3.5,-11) {Cosmological\\ embedding};
 \node[align=center] (es) at (3.5,-11) {External\\training};
 
 \draw[latex-latex,ultra thick] (vs) -- (qmf);
 \draw[latex-latex,ultra thick] (qmf) -- (l);
 \draw[latex-latex,ultra thick] (l) -- (vs);
 
 \draw[latex-latex,ultra thick] (m) -- (qggf);
 \draw[latex-latex,ultra thick] (qggf) -- (mlw);
 \draw[latex-latex,ultra thick] (m) -- (mlw);
 
 \draw[latex-latex,ultra thick] (tq) -- (ce);
 \draw[latex-latex,ultra thick] (ce) -- (es);
 \draw[latex-latex,ultra thick] (es) -- (tq);
 
 \draw[thick,loosely dashed] (vs) -- (m);
 \draw[thick,loosely dashed] (m) -- (tq);
 \draw[thick,loosely dashed] (qmf) -- (qggf);
 \draw[thick,loosely dashed] (qggf) -- (ce);
 \draw[thick,loosely dashed] (l) -- (mlw);
 \draw[thick,loosely dashed] (mlw) -- (es);
 
 \end{tikzpicture}
 \end{center}

 The action is cubic in the matrices~\cite{cubicm1}, so the equations of motion
 are quadratic equations.   With the imposition of various identifications and constraints, the matrix degrees of freedom can be shown to be isomorphic to the degrees of freedom of a large class of gauge field theories, on various background geometries, including Yang-Mills theories 
 and general relativity~\cite{universal}. To some extent, the spatial dimensions, gauge groups, spatial topology and matter representations can be chosen.
 
 These can thus be seen as quantum gauge or gravity theories,
 on which a variety of infrared cutoffs (also known as
 compactifications) have been imposed.

There is thus at least one more triangle:

 \begin{center}
 \begin{tikzpicture}[scale=1]
 \node[align=center] (v1) at (0,0) {Choice of compactification,\\ constraints, identifications\\ under translations.};
 \node[align=center] (v2) at (-3.5,-3) {dimension, gauge\\ group, reps};
 \node[align=center] (v3) at (3.5,-3) {hypothesis as to\\ symmetries in the data};
 \draw[latex-latex,ultra thick] (v1) -- (v2);
 \draw[latex-latex,ultra thick] (v2) -- (v3);
 \draw[latex-latex,ultra thick] (v3) -- (v1);
 \end{tikzpicture}
 \end{center}

Most of these correspondences have yet to be worked out.

 In this paper we work with a particularly minimal set of matrix models, which have at most cubic actions\cite{universal,cubicm1,cubicm2,cubicm3,cubicm4}.   
 
 We note that these are in a different universality class then some of the matrix models studied previously, which are explicitly dependent
 on a background geometry specified 
 by a constant metric~\cite{matrix-general,BFSS,HoppePhD}.  We see an
 example of this in equation (1) where the fixed metric is
 $g_{ab}$.   In the limit $N \rightarrow \infty$   these can be
 interpreted either as an $N= \infty$ Yang-Mills theory~\cite{dWHN,BFSS} or a membrane theory~\cite{HoppePhD,Hoppesolves}.   The latter construction is based
 on the fact that the $N \rightarrow \infty$ of $SU(N)$ goes to the group of volume preserving diffeomorphisms on the 2-torus~\cite{HoppePhD}.
 
 On the other hand the cubic matrix models do not depend on any background invariant structure, save the trace of the algebra.   So these
 appear to be parts of other universality classes, defined algebraically,
 which include the topological field theories, general relativity, and perhaps other diffeomorphism-invariant theories.

On the other side, we will show here that there exists maps
between a large class of neural network models of machine learning and the cubic matrix models.  By combining maps we get a map between   machine learning models and gauge and gravity theories.  The basic maps extend between the transformation matrices that take the data among the several
 screens or layers of the neural network model and the degrees of freedom of the gauge theories. In the case that the
 dimensions are the same-otherwise they may correspond to renormalization group transformations.
 
 We should note that the correspondences we have defined are stated for finite $N$ -- finite matrices $N\times N$, finite screens, as well as a finite number of loops through the sequence. Let us comment on the objects and correspondences in the triangles in this case. First, the matrix models are all well defined. The universality, i.e. being able to model gauge fields SO(N) or U(N), in any dimension $d$, and the map between different $d$ and $N$ is well defined at this stage~\cite{universal}.  In the mapping to the QFT on $d$-torus, for instance, there is both infrared and ultraviolet regularization which can be seen by the discrete rotations around each $S^1$ having a finite number of states.  So it takes a finite amount of energy to transition or tunnel between any pair of related ground states.  The corresponding cut off gauge field theories are well defined.  Indeed QFTs are often defined by taking limits
on such cutoff QFTs.

The restriction to finite $N$ for neural networks is standard.  It is the finite neural network that has been shown to learn as well as to satisfy Chiara Moletta's criteria to carry information, that is, to have at least two distinguishable states which can be swapped and copied~\cite{moletta2021science}.  The limit $N\to\infty$ is subtle and requires great  care, even in the best known case -- that of QFT -- a study that has involved many of the best mathematical and theoretical physicists over most of the past century.   It will be interesting to try to use the correspondence to study what the infinite $N$ limit of a neural network is, but that is left for later work\footnote{We note there are a small number of papers in the literature that study a class of neural networks that are called ``infinite'', see~\cite{cote2016infinite}.}.
 
 The behavior of a learning machine may include
 steps where the degrees of freedom are thermalized or quenched.
 These may correspond to putting the gauge theory in a cosmological
 model.
 
 Through these maps, one can then imagine building a future
 learning machine out of the very gauge and gravitational fields with which our world is constructed.  The coding will involve choices of topology of the compactifications of the matrix models.
 
 But if we can build a computer out of parts which  amount to choices of topologies, boundary conditions and initial conditions, couldn't they arise naturally?  Then we would have arrived precisely at an understanding
 of quantum geometry as an  autodidactic  system, able to learn in an unsupervised context.
 
 As we will see, once we can describe a means for the physics we know to have come about through an adaptive process, we can also describe the means for that process to take on properties of self-reference, modeling, and what we can call learning.

We will then seek a general paradigm of systems, not necessarily
modeled on biological systems, that can "teach themselves"
how to successfully navigate landscapes with increasing robustness, allowing for a space of future contingencies.  These {\it autodidactic} systems will provide new paradigms for how physical systems may explore landscapes of theories.

In order to examine this idea, we can consider contemporary, engineered systems that appear to be
able to modify their behavior in a way that is generally understood to be 
"learning": machine learning algorithms.  
 
 Roughly speaking, contemporary computer science distinguishes two kinds of learning  algorithms: supervised and unsupervised.  The first kind are operated by engineers to match new inputs to desired outputs.  It is impressive that we can build such devices.  But the unsupervised models are perhaps more impressive, because they have only simple starting criteria and require few interventions. They have architectures that apply principles like internal adversarial processes to propel themselves.

Unsupervised systems are currently harder to predict; they appear to be more likely to produce novel output such as chess tactics that are otherwise difficult to discover.  Thus far, such systems have not generated important new theories in physics, but they might. 

Machine learning ideas \textit{have} been successfully used in a variety of ways in physics, such as a new framework for stating accepted laws~\cite{Vanchurin_2020}. When endowed with cleverly chosen priors, machine learning models have been able to "learn" aspects of physics such as the Lagrangian and Hamiltonian formalism~\cite{cranmer2020lagrangian, Schmidt81, greydanus2019hamiltonian}, renormalization group transformations~\cite{optimal_rg_information, mutual_information}, and have ventured into the realm of holography~\cite{Hashimoto_2018, Hashimoto_2019}.  There are also instances where field theoretic concepts have been used to elucidate the foundations of neural networks; see~\cite{schoenholz2017correspondence, Lin_2017}.

When a model becomes sufficiently good at mirroring observable reality, it becomes natural to ask if it could be considered as if it  were an aspect of reality, not just an approximation.  A sense that models are  substantial motivated the discovery of previously unsuspected phenomena, such as antimatter, which was predicted because of the available solutions to an equation.  We are extending Wigner's trust given to the "unreasonable" success of theory.  If neural networks can predict or rediscover the theories we know about, might nature not be as similar to the neural networks as to the theories?

\section{Matrix models as learning systems}

In this section we will provide an example of a mapping between the degrees of freedom of two well studied theories.   On the one hand we have the deep neural network - the two layer network familiar from the Boltzmann and restricted Boltzmann models.

On the other hand we have a theory also very studied - General Relativity in 3+1 dimensions - in a chiral form due to
Plebanski.

\subsection{Matrix models}

Let us first talk about matrix models in general. These are models whose degrees of freedom are large $N$ matrices.
They can be of various types, say orthogonal, unitary, etc.   The matrix elements represent the physical degrees of freedom.

A common example is~\cite{HoppePhD,dWHN,BFSS}
\f
H=  \Tr  \frac{m^2}{2} \{ X^{a} \circ X^{b} 
\}  g_{ab} 
-\frac{1}{24 g^2}  \Tr 
\{ [X^a, X^b ] [X^c, X^d ]   g_{ac}  g_{bd}\}\,.
\ff
This depends on a fixed constant background metric for space  $g_{ab}$.

The basic idea behind many constructions in high energy theory and condensed matter physics is that of
{\it effective field theory}   with an ultraviolet cutoff, $E_P$.
The idea is that  observing some phenomenon at
an energy $E << E_p$ , you will measure the influences of all operators that could be in the action or Hamiltonian consistent with the symmetries of the problem.   These will organize themselves in a power series expansion organized in powers of the ratio $r= \frac{E}{E_p}$.

In this kind of scenario we only need specify the simplest action or Hamiltonian that displays the global symmetries thought fundamental.  
The other relevant terms will be generated by renormalization
group transformations.
Effective field theory is an especially transparent method to develop the low energy consequences of spontaneously broken symmetries.

In addition, it is natural in a matrix model to pack the multiplets of the remaining symmetry into multiplets of the higher symmetry, while using the Lagrange multiplier trick to reduce the degree of the equations of motion, or Lagrangian, while increasing the size of the matrices.  Using this trick repeatedly you can always reduce the action of a non-linear theory to a cubic, as you cannot express non-linear dynamics as linear equations.  In this simplest form the equations of motion are at most quadratic equations.  

Thus, if we take a single very large matrix, $M_a^b$ we can write a universal action~\cite{universal,cubicm1,cubicm2,cubicm3,cubicm4}:
\f
S_{Univ} = -\frac{1}{g^2 N^2}
\Tr M^3\,,
\ff
whose equations of motion are
merely
\f
M^2 =0\,.
\ff

This has originally U(N) symmetry, which can be broken by a choice of solution in a very large number of ways.   By choosing a diverse set of solutions and expanding around them, we can find to leading order a very large number of gauge and gravitational theories, including Yang-Mills theory and general relativity, invariant under a large choice of symmetries, in a variety of space-time dimensions. Some of these are described in~\cite{universal}   One
example, a first order form of general relativity, is the subject of Sec.~\ref{sec:plebanski}.

We first study a simple example of a continuous theory generated from
this matrix model, which is Chern-Simons theory.

\subsection{Learning systems, gauge theories and laws that learn}

  The question we want to investigate here is whether learning systems,
  might be useful as models of,  or frameworks for
  novel formulations of fundamental theories.  There is an ambiguity we accept about whether such models of frameworks might be similar to what goes on in nature, or whether they are only a rough, remote sketch.  In either case, these will be 
  {\it laws that learn}, and the setting for that learning will be cosmology.
  
An important clue for us is that all of the frameworks we currently have for
fundamental theories, namely Yang-Mills theories, general relativity and string theories, have local gauge invariances, both internal  
and diffeomorphisms.  They are theories whose degrees of freedom are 
expressed in terms of parallel transport.   Can theories with these
 kinds of gauge invariances be coded as learning systems?  
 
 Another question immediately appears:  what will they learn?  These fundamental theories have several mechanisms for changing the effective laws, including spontaneous symmetry breaking, phase transitions, compactifications and decompactifications, dimensional reductions, and enhancement, and others.   

 These various moves gives us the beginnings of a vocabulary, which a theory may learn
 to think in.
 
 All of these can be  easily described in the language of matrix models, which means that if we can establish the kind of correspondence we seek, we will be
 able to learn how to speak about them in the language of machine learning. 
 We should also emphasize that in this emulation of the idea, changing laws does not necessarily mean a changing
 learning architecture, though more generally it could. The parts that we suppose do change via learning are the matrix weights,
 the optimization strategy, and the cost function.

Let us look at what we know about these relationships.

\subsection{Neural network architectures}
\label{sec:nnarch}

Let us then get to know the basic architecture of a class of learning systems.
We  begin by reviewing the structure of the simplest class of deep neural networks - those with a single internal or hidden layer.  These are 
similar to the restricted Boltzmann machines - but we consider several variants.   We will then comment on several special cases: the Boltzmann machine, both ordinary and restricted (RBM), and recurrent neural networks with memory.

The neural network models have variables that live on nodes, representing neurons, which are organized
into sheets or layers.  These are connected, and the strength of each
connection is given by a variable weight, which are among the degrees of 
freedom.   

In Restricted Boltzmann Machines (RBMs) there are just two layers of nodes,  $N$
visible nodes, $v^a$, $a=1, \ldots N$, and a hidden layer of $M$ 
hidden nodes
$h_b$, $b=1, \ldots M$. 
We will consider first the case $M=N$.

In the case of the RBM each neuron is connected with all the neurons in the other layer - but there are no  connections within layers.  We adopt
that architecture here.

The matrix of weights defines a linear transformation from the visible
layer to the hidden layer:
\f
h_b = W_b^a v_a\,.
\ff

External (visible) layers interact
 directly with the environment and internal, or hidden layers, do not. A neural network with one or more hidden layers is called a \textit{deep} neural network. Hidden layers improve the expressive power of a neural network, since the sequential layers of linear units stack together in a way that allows the entire system to model complex non-linear phenomena. Transformations between adjacent pairs of layers are linear maps between sub-spaces. A transformation that maps a layer with $N$ nodes to one with $N'<N$ nodes is,
from the physics perspective, a renormalization.

There can also be
maps which take a layer to one with the same number of nodes, which
can be thought of as linear transformations
that permute the nodes. 

In the case of a circle of layers, 
we may denote these maps as
\f
h_b^{I+1} = [W^{I,I+1} ]_b^{\ \ a} h_a^{I}\,,
\ff
up until the last which is
\f
v_b^{\prime} = [W^{I,0} ]_b^{\ \ a} h_a^{I}\,.
\ff
Of course we can imagine more complicated architectures, with
branchings of signals, but we will not study them here.

We return to the simplest case of one external and one internal layer.

The following is a variant of the restricted Boltzmann model, where
we substitute real continuous variables for the discrete binary
variables of the original model.   This also lets some of the
dynamics be deterministic rather than stochastic.  This will
not prevent us from identifying correspondences.
The alternative is to modify the matrix and field theory models
to depend on binary variables.

\subsubsection{Review  of the RBM, with continuous variables}

The usual Boltzmann model has discrete binary variables.  We modify the theory by using continuous variables.  These can be thought of as  coarse-grained averages over regions of the layers and over time of the binary variables.

The reason we do this is that we are seeking to make a correspondence between intelligent machines, or more properly, designs or mathematical models
of such machines, and the dynamics of space-time and gauge fields.  But before
we search for such a correspondence there are some basic issues to be mentioned. One is the difference between smooth fields and smooth evolution equations and discrete variables. Another is   the irreversibility of almost all computers, while the field theories are most often reversible.  Still another is that some of the machines 
are stochastic while the classical field theories satisfy deterministic equations
of motion.

These are subtle issues and we will go into them in detail later, in section 5.   

The last issue suggests that we see the classical field theories as effective field theories that hold in the thermodynamic limit, in much the same way that the semi-classical limits of quantum field theories are described by classical equations.

For the moment, we will be proceeding by constructing an extension of the RBM in which the degrees of freedom are continuous variables, which evolve continuously
in time.  We will formulate this as a Hamiltonian system, defined by a Lagrangian,
(\ref{Ham1}) below.  The resulting equations of motion below, are time reversal invariant.

 We will then look for solutions in which the time derivatives all
vanish - or are very slow.   This may break the time reversal symmetry.   There is of course nothing wrong with solutions to a theory not having all the symmetries of the dynamics - this is a very common situation.    In the present case, this
 will lead to the recovery of a set of  equations very similar to those that describe a Restricted Boltzmann Machine.   The effective equations which define this
RBM sector break time reversal invariance.

\noindent The $RBM$ model evolves by iterating the following discrete steps:

\begin{enumerate}
\item{}{\bf   Initialization:}     The visible nodes, $v^a$,  are set, as in a photograph or another form of input.   We also choose the density weight matrix, $W_a^{ \ \ b}$ and bias $r_b$.

\item{} During the {\bf forward pass} the states of the hidden screen $h_b$ are set by the following mapping.  We also include a bias, $r_b$
\f
h_b (t)  = W_b^a v_a (t)   + r_b\,.
\label{forward}
\ff

Note that the $v^a, h_a $  and  $W^a_{\ \ b}$  are dimensionless.

\item{} Compute the outer product of $v_a$ and $h_b$, the result of which is the \textbf{positive gradient}.
\f 
p_a^{\ \  b} = v_a h^b = v_a ( W_b^a v_a + r_b)\,.
\ff

\item{} There follows the 
{\bf backward pass}, during which we reset the visible nodes by mapping the
hidden nodes onto them, plus another bias, $u_b$.   We denote
the new values of the visible layer, $v_a^\prime $.
\f
v'_a = (W^T)_a^b  h_b + u_b\,,
\label{back}
\ff
Notice that we use the transpose of the weight matrix to make the backwards
pass.  

\item{}{\bf Repeat the forward pass} to
\f
h^{\prime \ \ b} = W_a^b v'_a + r^b\,.
\ff

\item{} Compute the outer product of $v'_a$ and $h'_b$, the result of which is the \textbf{negative gradient}.
\f 
N_a^{\ \ b} = v_a^{\prime b} h^\prime_a\,.
\ff

\item{} This is followed by an \textbf{update of the weights 
and the biases}, via some learning rate $\epsilon$ times the positive gradient minus the negative gradient. The biases are also updated analogously,

\f
\Delta W_a^b = \epsilon(v_a h^T_b - v'_a h'^T_b)\,,
\ff
\f
\Delta r_a = \epsilon (v_a - v'_a)\,,
\ff
\f
\Delta u_b = \epsilon (h_b - h'_b)\,.
\ff
\end{enumerate}

\noindent The whole process is then iterated until a convergence criteria is met.

To understand these a bit better, we combine (\ref{forward}) and (\ref{back}) to find
\f
\dot{v}^a (t)= \left [W^T (t) \circ  W (t) - I \right ]^{\ \ a}_b   v^b (t)
+W^{a T}_b  r^b +  u^b \,,
\label{dotvb2}
\ff
 
 To see the structure let us for a moment set  the biases, $r_b$ and $u_b$ to zero.
 
 \f
\dot{v}^a (t)= \left [W^T (t) \circ W (t) - I \right ]^{\ \ a}_b   v^b (t)\,,
\label{dotv}
\ff
 The weights evolve subject to
 \f
\dot{W}_{ab}  = -\frac{\alpha}{2} \left ( \dot{v}^T_a {h}_b +  {v}^T_a \dot{h}_b \right )
 = -\frac{\alpha}{2} \frac{d}{dt}  \left ( {v}^T_a {h}_b  \right )
\,   \label{dotW2}
\ff
 We see that there are limit points and limit cycles given by
  $W_a^b = W^{T \ \ b}_a $ is orthogonal,  which implies both

\f
\dot{v}^a = 0   \ \ \   \mbox{and}   \ \ \ \ \ \dot{W}^a_b= 0
\label{fixedpd}
\ff
This implies
\f
0=  -\frac{\alpha}{2}   
\frac{d}{dt}\left ( {v}^T_a {h}_b    \right  )
\ff

Thus, starting with a general $W$ and  $v$,  we get a trajectory through the space of possible ``laws" that settles down when it runs into a orthogonal $W$.  As these are
unitary, we have a system that converges to a real version of quantum mechanics.

One can show that $\ddot{W}_a^b$ and $\ddot{v}^a$ are functions of the initial
data $W_a^b$ and $v^a$.

Now we include the biases and we find (\ref{dotvb2}), which gives us a real,orthogonal form of quantum mechanics

\begin{eqnarray}
\dot{W}_{ab} & =& -\frac{\alpha}{2} \left ( \dot{v}^T_a {h}_b +  {v}^T_a \dot{h}_b
\right )\,     =  -\frac{\alpha}{2}\frac{d}{dt} 
 \left ({v}^T_a {h}_b   \right )\,,    \nonumber    
\\
&=&  -\frac{\alpha}{2} \left ( \left [W^T (t) \circ  W (t) - I \right ]^{\ \ c}_a   
+W \circ b^T +u^T_b 
v^T_c  \right )
 {v}_b 
 \nonumber 
 \\  
 &+&
 {v}^T_a \left (   \left [W^T (t) \circ  W (t) - I \right ]^{\ \ c}_b   
v^T_c
+W \circ b^T +u^T_b 
\right )\,
\label{dotW3}
\end{eqnarray}

The orthogonal  fixed points are, as before, at~\eqref{fixedpd}

\begin{eqnarray}
\dot{W}_{ab} &=& 0  \,,
\\
 \dot{v}_c &=& 0 \,, 
 \\
  W \circ W^T &=&I\,,    
\\
 W \circ b^T  &=& -  u^T \,
\end{eqnarray}

Thus, this is generically a time-asymmetric non-linear dynamical system,  
which has fixed points (or rather limit sets) which are the orbits of orthgonal
transformations.   The dynamics is time reversible and orthogonal on the limit sets.
Associated to each limit set will be a basin of attraction of configurations that
converge to it.

Each limit set defines a quantum dynamics whose time evolution operator is an
orthogonal
matrix.     

Note that had we used complex weights and values, we would have gotten to
the same place, but with unitary dynamics governing the limit sets: ie quantum mechanics.

This may be useful for talking about the evolution of laws as well as the 
measurement problem.

We would like to go one more step, to formulate the learning machine in terms
of continuous time evolution.  We have two stages of dynamics: the first where
we hold the weights and biases fixed and evolve the layers, which alternates
to the gradient flow stages where we do the opposite.   Can we unify these
into a single dynamical system?

One approach is the following.   We unify the two stages within a single Hamiltonian
system, which is governed by an action principle of the form.

\f
S= \int dt   \left (    
- {\cal H}[v,h,W,r,b] 
\right )
\label{Ham1}
\ff
Where the Hamiltonian, ${\cal H}$ is,
\f
{ \cal H}[v,h,W,r,b] = v^a W_{a}^{\ \  b}  h_b   -    r^a h_a -  b_a v^a 
+ \frac{\gamma}{2} ( v^a v_a + h^a h_a )
+ \frac{\omega^2 }{2} (  W^a_{\  b} W^a_{\  b}  )
\label{Ham31}
\ff
Let's look at the equations of motion:

\begin{eqnarray}
- W^a_{ \ b} v^b & = & \frac{\delta S}{\delta h_a}
\nonumber \\
&=& \gamma h^a + r^a 
\\
W^{T a}_{ \ b} h^b & = &  \gamma v^a + b^a 
\\
 -\omega^2  W^a_{\  b} & = & h^a v_b 
\end{eqnarray}

We get the equations governing the RBM:
\begin{eqnarray}
h^a &=& \gamma^{-1} \left (     W^a_{ \ b} v^b          -r^a \right )
\\
v^a &=& \gamma^{-1} \left (     W^{T a}_{ \ b} h^b          -b^a \right )
\\
 {W}^a_{\  b}  & =& -\frac{1}{\omega^2}   h^a v_b
 \label{W17} 
 \end{eqnarray}
 
The time derivative of 
(\ref{W17})
gives us the gradient rule of the RBM,  which is (\ref{dotW2}).

\subsubsection{Features of learning systems}

\begin{itemize}

\item{} To each way of connecting the outside layer to the hidden layer, there corresponds an adjacency matrix. Dynamics of the networks, such as those which occur in machine
learning, are expressed in terms of operations on these matrices.

\item{}But large matrices are often used to represent  gauge fields and their 
dynamics, including their self-coupling and their coupling with matter fields.

\item{}The question is whether we can take these correspondences all the way across,
to represent the dynamics of gauge fields in terms of the dynamics of weights
in machine learning?  We will show in this section that we can.  The main reason is
that we can restrict the matrices of connections and weights so that they code transformations from one layer to another.  That is, they describe parallel transport among the layers.
So they are about the same kind of objects that gauge theories are based on.

\item{}The matrices of weights, defined on networks, represent transformations
of a separate set of degrees of freedom, which are various layers
in which are encoded the patterns that learning machines learn to recognize.

\item{} By combining sequences of such transformations, we see that they
satisfy the group axioms.  What sorts of groups are they?  Because we want their
properties to scale as the sizes of the layers increases (indeed to 
very large $N$), it seems these are
not finite groups, they must be representations of Lie groups.  This implies they will have generators which are in the adjoint of some Lie algebra, ${\cal A}$.

\item{}The layer variables seem to transform linearly under the action of the
weight matrices.   They are thus directly analogous to the matter fields of
conventional gauge field theories.  They may then be decomposed 
into combinations of representations of ${\cal G}$.

\end{itemize}

\subsubsection{Features of matrix models of gauge theories}

\begin{itemize}

\item{} Let us consider a matter field, $ \Phi $, or a gauge field,  $A_a $, 
which is 
valued in a representation of a Lie group.  Typically matter fields are valued in a fundamental
representation while gauge fields are valued in the adjoint representation.

Given a field theory, we can construct a matrix model by setting all the
derivatives to zero.  This gives us a reduction to constant fields, i.e.
 \f
  \partial_b \Phi =0    , \ \ \ \ \ \ \partial_b   A_a =0\,,
  \ff
under which the  $\Phi$ and $A_a$ are matrices.   

The mapping involves sending fields to $N \times N$  matrices.  That is, we set all derivatives to zero so that,
\f
F^{AB} \rightarrow  [ A,A ]^{AB}  , \ \ \ \ \ {\cal D} \wedge B^{AB}
\rightarrow  \epsilon^{abcd}  [A_a ,  B_{bc}]^{AB}\,.
\ff
The resulting action is invariant under two  kinds of gauge invariances.

\item{} First, under the homogeneous $SO(N)$ gauge transformations:
\f
A_a \rightarrow g^{-1} \circ A_a \circ g, \ \ \ 
\Phi \rightarrow g \circ \Phi\,,
\label{hom}
\ff
depending on the representation.

\item{} In addition,  gauge invariance of the  continuum theory is,
unlike (\ref{hom}), inhomogeneous.
This implies a translation gauge invariance
in which the action $S$ or Hamiltonian $H$ is invariant under translations
of the form,
\f
A_{a i}^{\ \ j} \rightarrow A_{a i}^{\ \ j \prime} 
=A_{a i}^{\ \ j}      + I^{\ \ j}_{i} c_a\,,
\label{inhomogauge}
\ff
where $c_a$ is time invariant.

This means that the dynamics of $A_{a i}^{\ \ j}$ must be expressed through
commutators, i.e., $ [A_a, A_b ]$, which is the constant-field limit of
$F_{ab}$, and $ [A_a, \Phi ]$, which is the constant field limit of
${\cal D}_a \Phi $.

\item{}There is also an inverse transformation from matrices to gauge fields on
compact manifolds, described in~\cite{matrix-general}.

\end{itemize}

\subsection{Plebanski gravity as a learning system}
\label{sec:plebanski}

To illustrate the points we have made, we give a quick example of
a physical theory that can be expressed as a matrix model, and through
that, mapped to a corresponding neural network model - 
general relativity.

We described dynamics of the restricted Boltzmann learning models in Sec.~\ref{sec:nnarch}.

Plebanski's version of general relativity~\cite{plebanski1977separation,Capovilla_1991} is not known to many people outside of
specialists in quantum gravity.  This is unfortunate, as it is both elegant and compact; the action is made of quadratic and cubic terms, and so the equations of motion
are quadratic equations.  Remarkably, this is the case also with the equations
and Hamiltonian of the RBM.

The degrees of freedom for Plebanski gravity are a left handed SU(2)
connection $A_a^{AB}$ whose curvature two-form is written
$F^{AB} = dA^{AB} + ( A \wedge A )^{AB}$, and a two-form $B^{AB}$  (also far
valued in the vector of SU(2) and hence symmetric in the spinor indices).
In addition, there is a scalar field $\Phi^{ABCD}$ which is pure spin two in SU(2)
and hence totally symmetric in spinor indices.

There is a constraint on the two forms $B$,
\f  
B^{(AB} \wedge B^{CD)} =0\,,
\label{eq:bconstraint}
\ff
which is satisfied if and only if there is a frame field, $e^{AA'}$
such that
\f 
B^{AB} = e^{(A}_{A'} \wedge e^{B) A'}\,.
\ff

We are going to make the matrix version, which sets all the derivatives  to zero.  For simplicity we will study the Euclidean 
signature case in which all fields are real.   We will also restrict to cases where the time variable is periodic,  with period
$2 \pi \beta$ with $\beta$ inverse temperature.

We will use the simple action (\ref{m3-1}) and take for the single matrix
\f
M=   \Phi  \otimes \gamma_5 + A_{a I} \otimes T^I \otimes \gamma^a  + B_{ab I} \otimes T^I \otimes \gamma^{ab}\gamma_5\,.
\label{ansatzP}
\ff

Here let $\gamma^a$ be the four Dirac gamma
matrix,  $a,b,c= 0,1,2,3$ and let us write $\gamma^{ab}= [\gamma^a , \gamma^b]$.

The matrix theory action is (\ref{m3-1}).   
\f
S = \Tr M^3=  \Tr (A_{aI} A_{bJ} B_{cdK} )  \epsilon^{abcd} f^{IJK} 
+ \Tr (\Phi_{IJ} B_{abI} B_{cdJ} )  \epsilon^{abcd}  \,,
\ff
where

\f
\Phi^{IJ}_{mn}= [ \phi_{mn}]_i^j (T^I T^J)^i_j\,.
\ff
We will also impose the constraint
\f
 [ \Phi_{mn}]_i^i = -\Lambda \delta_{mn}\,,
\label{cosmo}
\ff
where $\Lambda$ is the cosmological constant.

We now can go back to a theory of smooth functions on a four manifold.   To
do this we invert the reduction to matrix variables,
giving us back fields.

Using the compactification trick when we take $N\to \infty$ we have the emergence of a four-torus,
\f
\Tr \rightarrow \int_{T^4} \Tr\,,
\ff

the action in  terms of smooth fields is  
\f
S^{Pl} = \int_{\cal M} B^{AB} \wedge F_{AB} - 
\frac{\Lambda}{3} B^{AB} \wedge B_{AB} -
\frac{1}{2}\Phi_{ABCD} B^{AB} \wedge B^{CD}\,.
\ff

There is of course much more to say about Plebanski's formulation
of general relativity.

We want to show the correspondence to neural networks.  So our
next step is to show that
the equations of motion of Plebanski gravity can  be mapped to the equations of a quadratic neural network -- namely, one with two layers.

To do this we employ the fact that each can be mapped onto a neural
network mode.   This is in particular a two layer network, it is
essentially the one employed in restricted Boltzmann models.

The mapping involves sending fields to $N \times N$  matrices, as in matrix models of gauge theories.  That is, we set all derivatives to zero so that,
\f
F^{AB} \rightarrow  [ A,A ]^{AB}  , \ \ \ \ \ {\cal D} \wedge B^{AB}
\rightarrow  \epsilon^{abcd}  [A_a ,  B_{bc}]^{AB} \,.
\ff

Next we construct a correspondence from the matrix representation of Plebanski to
the degrees of freedom of the neural network.

We make the following identifications:

\begin{itemize}

\item{} \textit{Visible Layer}: The $B$ field,
\f
B= B_{ab}^{AB} P^{a} \wedge P^{b}  \otimes  \tau_{AB}\,,
\ff
or, equivalently, the frame field, $e^{AA'}$, such that
\f
B^{AB} = e^{(A}_{B'} \wedge e^{B^\prime B)}\,.
\ff
Note that the last relation is a quadratic equation.   Solving
and inverting quadratic matrix equations play the role here of the gradient flows in an RBM.

\item{}\textit{Hidden Layer}:    the left handed SU(2) connection matrix is represented,
\f
A_a^{AB} (x)   \rightarrow    A =  A_{a}^{AB} P^{a} \otimes  \tau_{AB} \otimes SU(N)\,.
\ff
Note $\tau^{AB}=\tau^{BA}$ are the SU(2)  Pauli matrices, $P^a$ are the translation generators in $R^3$. $N$ is taken very large; the QFT is claimed to be
reproduced in the limit $N\to \infty$

The visible layer is connected to the hidden layer by the weights $W$,   which are maps from
$ Sym \otimes (0,1)$ to $ Sym \otimes (0,1)$.   We now call them $\Phi_{ABCD}$, which 
are valued in the spin-two representation, so they are completely symmetric in spinor indices,
\f
\Phi^{[AB]CD}=0\,.
\ff

\end{itemize}
The dynamics consists of a sequence of a ``forward pass", a ``backwards pass",
each followed by an extraction of a matrix square root (rather than a gradient flow).  We note that while RBM training can involve a non-zero temperature, which allows for stochastic training, here we consider the deterministic $T=0$ variant.  For the time being the two biases are
turned off (these are the currents).   The cosmological constant $\Lambda$ could
be regarded as playing the role of a bias.

\noindent We now iterate the following steps, which are parallel to the
steps that define the neural network model.

\begin{enumerate}

\item{}{\bf Initialization}:  We set $n=1$ and choose initial guesses for
$\Phi^{AB}_{(1) \ CD}$ and $e^{AA'}_{(1)}$. 
Notice that the latter is equivalent to choosing the metric
\f
g_{\mu \nu} = e^{AA'}_\mu \otimes e_{\nu AA'}\,,
\ff
or self-dual two form  subject to constraints~\eqref{eq:bconstraint},
\f 
B^{AB} = e^{(A}_{B'} \wedge e^{B) A' }\,.
\ff

\item {\bf Forward Pass}: We compute $ F^{AB}_{(n)}  $ by
\f
  F^{AB}_{(n)} \rightarrow  F^{AB}_{(n+1)} =
   \Phi^{ AB}_{(n) \ \ CD}  B^{CD}_{(n)}
-\frac{\Lambda}{3}  B^{AB}\,.
\ff

\item{} {\bf First Inversion}:
Update $A$ 
\f
A_{(n)} \rightarrow A_{(n+1)}\,.
\ff

using the new $F_{(n+1)}$ to satisfy the two constraints:

\f
F^{AB}_{(n+1)} = [ A_{(n+1)},A_{(n+1)}   ]_{(AB)}\,,
\ff

\f
\epsilon^{abc} [A_a^{(n+1)} , B^{bc}_{(n)}] = 0 \,,
\ff
holding $B_{(n)}$ fixed.

Next, update the $ \Phi_{BCDE}$, 
\f 
\Phi_{BCDE}^{(n)} \rightarrow \Phi_{BCDE}^{(n+1)}=0\,,
\ff

holding $A_{(n+1)}$
fixed
\f
[A_{E (A}^{(n+1)} \Phi_{BCD)}^{E (n+1)} ] =0\,.
\ff

\item{} \textbf{Backward Pass}:

Finally, we update the self-dual two form, $B^{AB}$,

\f 
B^{AB}_{(n)} \rightarrow  B^{AB}_{(n+1)}\,,
\ff

by setting

\f
B_{(n+1)}^{AB} = \Phi^{-1 AB CD }  \left ( [A_{(2)},A_{(2)} ]^{(CD)}      
 + \frac{\Lambda}{3} \delta^{ AB}_{ \ \ CD}   \right ) \,,   
\ff
together with the quadratic constraint
\f 
B_{(n+1)}^{(AB} \wedge B^{CD)} =0\,.
\ff

At this point we have updated each of the fields once
\f 
(A^{AB}_{(n)} ,  B^{AB}_{(n)}, \Phi^{ABCD}_{(n)}    )   
\rightarrow (A^{AB}_{(n+1)} ,  B^{AB}_{(n+1)}, \Phi^{ABCD}_{(n+1)}    )\,.
\ff

We next impose a test of closure on these matrix fields. We may go around this loop however many times it takes to converge to a 
solution which is when
\f 
(A^{AB}_{(n+1)} ,  B^{AB}_{(n+1)}, \Phi^{ABCD}_{(n+1)}    )   
\approx  (A^{AB}_{(n)} ,  B^{AB}_{(n)}, \Phi^{ABCD}_{(n)}    )\,.
\ff

\item{}\textbf{Second Inversion}:
Finally we update the frame field, $e^{AA\prime}$ by solving for 
\f
e^{AA\prime}_{(1))} \rightarrow e^{AA\prime}_{(2)} =  B^{AB}_{(2)}  = e^{(A}_{B' (2) } \wedge e^{B^\prime B)}_{(2)}\,.
\ff
\end{enumerate}

\section{Cubic learning systems}

We now define a new class of intelligent machines called
cubic learning systems; these have three layers in total, one
external and two internal.  We find direct correspondences to a
set of cubic matrix models, and through the latter's  correspondence with gauge and gravitational field theories,   to those theories as well.

Related to cubic matrix model is the idea of triality and quantum reference frames. This leads from a principle of background independence in which our fundamental theories should be formulated independently of backgrounds they are expanded around. Furthermore, the backgrounds themselves should be subjected to dynamics and not be fixed. Consider now the Born duality $x \rightarrow p$, $p \rightarrow -x$. What is clear is that in this case time can be considered rigid, independent of the $x$ and $p$ and generally non-dynamical. Consider now,  instead of a duality, a triality between position, momentum and time. Concretely, the standard Poisson bracket is now replaced by a triple product. One can make a connection to the cubic matrix model by promoting $x,~p,~\frac{d}{d t}$ to large matrices from which you can get out the symplectic structure.

For a large number of degrees of freedom, you can get out dynamics of Heisenberg operators. In the Heisenberg picture you are in a dynamical picture of the operators - these are matrices where matrix elements can be thought of as hidden variables. One may translate the Born rule in quantum mechanics $P = |\langle \alpha | \beta \rangle|^2$ to the form $P = \Tr(A B I)$ where $A_a^b = \alpha_a \alpha^{\dagger b}$, $B_a^b = \beta_a \beta^{\dagger b}$ are the density matrices corresponding to states $\alpha$, $\beta$ and $I$ is the identity matrix. In the context of the cubic matrix model, we can think of reinterpreting the concept of triality as a statement for the laws. For example, consider the matrices $A$, $B$, \& $C$ which we can think of as gauge fields. The dynamics for this theory without fermions is given by,
\begin{equation}
    \dot{A} = [B,C]\,.
    \label{dynCMM}
\end{equation}
An interesting question to ask is whether the quantum dynamics of the matrix model could be realized in the context of a machine learning algorithm.

Inspired by this rich space of theories, we formulate in this section a learning architecture for the general class of cubic matrix models. An autodidactic system then has the ability to learn its laws by exploring the space of effective theories using moves which are (de-)compactifications, dimensional reductions and expansions, symmetry breaking and restorations, and possibly others.  We begin by introducing general architectures for neural networks, including some examples which include consequencers in the form of memory modules. We then describe the precise relationships between discrete topologies, learning systems, and gauge field theories, after which we propose a set of axioms for the architecture and dynamics for cubic learning systems. We show that cubic learning systems provide a concrete bridge between matrix descriptions of topologies and gauge fields, which will be a critical step used in the next section, where we explore a handful of examples of autodidactic dynamics.

\subsection{Learning architecture of cubic matrix models}

We seek an evolution rule for the weights of a neural network that can correspond to the dynamics of a gauge theory, which in turn can evolve in time in a way that would reflect nature learning its laws.    

The matrices, or weights, are represented by an $N \times N$ matrix, with $N$ taken very large.   These may be viewed either as matrices of weights or as adjacency matrices which specify a large decorated graph.     

The dynamical law, whether evolving or static, generates a sequence
of such matrices,
\f
\{X_1, X_2, X_3,\ldots,\}\,,
\label{m3-1}
\ff
which represents their evolution in time.
\\[0.2cm]
\noindent As shown in~\cite{unifysl}, a large class of these can be realized with the following assumptions:

 \begin{enumerate}
 
 \item{} The evolution rule should mimic second-order differential equations, as higher order equations can breed instabilities.  Moreover, no higher than second derivatives appear in any of the field or particle theories we know. So two initial conditions should be required to generate the evolution.  We should then need to specify $X_0$ and $X_1$ to generate the full sequence.  Therefore, we are interested in rules of the form,
 \f
 X_n = {\cal F}(X_{n-1}, X_{n-2})\,.
 \label{m3-2}
 \ff
\item{}  For massive matter fields, the  changes should be small from matrix to matrix, at least given suitable initial conditions.  This is needed so that there can be a long timescale on which some of the information in the matrices is slowly varying.  This makes it possible to extract a notion of slowly varying law acting on a faster varying state.  We will ask that for matrix representation of matter fields,
 \f
 X = {\cal F}(X, X )\,.
 \label{m3-3}
 \ff
For gauge fields, however, this is ruled out by the translation gauge invariance 
 (\ref{inhomogauge}).
 
\item{} We require that the evolution rule be non-linear, because non-linear laws are needed to encode interactions in physics.  But we can always use the basic trick of matrix models of introducing auxiliary variables, through the use of repeated instances of the tensor product decomposition to expand the matrix, in order to lower the degree of non-linearity.   As we take $N$ larger and larger, there is always room for more. 
This accords with the fact that the field equations of general relativity and Yang-Mills theory can, by the use of auxiliary variables, be expressed as 
quadratic equations\footnote{As in the Plebanski action, for instance.}~\cite{unifysl}. The simplest non-linear evolution rule will then suffice, so we require a {\it quadratic evolution rule.}  

\item{}The basic theory should have a big global symmetry group, $G$, that can be spontaneously broken in various ways to reveal different fundamental theories.
Our theory will then unify gauge theories with diverse continuous and discrete symmetries~\cite{unifysl}.

\end{enumerate}

A simple evolution rule that realizes our desired dynamics is 
 \f
 \boxed{
 X_n = X_{n-1}  +  [ X_{n-1}, X_{n-2}]\,. }
 \label{e1}
 \ff
 This rule is not unique, but it is nearly so.  It is easy to derive the general rule satisfying the four requirements just mentioned.
 
 On the other hand, if we drop the translation symmetry (13), we find a more
 restricted solution that contains just the commutator term,
 \f
 \boxed{
 X_n =     [ X_{n-1}, X_{n-2}]\,. }
 \label{e3}
 \ff

As each move has only a single output and two inputs, we need no more
 than three matrices at once.
We relabel the time counter $n$ to
be 
\f
n\to\left\lfloor\frac{n}{3}\right\rfloor\,,
\ff
so that the set of three matrices form the sequence,
\f
\{A_1, B_1, C_1, A_2, B_2, C_2, A_3, \ldots \}\,.
\ff
Then there is a simple evolution rule that preserves the forgoing as well as the permutation group on three elements:
 \f
 \boxed{
 A_{n+1}= A_n + [B_n , C_n]\,.}
 \label{e2}
 \ff
 and cyclic,
 \begin{eqnarray}
 A_{n+1}= A_n + [B_n , C_n]  \,, 
 \nonumber \\
  B_{n+1}= B_n + [C_n , A_n]   \,,
  \nonumber \\
   C_{n+1}= C_n + [A_n , B_n]   \,. 
 \end{eqnarray}
Again, in the case that we also impose the gauge/translation invariance conditions,
 we find
  \begin{eqnarray}
 A_{n+1}=  [B_n , C_n] \,,  
 \nonumber \\
  B_{n+1}=  [C_n , A_n]   \,,
  \nonumber \\
   C_{n+1}=  [A_n , B_n]    \,.
 \end{eqnarray}

 Thus the various restrictions we have imposed have led to a small
 class of matrix theories.  The degrees of freedom are three 
 $N \times N$ matrices,   
 \f
 A_n \,, B_n\,  ,   C_n\,,
 \ff
 where $n$ represents a discrete time.
 The {\it matter fields} are in some representation of
 $SO(N)$. 
 The possible terms that are invariant,
  to leading order,  under the full set of gauge invariances,
 make up a simple Hamiltonian:
 \f
 {\cal H} =  \sum_n   \left ( \Tr  ( A_n B_n C_n  ) + \Phi^{T} X_n \circ \Phi 
 \right )\,.
 \label{cmm}
 \ff
 
 This simplest possible matrix model contains (by virtue of various
 compactifications, symmetry breaking, and so forth) most of the theories
 of interest for fundamental physics~\cite{unifysl}. 
 Concretely, one can obtain different physical theories from~\eqref{cmm}
 using a combination of tensor product decompositions, $M = A \times B$, and  circle compactifications realized by the substitution $A = [i \partial_{\theta}] + [a(\theta)]$. Notice that~\eqref{cmm} does not include any spatial or temporal dependence, thus this step serves as a means to introduce dimension. 
 
 This is our Rosetta stone.
 Next we demonstrate how to translate these, plus a specification of the cosmological
 setting, into a learning system.

\subsection{Dynamics of cubic learning systems}
\label{sec:dyn_cls}

We can now sketch the architecture of a type of recurrent neural network that captures 
the dynamics of the cubic matrix models.

Our model contains three layers, each represented by an $N$-dimensional vector, 
\f
Z_n^a = (a_n , b_n , c_n )\,,
\ff
where $n = 1,2,,3 , \ldots $ is an integer valued time variable (the clock).
The three layers are arranged in a loop as in Fig.~\ref{fig:abc_cycle}. 

\begin{figure}[h!]
    \centering
    \includegraphics[scale=0.5]{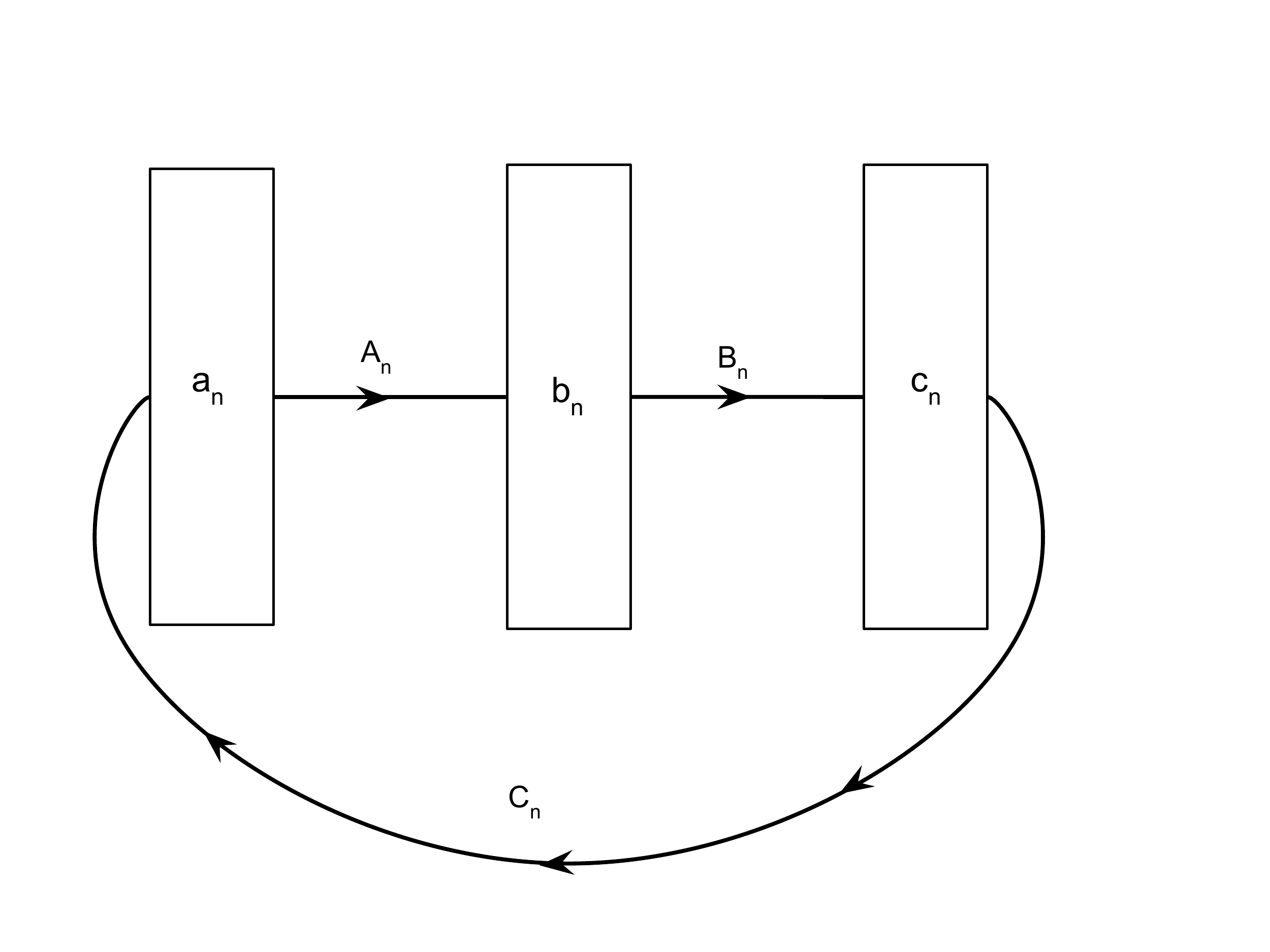}
    \caption{Cycle symbolizing the relations~\eqref{eq:abc_cycle},~\eqref{eq:abc_cycle2} }
    \label{fig:abc_cycle}
\end{figure}

There are also three weight matrices. As discussed above these form or generate
a group, which we will take to be SO(N).  We choose to represent them by
 their Lie  algebra generators. 
 \f
 L^i_n = (A_n , B_n , C_n )\,,
 \ff
 which are antisymmetric matrices.  At each time $n$ we carry out the following
 update rules:
 \begin{eqnarray}
 \label{eq:abc_cycle}
 A_{n+1} =  [B_n, C_n ] + b^T_n a_n  \ \ \mbox{and  \  cyclic}\,,
 \\
 \label{eq:abc_cycle2}
 b_{n+1} = e^{A_n} \circ a_n   \ \mbox{and  \  cyclic}\,.
\end{eqnarray}

If we want to thermalize to learn, then every three moves, which corresponds to
one time round the loop, we apply one of the thermalization algorithms to 
induce
\f
Z_n^a = (a_n , b_n , c_n ) \sim
Z_{n+3}^a = (a_{n+3}, b_{n+3} , c_{n+3} )\,.
\ff
We  do this by thermalizing with  respect to a Hamiltonian
(actually, just the potential energy; there is no kinetic energy in  this
quenched version),
\f
{\cal H}_n = \Tr e^{A} e^{B} e^{C} 
+ b_{n+1}^T e^{A_n} \circ a_n
+ c_{n+1}^T e^{B_n} \circ b_n
+ a_{n+1}^T e^{C_n} \circ c_n\,.
\ff

We may want to study a weak field expansion, so we scale by a small
dimensionless coupling constant $g$, to find, to leading order in
$g^2$,

\f
{\cal H}_n = g \Tr (A_n B_n C_n  )
+ b_{n+1}^T [ 1+  g{A_n} \circ  ] \circ a_n
+ c_{n+1}^T  [ 1+  g{B_n} ] \circ   a_n
+ a_{n+1}^T [1+  g{C_n} ]  \circ c_n\,.
\label{eq:cls}
\ff

\subsection{The correspondence}

We are now in a position to claim the following correspondence result.

We specify a solution to a cubic matrix model by the following parameters

\begin{itemize}

\item{} Pick a gauge symmetry, ${\cal H}$ to be the gauge symmetry after the first level of spontaneous symmetry breaking.  For example, to incorporate the standard model through an extended Plebanski~\cite{CITE1,CITE2}  embedding, we must pick
\f
{\cal H} = SU_L (2) \otimes SU(3) \otimes SU(2) \otimes U(1) \otimes O(M)
\ff

The first $SU_L (2)$ gives the chiral half of the local Lorentz gauge 
invariance, while the final $M$ dimensions code the translations around a  set of $D$ tori.   Here  the radius of each torus $R_i$ is
given by
\f
R_i = M_i l, \ \ \   \sum_{i=1}{D} R_i l = M \,.
\ff

\item{}  Pick a large, but finite  $N$ such that
\f
{\cal H} \subset SO(N)
\ff
    \item     Choose the topology, such as 
    \f
    [S^1]^x \otimes [S^2]^y
    \ff  This leaves a global symmetry, $\cal R$.
    
    \item{}   We break the SO(N)  symmetry to $local {\cal H}$ and global $\cal R$  by finding a classical solution of the following form that
    preserves them.
    
    The background geometry is specified by the three matrices
    $A_0, B_0, C_0$ such that
    \f
    [A_0, B_0]=0, \ \   
    [B_0, C_0]=0, \ \      [A_0, C_0]=0 \,.\label{eq:threematrices}
    \ff

\end{itemize}

We next expand around the solution~\eqref{eq:threematrices}
\f
A= A_0 + \alpha   \ \ B= B_0 + \beta   \ \  C= C_0\,.
\ff

The perturbations satisfy
\f
[A_0, \beta] =0   \ \ [C_0, \alpha] = \beta \ \  [C_0, \beta] = -\alpha\,.
\ff
These describe a regulated gauge-gravity theory, on a closed $4+D$ dimensional space-time.
with spatial topology fixed.

We showed in the previous sub-section that the same data proscribes
a cubic learning system.    Thus we have a {\it correspondence}
between the three theories.   In particular the correspondence
connects a learning machine of the class defined in Sec.~\ref{sec:dyn_cls}
with a gauge invariant QFT.

Before we close this section we should also make it clear
what we {\it do not have.}

\begin{itemize}
    \item{} We do not have an equivalence.  The gauge and gravity theories are only complete in the limit $N \rightarrow \infty$.   But
    we don't have a definition of the learning architecture in that limit.
    
    \item{} The learning machines employ thermal effects to learn effectively in which the degrees of freedom are heated and then
    quenched.  We conjecture that would employ embedding our learning
    systems in cosmological models.  But we have yet to work
    out the details of this.

\end{itemize}

\subsection{Cubic matrix model and Chern-Simons theory}

The cubic matrix model may be reformulated as follows.

Define the metamatrix

\f
M =\tau^a  \otimes A_a\,,
\ff
where $a=1,2,3$, $\tau^a$ are the three Pauli matrices and the three
$A_a$ are $(N-2) \times (N-2)$ dimensional matrices.

The action is then

\f
S= Tr M^3 = \epsilon^{abc} Tr_r A_a A_b  A_c \,,
\ff
and the equations of motion are
\f
[A_a , A_b ] =0\,,
\label{[]=0}
\ff

We specify a solution to a cubic matrix model by the following parameters

\begin{itemize}

\item{} Pick a gauge symmetry, ${\cal G}$.

\item{}Pick a three dimensional topology, the simplest is the
three torus, $T^3$.

\item{}Pick three matrices, $D_a $ that solve for every $N$ the equations of motion  (\ref{[]=0}).

\item{} Expand around each solution
\f
A_a = D_a+ a_a\,.
\ff

\end{itemize}

In the limit $N \rightarrow \infty $  the $D_a$ act as derivatives and the theory that
emerges is Chern-Simons theory
\f
S^{\infty} = \int_{T^3}   \Tr  ( a \wedge da + a \wedge a \wedge a )
\ff

One reason that this is interesting is that Chern-Simons theory regarded as a functional of embedded Wilson loops
provides a class of knot and graph invariants.  The connections we have sketched here suggest that machine
learning may offer a powerful tool to creating and evaluating knot invariants\footnote{For other 
approaches see also~\cite{lovesknots}  .}

Finally, we need to put the matter degrees of freedom $\Psi$ back into the picture.

We will represent matter by an $SU(2)$ spinor  valued matrix $\Psi_A$,   and dual spinor valued matrix,    
with $A= \{ 0,1), $ 

We write the full matrix Chern-Simons action
\f
S^{MCS} = \epsilon^{abc} Tr_r A_a A_b  A_c    + Tr \Psi^\dagger_{A } [M_a , \Psi_{ B} ] \sigma^{a AB} \,,
\ff

This matrix theory is well defined.  When we take a limit to infinite dimensional matrices this goes to the full continuum Chern-Simons
theory, 
\f
S^{\infty} = \int_{T^3}   \Tr  ( a \wedge da + a \wedge a \wedge a )].
\label{Sinf}
\ff

The most important thing to understand is the influence of a fixed background 
metric, $\eta$
\f
Tr \sigma^a \sigma^b = \eta^{ab}
\ff
As a result the Chern-Simons theory, including coupling to a spinor field is no longer topological.  

We can read off the neural network model from the form of
(\ref{Sinf} ); a diagram of it is shown in Figure 1.

The screen degrees of freedom are distributed as follows.
On the $\frac{a}{mod 3 } $ screen we have the variables
\f
\Psi_A = \sigma^{a}_{\ AB}   \Psi_a^{B} 
\ff
which are parallel transported to the next screen 
($\frac{a}{mod 3 } + 1 $) by taking a commutator with $A_a$.

\section{Protocols for autodidactic systems}
\label{sec:protocols}

We set out to explore whether the notion of a \textit{self-learning system} could be relevant to fundamental physics. Namely, we are interested in systems for which the rules governing time evolution are partly \textit{learned} from the features of the explored configuration space.

We have argued that the idea of laws that learn gives us a powerful framework for going beyond the earlier 
concept~\cite{LOTC,evolve} of laws that evolve in time.

When we speak concretely of laws that learn,
we realize that the usually strict lines between laws, theories, states, and solutions of theories seem to break down~\cite{unifysl}.

How would we recognize such a system?  One necessary, but not sufficient requirement is that the
late-time behavior of such systems will be highly sensitive to the initial conditions and early-time dynamics. 
Another is that the dynamics includes a feature we called a consequencer.  This may be a simple feedback loop or a highly elaborate set of hidden variables in a recursive network.

In this section we discuss several protocols for autodidactic systems, as well as some preliminary computational experiments. A protocol defines a dynamic architecture with rules for change that lend themselves to interpretations as learning strategies or reward functions. Generally speaking, the autodidactic paradigm suggests none of these may be fixed a priori, since one history guided by a set of rules might transform itself to be guided by altered rules, though in practice it is often helpful to limit the number of ways in which the system may evolve.  Autodidactic systems are distinct, however, from emergent cosmological models that apply a priori growth rules consistently, as in~\cite{wolfram}.

We describe experiments in which small autodidactic systems are allowed to develop in simulation so that we can observe emergent properties.  In all experiments, the free variables that describe the systems are expressed in terms of matrices.

Therefore, they can be considered in the context of the correspondence demonstrated in previous sections.  Compatibility with the correspondence leaves open a number of possible interpretations for how autodidactic systems, such as the ones in the experiments to follow, should be interpreted in the context of physics.

In the examples previously explored, there was an explicitly stated way that a law could also be a part of a learning process, and part of the history of a universe, but in some of the examples to come it might be argued that there is more room for interpretation.  We will do less to explicitly tie the models in this section to laws.

For instance, when we see structures that resemble deep learning architectures emerge in simple autodidactic systems (as shown in Figure 9) might we imagine that the operative matrix architecture in which our universe evolves laws, itself evolved from an autodidactic system that arose from the most minimal possible starting conditions?  That notion correlates with a more complicated hypothesis of histories for the early universe.  An alternative is to suppose that a matrix structure substantial enough to support law evolution was part of the starting conditions.  

Our goal in this section is not to express preference for one story or ontology over another, but simply to observe how emergent properties in small autodidactic models display properties that are relevant to learning universe ideas.

We have already discussed two protocols whose degrees of freedom can be put in correspondence with those of a gauge or gravitational theory.  We were able to find a formulation of general relativity and put it in correspondence with a two layer neural network.  We then invented a class of three layer neural networks and put them in correspondence with Chern-Simons theories, which are topological quantum field theories.

We will continue in Sec.~\ref{sec:rg_learning} by discussing learning using the renormalization group. Using either the renormalization group neural network architecture or the RBM architecture, one can construct a learning algorithm which attempts to maximize the mutual information between the learning system and its environment, without restriction to any particular learning strategy. This approach is inspired by information theory and is motivated by the 
Wilsonian picture according to which quantum field theories which have good ultraviolet completions do so because their
high energy behaviour is dominated by an asymptotic scaling governed by the renormalization group.

We then discuss in Sec.~\ref{sec:precedence} another protocol which uses the RBM architecture with a novel learning strategy: the Principle of Precedence. Precedence describes an optimization technique in which the future behavior of a system depends not just on its cost function, but also on its prior set of states. Precedence can be implemented in a number of ways -- as an attention layer, a memory module such as the LSTM or GRU, a set of hidden variables, etc. We give an example of the hidden variable version using an RBM and then provide an example of a continuum limit of a learning process.

The fourth method we study in Sec.~\ref{sec:self-directed-graphs} likewise uses self-attention via self-sampling procedures. While the Principle of Precedence uses a measure on prior states, self-sampling is used to grow graphs, so that prior states are described by subgraphs and the measure on the prior is encoded in the measure on subgraphs. This compact representation is particularly useful when we want to model a growing discrete system using a recurrent learning architecture.

Another protocol which uses a recurrent learning structure is described in Sec.~\ref{sec:variety}. We introduce a cost function inspired by graph theory and quantum foundations called the \textit{variety}, which is, loosely speaking, a measure of topological heterogeneity. We use a learning strategy based on simulated annealing to generate graphs which maximize variety and show these graphs are distinct from the set of random graphs.

Finally, in Sec.~\ref{sec:geometry} our last protocol describes another recurrent learning system that uses annealing. We consider the components required for a discrete system to self-assemble into a discrete manifold, either a simplicial manifold or a random geometric graph. Because it is not clear a priori whether geometry is bound to emerge from a pre-geometric system, we mainly focus on a learning procedure in which the system optimizes the parameters of its cost function in order to generate a geometric system.

An autodidactic protocol must give rise to a consequencer, as defined earlier; a reservoir functioning as an accumulator of information in a learning process.  Here is how consequencers can form in the above examples:

\begin{itemize}
    \item In the RG section:  Renormalization is by definition a consequencer generator provided it is part of a feedback dynamic.  If renormalizations are relevant to the ongoing evolution of a system, then that system is driven by an exemplary consequencer.  Renormalization in that case partitions the most causally relevant features of the system.

In the Precedence section:  Precedence is also by definition a consequencer generator; that is the very notion.  \item
In the Self-sampling section:  The example of the “persistent hub” shows how a consequencer emerges in self-sampling autodidactic systems.  \item
In the Variety section:  In this case, the consequencer emerges negatively, as a progressive refinement of adjacent configurations that have not yet appeared.  \item
In the Geometric Self-assembly section:  Here the  self-attention or precedence function is a form of consequencer, but so is the evolving optimization of the annealing schedule.  

\end{itemize}

The consequencer is no more and no less than the information that must change when learning occurs.

We defined "learning" earlier in physical terms. Learning includes {\it adaptive processes that become anticipatory, doing "more than they need to" based on any isolated instance of feedback} or perhaps we can choose a definition that \textit{maximizes the causal impact of an adaptive subsystem}.  As stated earlier, there need not be an abstractable ``thing learned"  - anything symbolic or semantic - for a system to learn something, even according to a more casual and causal sense of information, as is often attributed to Gregory Bateson, as in the already quoted phrase, 
{\it a difference that makes a difference}.  \footnote{We will not wade into the question of what should or should not be properly attributed to Bateson, but we will speak of Batesonian information here since that is almost a common usage.  Alternative terminologies have been proposed ~\cite{Gordian}.}

This interpretation of learning in the Batesonian sense was explored in~\cite{cardenas2020process} as \textit{info-autopoiesis}, which describes how information about a system may be created by the system itself. Rather than subscribing to the notion that information exists outside matter and energy, the Batesonian concept of learning is that information is described by \textit{differences} in matter and energy variables; hence, recursive dynamics characterize a fundamental mechanism for an autodidactic system to learn its laws and self-organize accordingly.

The idea of Batesonian information is similar to, but not identical to the idea of a consequencer.  A "lucky" cosmic ray that alters a gene is Batesonian, but not part of a consequencer.  Consequencers persist as structures in time even when physical components are replaced, and are not random, while Batesonian events can be singular and random.  However, a consequencer is made of Batesonian information.

\subsection{Renormalization group learning}
\label{sec:rg_learning}

A key aspect in many of the scenarios for emergent growth rules is the identification of relevant degrees of freedom, which may serve as a basis for such rules. One of the idea threads stretching through our work is the notion that dynamical rules can be found implicit in the properties of some substructure, which turns out to be both influential and persistent. For example, at the end of  Sec.~\ref{sec:self-directed-graphs}, we have specified a number of proposals for growth using self-sampling, which we hope may realize this idea.

This topic sits on the interface between physics and machine learning, where important progress has been made in both directions. Let us briefly review some of the key ideas available in the 
literature\footnote{For another approach to the relationship
between deep learning and the RG see~\cite{Kochchen}
Our view is that it is that and much more.}

\subsubsection{Review of basic concepts}
\label{sec:basicrg}

The position-space renormalization group (RG) procedure of Kadanoff~\cite{kadanoff_rg} works by rewriting a model in terms of coarse-grained degrees of freedom. In principle, any possible coarse-graining can be chosen; the RG transformation will result in an effective Hamiltonian, which will in general contain all possible terms allowed by the symmetry of the problem, along with scale-dependent coefficients (coupling constants). The ``right" coarse-grained degrees of freedom are ones which result in a simple Hamiltonian with a finite set of relevant couplings; for example, in the Ising model we start out with only nearest-neighbor interactions, and we want to preserve the locality in the renormalized Hamiltonian. 

In~\cite{mutual_information,optimal_rg_information}, the authors propose how one can formulate this procedure in a way that can be translated into a machine learning algorithm. They start by promoting all physical degrees of freedom to random variables. As we discuss below, some of this work~\cite{original} is inspired by tensor network methods for representing quantum states such as MERA~\cite{mera}; when trying to apply a quantum mechanical method to classical physics it is natural that we end up working with probability distributions. In~\cite{mutual_information}, the system is partitioned into a subsystem $\mathcal{V}$, from which we want to capture several coarse-grained degrees of freedom, and the ``environment" $\mathcal{E}$, which is the complement\footnote{In practice, one can also partition the environment $\mathcal{E}$ into the ``buffer" $\mathcal{B}$, which contains the degrees of freedom closest to $\mathcal{V}$, and the remainder.}. In the usual real-space renormalization group story for, say, the 2D Ising model, $\mathcal{V}$ corresponds to square subsets that are coarse-grained. Then, the ``best" choice of coarse-grained variables $\mathcal{H}$ are those functions of degrees of freedom in $\mathcal{E}$ which maximize the mutual information between variables in $\mathcal{H}$ and variables in $\mathcal{E}$. 

The mutual information between two probability distributions $P_X$ and $P_Y$ is defined as

\begin{equation}
\label{eq:mutualinf}
    I(X:Y) = \sum_{x \in X} \sum_{y \in Y} P_{X,Y}(x,y) \log \left(\frac{P_{X,Y}(x,y)}{P_X(x) P_Y(y)} \right)\,,
\end{equation}

where $P_{X,Y}$ refers to the joint probability distribution. This is an information-theoretic measure which characterizes the uncertainty of a variable sampled from one distribution given a variable from the other. Somewhat vaguely, it can also be thought of as a generalized measure of correlation between variables $X$ and $Y$ that can also capture nonlinear dependencies. In the context of the renormalization group, the best choice of coarse-grained variables of a given cell are those which are maximally correlated with the rest of the system. In~\cite{mutual_information}, the renormalization group transformation is implemented by a kind of restricted Boltzmann machine whose latent variables correspond to coarse-grainings, such that~(\ref{eq:mutualinf}) is minimized. 

When working with probability distributions, it will be useful to remember how mappings between them work. Given a bijection between the two sets of random variables $f : X \rightarrow Y$, then for $y = f(x)$ we have

\begin{equation}
    P_X(x) = P_Y(f(x)) \left|\det\left(\frac{\partial f}{\partial x}\right)\right|   \,.
\end{equation}

\begin{figure}[t]
\centering
\includegraphics[scale=0.8]{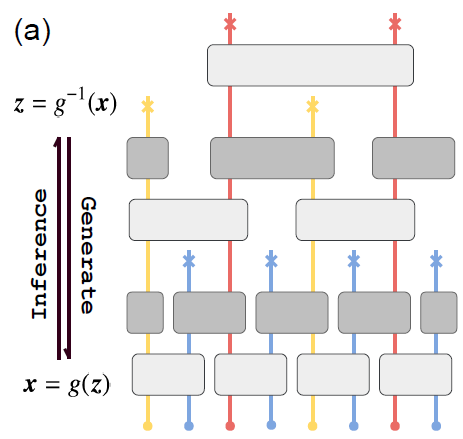}
\caption{Neural Network Implementing an Exact Renormalization Group. Courtesy of \cite{neural_rg}.}
\label{fig:fig1}
\end{figure}

A class of real-valued non-volume preserving flows ("normalizing flow") modules parametrizing bijections between distributions have a particularly tractable computation of Jacobians~\cite{nvp}. They are used in~\cite{neural_rg} to build a MERA-inspired neural network of the form shown in Fig.~\ref{fig:fig1}, where blocks correspond to NVP modules, and crosses correspond to latent variables that contain the coarse-grained degrees of freedom at different scales. Note that one novelty here is that the decimated degrees of freedom are also kept, and since all operations are invertible, one may also perform the inverse of the RG transformation.

\subsubsection{Variational RG}

In~\cite{variational_rg}, the authors propose that the action of the variational renormalization group of Kadanoff can be identified with a class of restricted Boltzmann machines. In the variational renormalization group, a transformation between the fine-grained spins $\{ v_i \}$ and the coarse-grained variables $\{ h_i \}$ is given by a function $T_{\lambda}(\{ v_i \}, \{ h_i \})$, with a number of adjustable parameters $\lambda$. The renormalized Hamiltonian $H_{\lambda}(\{ h_j \})$ is given by

\begin{equation}
    e^{- H_{\lambda}(\{h_j\})} \equiv \Tr_{v_i} e^{T_{\lambda}(\{v_i\},\{h_j\}) - H(\{v_i\})}\,.
\end{equation}

The original free energy is given by

\begin{equation}
    F = - \log Z = - \log \left( \Tr_{v_i} e^{- H(\{ v_i \})} \right)\,,
\end{equation}

while the renormalized free energy is given by

\begin{equation}
    F_{\lambda} = -\log Z_{\lambda} = - \log \left( \Tr_{h_i} e^{- H_{\lambda}(\{ h_i \})} \right)\,.
\end{equation}

The parameters $\lambda$ are then chosen to minimize the difference between the fine and coarse free energies:

\begin{equation}
    \Delta F = F_{\lambda} - F\,.
\end{equation}

Boltzmann machines start with a coupling between the visible neurons $\{v_i\}$ and the hidden neurons $\{h_i\}$, of the form (see section \ref{sec:nnarch} for details) 

\begin{equation}
    E(\{v_i\},\{h_i\}) = \sum_j b_j h_j + \sum_{i j} v_i w_{i j} h_j + \sum_i c_i v_i\,,
\end{equation}

for some set of "weights" $b_i$, $c_i$, $w_{i j}$. Then, the probability of finding the system in a state given by a set of values for $v_i$ and $h_j$ is

\begin{equation}
    p = \frac{e^{- E(\{ v_i \}, \{ h_j \})}}{Z}\,.
\end{equation}

In order to make contact between this restricted Boltzmann machine and variational renormalization group, visible neurons $v_i$ are interpreted as spins in the fine-grained model, and hidden neurons are interpreted as course-grained spins. Weights and biases of the RBM play the role of variational parameters $\lambda$. The renormalization group can then be conceived of as a sequence of successive coarse-graining transformations, which is implemented using a deep neural network. 

\subsubsection{RG-guided graph growth}

We now propose a kind of information-theoretic mechanism to study and refine graph growth. This complements our exploration of different notions of graph growth in Secs.~\ref{sec:self-directed-graphs} and~\ref{sec:variety} where we propose several simple algorithms that we conjecture might lead to the appearance of relevant features which can dictate the course of the growth.
By repeating an algorithm such as these many times, with different starting configurations, we obtain a statistical ensemble of graphs at each step in the growth process. This is a natural starting point for applying the tools of Sec.~\ref{sec:basicrg}; namely, we consider probability distributions over the space of graphs rather than individual graphs. Using methods from persistent homology, we can then identify features most consequential to the growth process by maximizing the mutual information between such features and their environments, i.e., the rest of the graph; this could also be implemented using Restricted Boltzmann Machines, as described in~\cite{mutual_information}.

Note that the process we just described is an ``\textit{inverse renormalization group transformation}''\footnote{Strictly speaking, renormalization group does not have an inverse}; we essentially started with a set of coarse-grained graph features and slowly added fine-grained features. The step of this fine-graining operation thus plays the role of time. Such probabilistic ``inverses'' of RG transformations have been studied in the context of the Ising model using deep Convolutional Neural Networks~\cite{super_resolving}. This work is inspired by the methods used in computer vision for creating image super-resolutions, and it applies similar tools to construct ``fine-grained'' Ising model configurations. These fine-grained configurations turn out to be sampled from related Boltzmann distributions so that they lead to correct thermodynamic quantities.

Stochastic growth processes can be specified by a probability distribution over potential graphs at different stages of the growth. The rules for making each following move then comes from conditional probabilities.

The space of such growth processes is vast. In order to narrow it down, we contemplate two questions:

\begin{itemize}
    \item How do we define a natural set of probabilities which specify a preferred growth process?
    \item Given such a set of probabilities, what do the emergent rules for making moves look like, and how do they depend on the state of the system?
\end{itemize}

Our intuition is that the two questions are intimately related. Namely, a good answer to the first question is one which allows the emergent rules to be easily captured by some highly relevant graph substructures. Note that while the statement that the rules depend on the state of the system is trivial, this requirement is quite constraining. For a particular growth process, such relevant structures may be identified as those which maximize the mutual information with the environment; the right growth process could be identified through the simplicity of relevant structures. Both criteria can be precisely defined and implemented through machine learning.

\subsection{Precedence}
\label{sec:precedence}

Let us consider the question of whether the laws of physics evolve over time.  Might they undergo some form of dynamics that leads to the laws we see today? An interesting proposal for how a {\it dynamics of  laws} may be realized is the {\it Principle of Precedence}~\cite{Smolin:2012mk}.

This idea can be set within an operational formulation of quantum mechanics such as that by
Hardy~\cite{hardy2016operational} or  Masanes and Muller~\cite{Masanes:2010tt}, but it is easy to informally state the general idea.  Quantum theory is envisioned as an example of a more general probabilistic dynamical theory.   A quantum process is described, from an
operational point of view,  as having three stages: 1)  a choice of initial state (i.e., preparation), 2) an evolution within an environment, 
described as an  example of a general framework  for probability preserving evolution, and 3) a final measurement, from which emerges one out of a finite number of answers to a question.

The three stages define a matrix of probabilities, $[P]^a_{\ b}$,  for $N$ inputs to evolve through the environment to yield any of the $N$ possible outcomes.  These probabilities are usually believed to not change in time because they reflect timeless laws.

The Principle of Precedence offers a different explanation for the probabilities
and their time independence.  Given each choice of preparation and measurement,
there is an ensemble of past quantum processes.  The principle says our system must pick out one of those randomly and copy its output. If that ensemble is large enough, the process of evolution via precedent converges to time-independent probabilities.

But what if there is no such past ensemble for a process defined by certain particular inputs and outputs?   This seems to be a question well adopted to investigation via
autodidactic neural networks such as the RBM.   As we now
explain, the notion of precedence seems naturally suited to be realized in a setting such as machine learning.

In this case the system has a clear reservoir of consequence, because  literally each future quantum process  has access to the whole ensemble of past processes with the same initial state and time evolution operator.
The access is through a random sampling of outputs, which is all the system needs to learn from the past.

\paragraph{Hidden layers can represent non-local hidden variables}

One reason precedence comes naturally is that the RBMs have degrees of freedom in their hidden layers that can represent the non-local hidden variables that are needed for
any realist completion of quantum mechanics.
This is shown in the model of Weinstein~\cite{weinstein2017learning}, where an RBM is trained to represent a hidden variable model of an EPR experiment.

One interesting aspect of Weinstein's model is that it violates Bell's inequality, not because it exhibits non-locality (meaning locality in the normal Bell sense, i.e., measurements at detector A should not impact those at B), as neurons of the visible layer are by construction non-local. That is, the RBM is a bipartite graph, but because it violates measurement independence, the distribution for hidden variables, $\rho(\lambda)$, is independent of the set up for the experiment. This violation is manifest from its construction, since the hidden layer of the RBM can be written in terms of the input layer and weight matrix, so $P(\lambda) \rightarrow P(\lambda | \alpha)$ where $\alpha$ represents some model parameters. 

\begin{figure}[!htbp]
    \centering
    \hspace*{2.8cm}\includegraphics[width=0.6\linewidth]{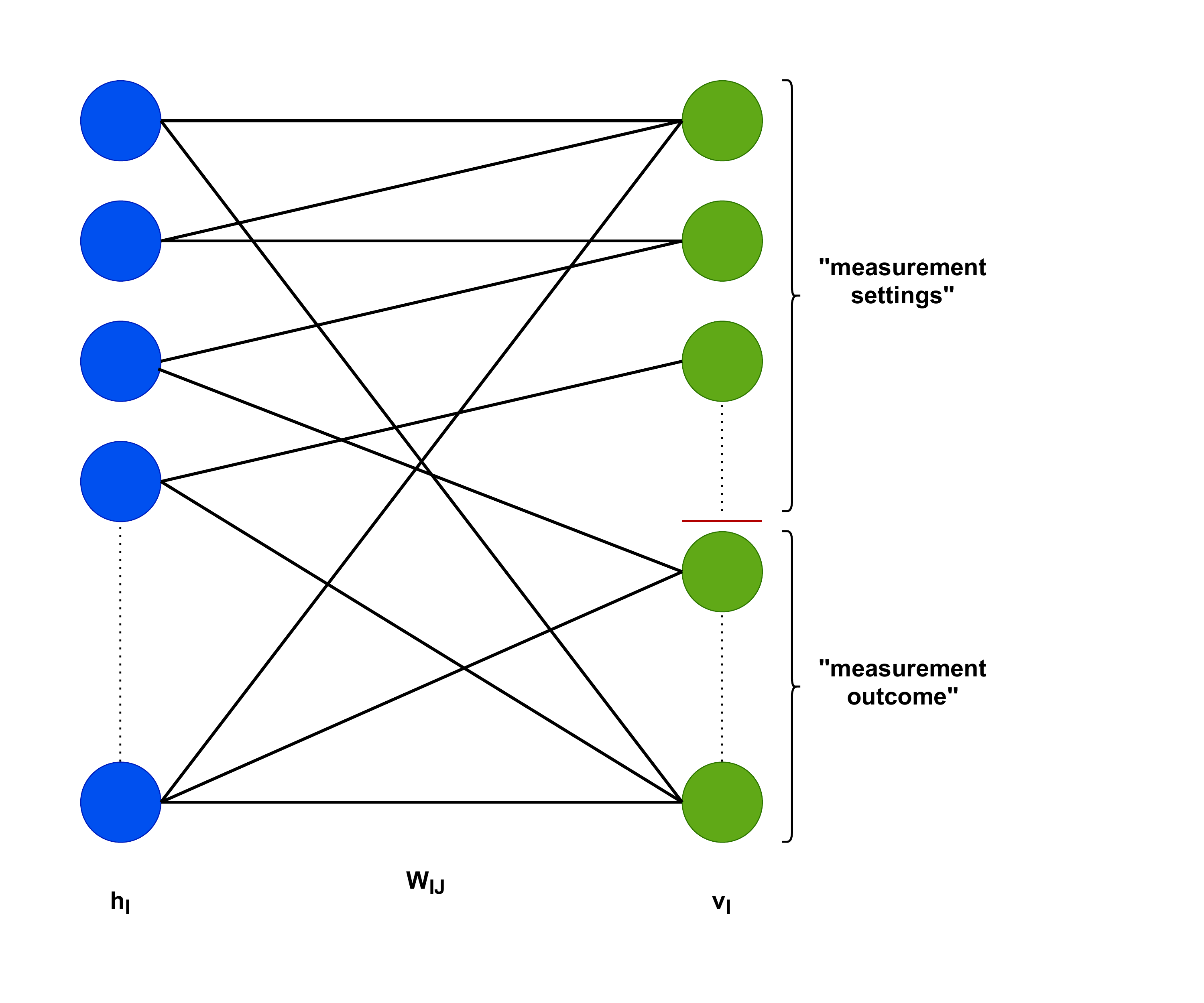}
    \caption{\textbf{Restricted Boltzmann Machines}. The RBM model is used to describe a physical theory whose laws may evolve in time. The architecture consists of two layers of neurons, one visible (green) and one hidden (blue). The visible layer is used both to specify the measurement details and to measure outcomes, while the hidden layer enables non-local connections between the visible nodes. As the training procedure iterates, the weights which connect the two layers update according to the Principle of Precedence until they converge on some limit set. We consider the case where the size of both layers very large, and where there is an equal number of neurons in each of the layers.}
    \label{fig:rbm}
\end{figure}

\subsubsection{Precedence, Weinstein, and machine learning}

The approach of Weinstein is well suited as a starting point for realizing the idea of precedence while encapsulating some of the other axioms for quantum theory. Imagine we want to train an RBM to reproduce the correlations of a  specific quantum state.  The visible layer $v$ consists of a mixture of the details for the measurement (say a choice of polarizer orientation in the standard Bell experiment) and the outcome for said measurement. The hidden layer $h$ corresponds to hidden variables which are connected to the visible layer via a weight matrix $W_{ij}$, see Fig.~\ref{fig:rbm}. As per the prescription for precedence, the first measurement of a given state in ``cosmic history'' will be completely random. Thus, we begin {\it training} our architecture by initializing the weight matrix  $W_{ij}$
to have random entries.    We next make a choice of measurements on the system,
$\alpha, \beta, ...$, and initialize their values to random numbers
 $x_\alpha, x_\beta, ...$.  This is the first pass.

We then update weights using  contrastive divergence, which provides new values for the matrix elements $W_{ij}$.

The reconstructed visible layer $v'$ can now inform our next input to the architecture. A simple idea would be to implement a basic rejection sampling routine based on the reconstructed probability for each neuron. One then repeats this procedure with the now {\it less random} state. Also notice that the influence of the past states persists, because  it informs how the new visible layer is constructed by means of the weight matrix. As one iterates this process, it should converge to a set probability distribution.

This approach is just an example and is likely extendable to other machine learning architectures, e.g., autoencoders. What should be clear though, is that if the universe were to exhibit a principle based on precedence, it seems inevitable that its physical manifestation would resemble a machine learning algorithm, though with one important difference.

The architecture just described is different from the {\it Principle of Precedence}, as it does not entirely realize the notion of being informed from past outcomes. In our RBM analogy, it is indirectly influenced by the past via updates to the weights and not prior outcomes explicitly.   Depending on how it turns out, as well as whether one believes the past exists, this may be a feature rather than a bug.

But nonetheless, whether the dynamics is continuous or discrete, the influence of the past states, and the presence of the hidden state, means that this system has consequencers built into the dynamics.

\subsection{Self-sampling learning methods}
\label{sec:self-directed-graphs}

In this section, we describe several proposals for using graph toy models which realize self-guided growth.

Let us consider some possible stochastic \textit{base rules} for growth which lead to interesting \textit{emergent rules} at larger scales. We are interested in growing structures that describe not only the state of a ``physical system'' but also the emergent laws, all of which are described using discrete data. Such laws can be determined by influential structures that are highly persistent but not eternal. Formation of such structures at early times constitutes Batesonian information~\cite{bateson2000steps}.

As a simple toy model, we start by considering stochastic growth of graphs.

We saw a class of dynamical systems that could be described this way, in the matrix models defined by~(\ref{m3-1}) to~(\ref{m3-3}).As discussed in~\cite{unifysl} the matrix degrees of freedom split, as $N \rightarrow \infty$ into fast
and slow degrees of freedom.  The former provide a   "background Hamiltonian" which over intermediate scales can be considered stationary. Over shorter times they dictate the evolution of the fast degrees of freedom.  We see how the slow degrees of freedom can be considered to function as a consequencer. 

Now we consider some dynamics of graphs, which give another example of the emergent splitting into "Effective laws" which dictate the evolution of faster degrees of freedom.  Again these splittings can be understood as creating consequencers. In the context of graph dynamics, simple updates to nodes/edges can be understood as fast degrees of freedom whereas changes in topology of the graph can be assigned to slow degrees of freedom.

\begin{figure}[!pt]
    \centering
    \includegraphics[width=0.80\linewidth]{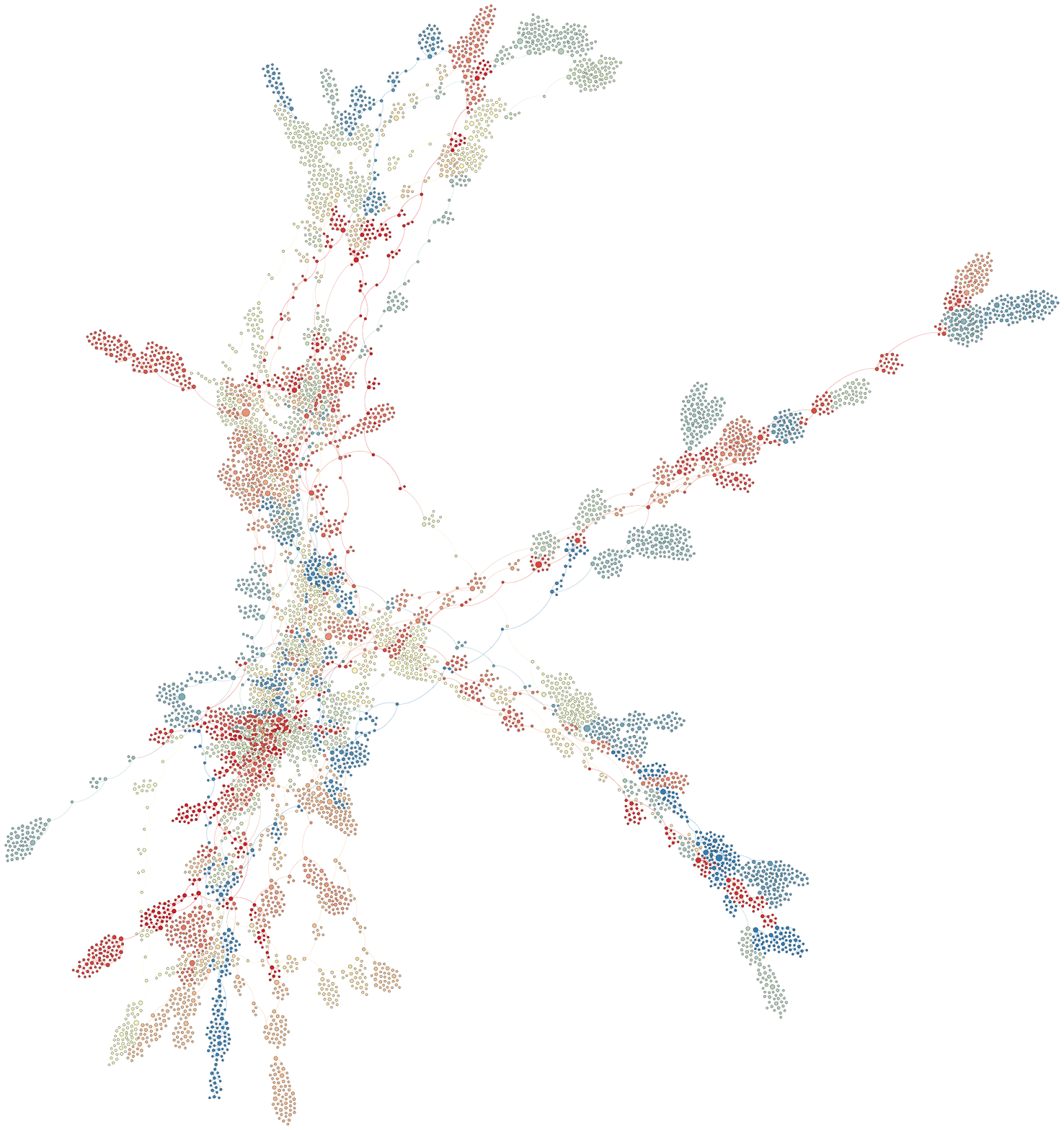}
    \caption{Graph with 7,088 nodes \& 7,304 edges constructed from sampling possible futures with a notion of stochastically optimizing the Wiener index - maximizing the sum of shortest distance between nodes. Graph is colored based on notion of modularity, with 73 total communities, and a very high modularity of 0.95 -- note that this is a structure very close to a tree.}
    \label{fig:weiner}
\end{figure}

Consider the time-evolution rule guided by a look-ahead algorithm which seeks to increase a particular quantity $Q$. Namely, at each step in the graph growth process, choose one from the subset of moves: we either add a new node and connect it to some subset of the rest of the graph, or we add an edge between an existing node pair. In order to pick the move, we perform a number of random walks several steps into the future and select the one that led to the optimal value of $Q$. Then, our move is selected as the first move in this winning walk. We note that the size of the search space grows super-exponentially with the graph size, so the accuracy of the look-ahead is limited by computational power. In this work we implement a toy example of this model, the details of which are as follows.\\

\noindent
{\bf Look-ahead algorithm}\\

Our algorithm consists of taking a starting graph, in this work two connected nodes, and applying the following procedure:
\begin{enumerate}
    \item Generate $N_C$ copies of the starting graph $G$.
    \item For each copy randomly add either an edge between existing nodes {\it or} add a new node and connect it to an existing node randomly. For each copy repeat this step a total of $N_F$ times - each copy now corresponds to a different future evolution of the original graph.
    \item  For each {\it future} calculate the value of $Q$ and identify the graph $G_f$ that extremizes $Q$.
    \item Now update $G$ with the graph formed by the first move in the history of $G_f$.
    \item Return to Step 1. and iterate, in principle, {\it ad infinitum}.
\end{enumerate}

\begin{figure}
    \centering
    \includegraphics[width=0.49\linewidth]{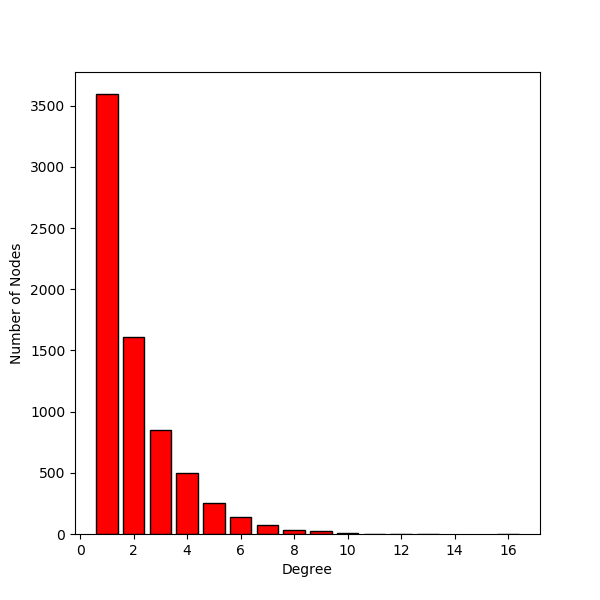}
    \includegraphics[width=0.49\linewidth]{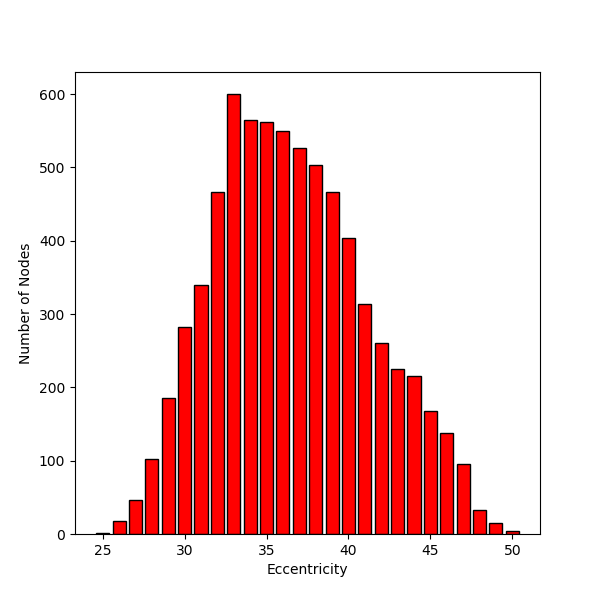}
    \includegraphics[width=0.49\linewidth]{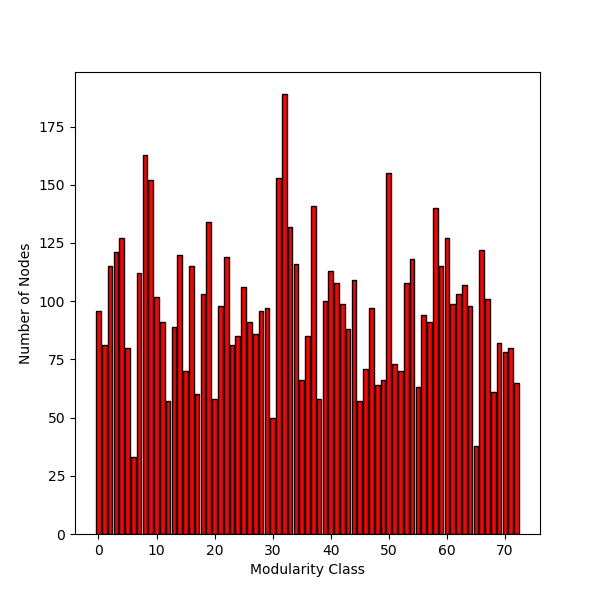}
    \includegraphics[width=0.49\linewidth]{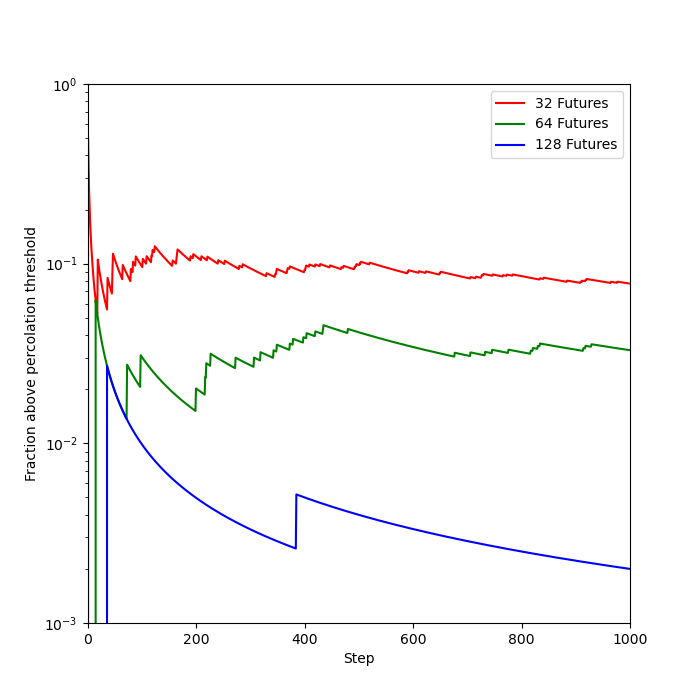}
    \caption{\textit{Clockwise from the upper left:} Degree distribution for the graph in Figure \ref{fig:weiner}, the eccentricity distribution, fraction that the connectivity is above the percolation threshold, and modularity class.}
    \label{fig:perc}
\end{figure}
    
We show numerical results for $Q$ as the Wiener index, i.e. sum of shortest distances between nodes, Fig.~\ref{fig:weiner}, for $N_f = 50$ \& $F_C = 64$. The graph has 7,088 nodes \& 7,304 edges and has diameter of $50$. We show the degree distribution, eccentricity distribution, and modularity class in Fig.~\ref{fig:perc}. One of the interesting properties of this graph is that it has a very high modularity of $\sim 0.95$ which implies  very sparse connections between communities -- of which there are 73 for this graph. Furthermore, we find that the graph is very close to the percolation threshold (i.e. that the connectivity of the graph is $\sim$ 2/\# nodes). One can appreciate this from Fig.~\ref{fig:perc} where we have calculated the fraction above the percolation threshold as a function of number of graph updates for different values of $N_C$. We find that considering more futures approaches closer to the percolation threshold. This result is not surprising since the high modality implies that the graph is close to a tree, i.e. a tree has a modularity of 1.

We also consider for $Q$ the graph diameter (largest distance among the graph nodes), which is depicted in Fig.~\ref{fig:dist1} for $N_f = 6$ \& $F_C = 64$ at two different snapshots - after 100 moves and after 1000 moves. The final iteration has 1,119 nodes \& 1,335 edges and has a modularity of 0.82 with 24 communities. Figure \ref{fig:diam-dist} shows the degree distribution, eccentricity distribution, and modularity class. This graph helps us appreciate that this algorithm doesn't amount to optimizing $Q$; instead, it is using $Q$ to guide the random walk. Indeed, we would expect for an algorithm that optimized $Q$ the final form would have a diameter much larger than $85$ given the size of the graph. One thing to note in Figure \ref{fig:dist1} is the presence of persistent structures -- for example the central hub of the graph and its arms which persist over many iterations of the algorithm. Since these structures influence decisions for future updates to the graph, they can be thought of embodying a form of Batesonian information.
    
Two other useful discrete topological objects are knots and braids. To model these, we label graph edge crossings with chiral variables.
Then, we apply the same guided random walk algorithm, while using a guiding quantity $Q$ that seeks to maximize both crossings and graph diameter. An example simulation is shown in Fig.~\ref{fig:chiral} where the graph has 91 nodes and 183 edges for $N_f = 6$ \& $F_C = 64$ and has attained a diameter of 14. Comparing the degree distribution for this $Q$ to only maximizing graph diameter, Fig.~\ref{fig:diam-dist}, one can see the impact of rewarding crossings.

Additionally, one may iteratively take tensor products of graphs with selected subgraphs to produce graphs with interesting structure, which are conjectured to posses fractal properties after a large enough number of iterations. While we were not able to verify this in simulation due to the fact that the size of such graphs grows exponentially or faster (depending on the details of implementation), we were able to produce some interesting example structures. We iteratively identified subgraphs containing neighborhoods of randomly chosen nodes, and took tensor products with the rest of the graph. We also considered a second prescription in which we identified subgraphs via random walks on the base graph, and then again took tensor products with the rest of the graph. The results are shown in Fig.~\ref{fig:randomnbhds}. Interestingly, this can lead to graphs with a layered structure, much like we find in deep neural networks.

\begin{figure}
    \centering
    \includegraphics[width=0.45\linewidth]{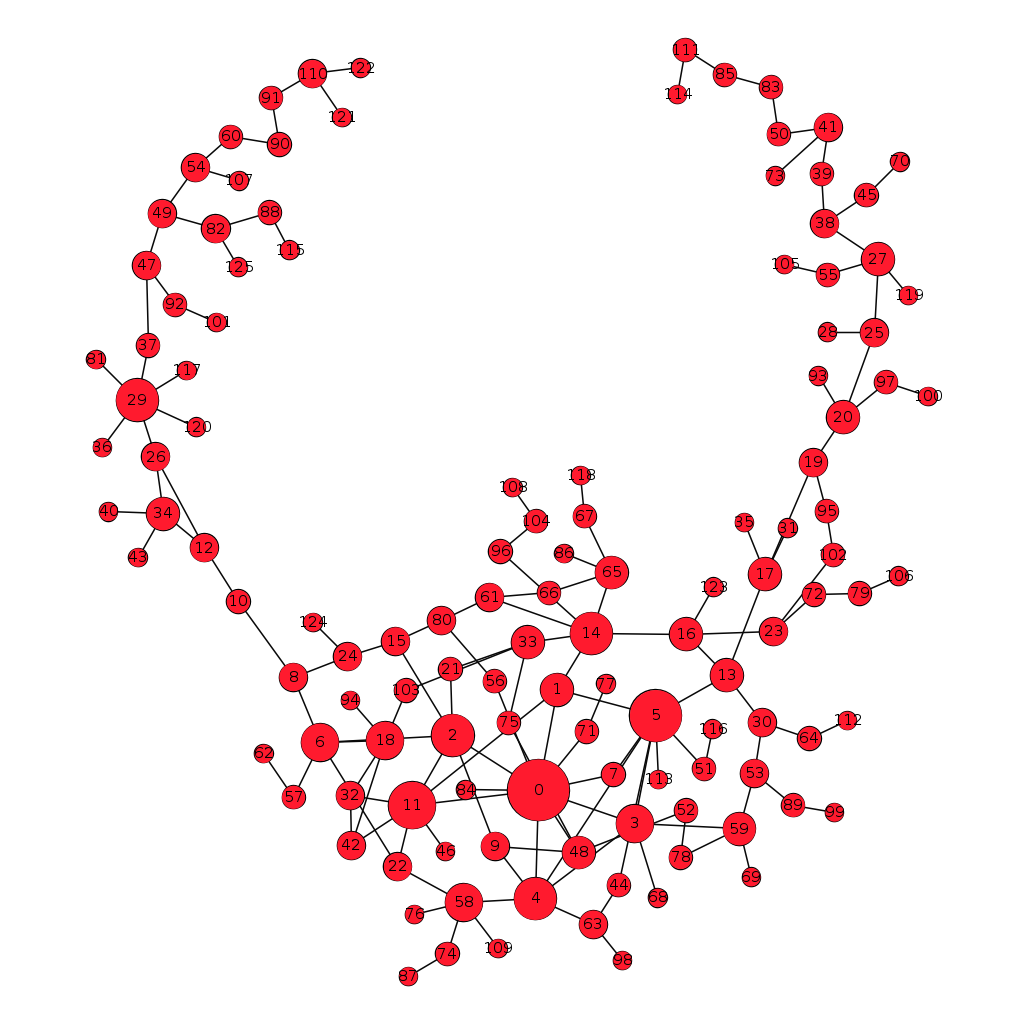}%
    \hspace*{1cm}%
    \includegraphics[width=0.45\linewidth]{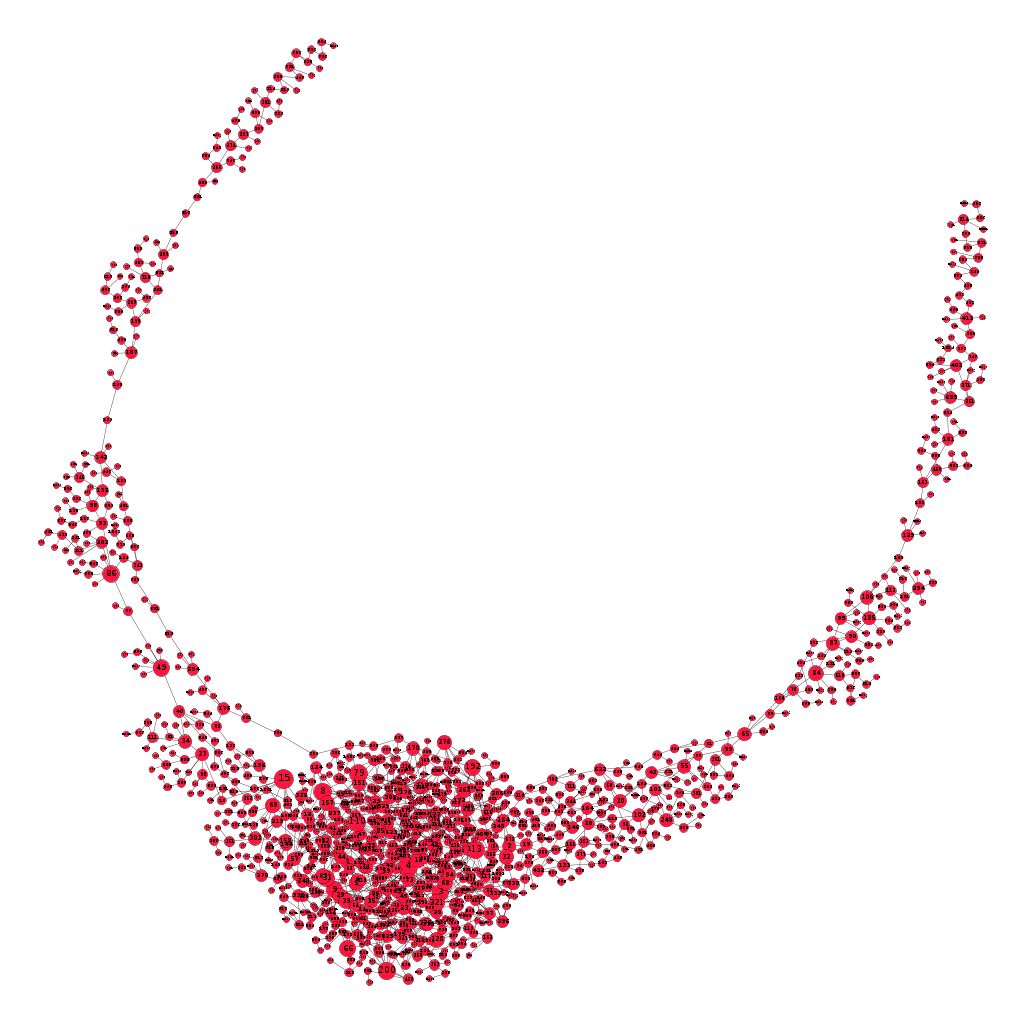}
    \caption{\textit{Left}: A graph grown with look-ahead algorithm where $Q$ is the graph diameter. \textit{Right}: The same graph after an order of magnitude more growth steps.}
    \label{fig:dist1}
\end{figure}

\begin{figure}
    \centering
    \includegraphics[width=0.49\linewidth]{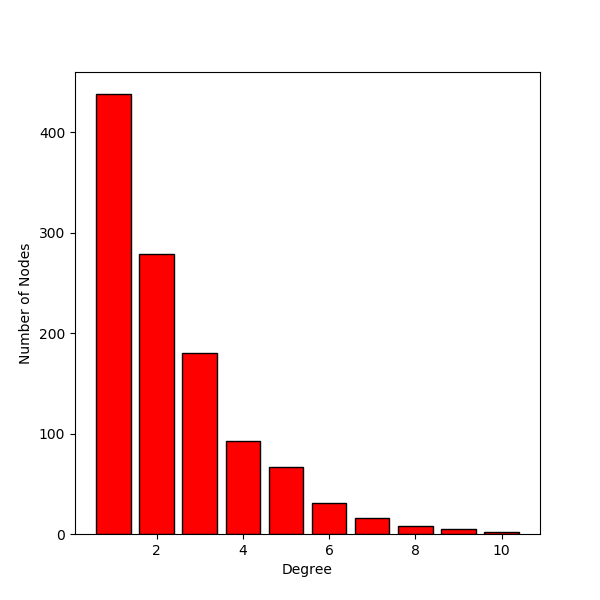}
    \includegraphics[width=0.49\linewidth]{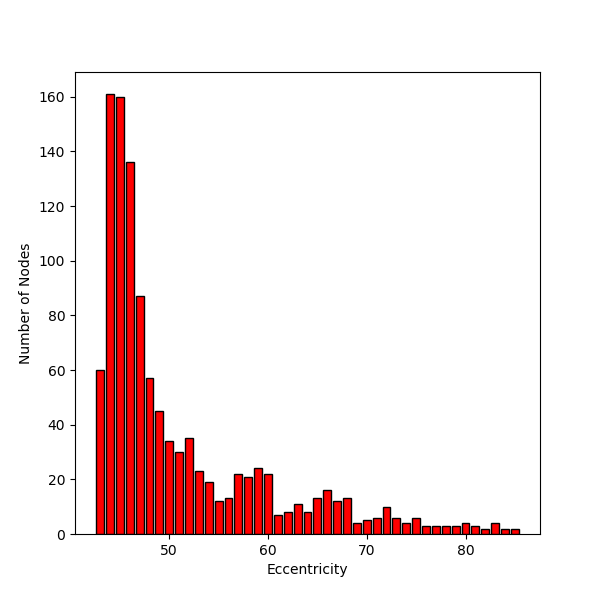}
    \includegraphics[width=0.49\linewidth]{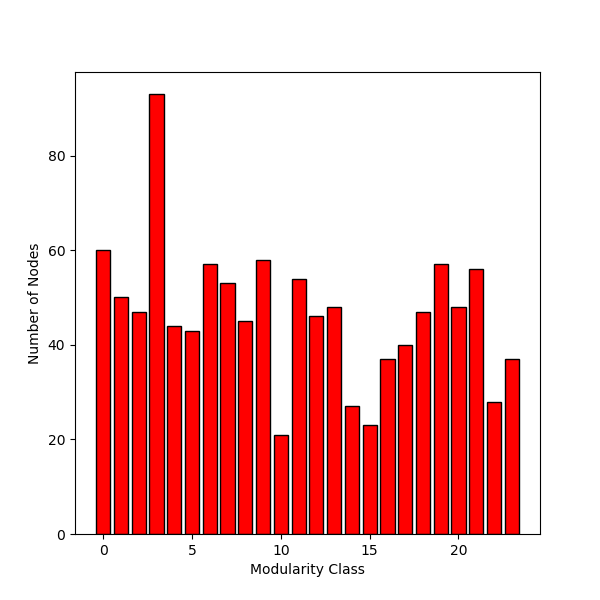}
    \caption{\textit{Clockwise from the upper left:} Degree distribution for the larger graph in Figure \ref{fig:dist1}, the eccentricity distribution, and modularity class.}
    \label{fig:diam-dist}
\end{figure}

The results here serve as a starting point for a more thorough study of graph growth directed by self-sampling. A more thorough study would seek to better understand the role of emergent substructures which become highly influential in determining the future graph growth, i.e., subsets which are frequently sampled. 
Though we do not have concrete numerical results yet, we can speculate about the form that emergent rules might take:

\begin{itemize}
    \item The emergent rule for graph growth is related to fractal properties of the graph. Structures present in the seed or formed early on in this process will determine the basic fractal structure. A growth algorithm similar to the iterated tensor products might satisfy this property.
    \item The emergent rule for graph growth is contained in a mechanical structure that imparts new shape. We may imagine a ``brewing vat'' embedded in a larger universe. Note that such structures are known to exist in Conway's Game of Life~\cite{Game_of_Life}.
    \item The emergent rule is related to the way information is transmitted through the graph. Rather than having a graph which constantly grows in all directions, like in our iterated tensor products algorithm, we may imagine graph growth happening only at specific places at one time. In order to determine such growth locations, we could in principle utilize something like the \textit{GraphWave} algorithm, in which we release a wavelet from a given node and see where it is most likely to land.
    \item The emergent rule consists of the graph itself interpreted in some machine language. This begs the question of what complexity of the ``external CPU" that interprets this  graph we are willing to allow.
    \item The look-ahead protocol is somewhat related to algorithms in reinforcement learning such as \textit{AlphaZero}~\cite{silver2018general}. Yet, we note the sample space here is much larger, which is likely to make learning more difficult.
\end{itemize}

\begin{figure}
    \centering
    \includegraphics[width=0.45\linewidth]{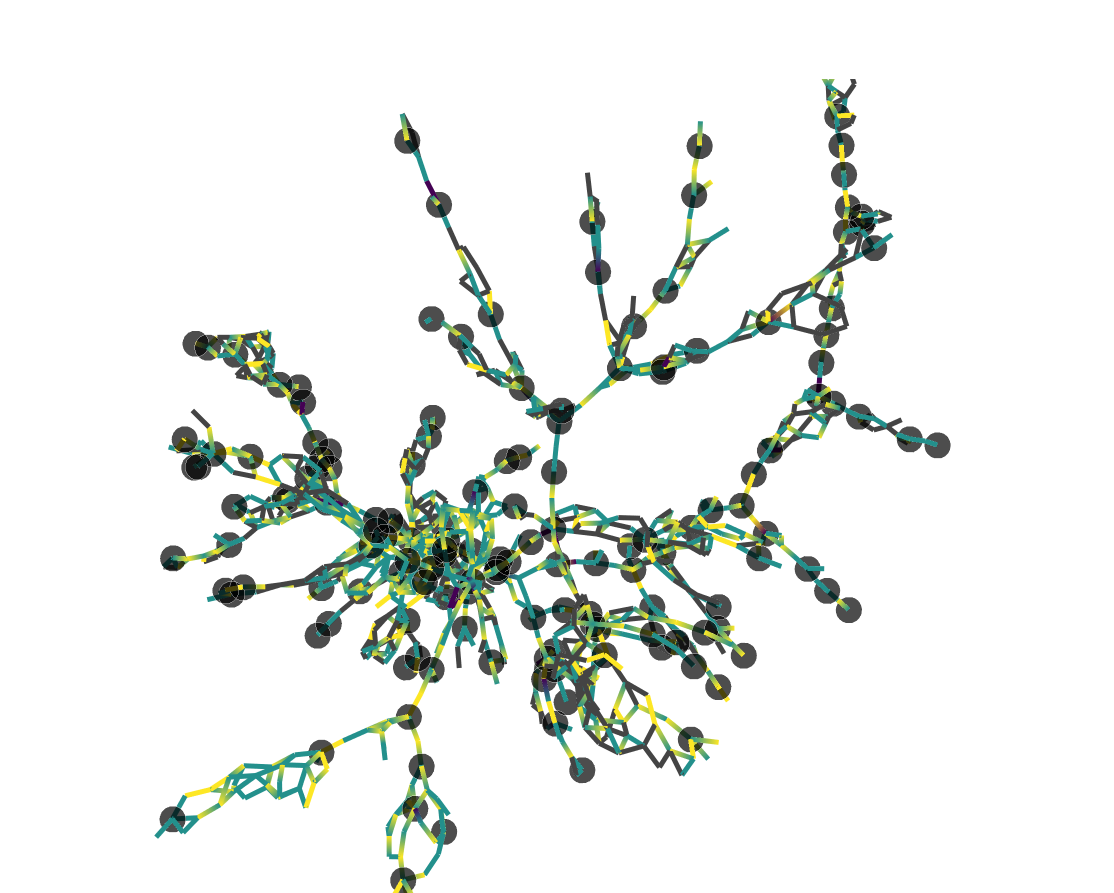}
    \includegraphics[width=0.45\linewidth]{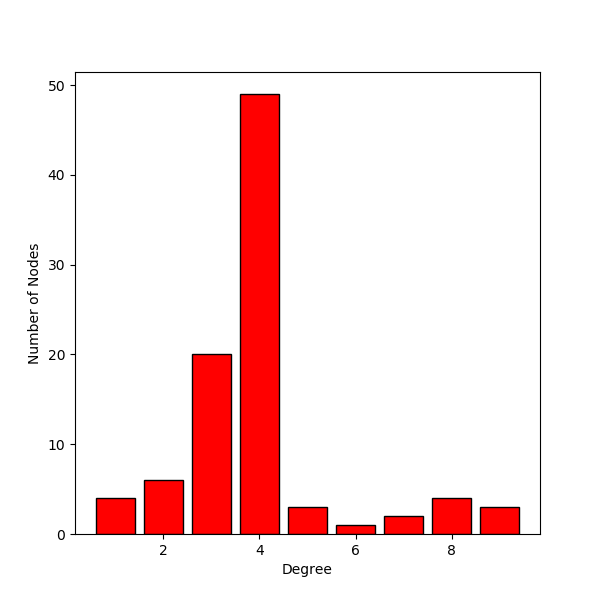}
    \caption{\textit{Left:} This graph is constructed such that it explicitly keeps track of edge crossings, where yellow (green) correspond to over (under) crossings, thus we can discuss knots and braids. Algorithmically, this graph is constructed with our look-ahead algorithm where we simultaneously try to maximize crossings and graph diameter. \textit{Right:} The degree distribution for the graph.}
    \label{fig:chiral}
\end{figure}

\begin{figure}
    \centering
    \raisebox{-0.5\height}{\includegraphics[width=0.4\linewidth]{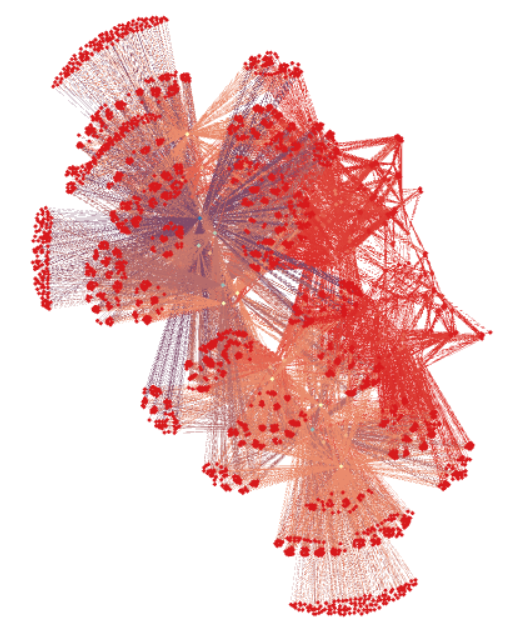}}%
    \hspace*{1cm}%
    \raisebox{-0.5\height}{\includegraphics[width=0.4\linewidth]{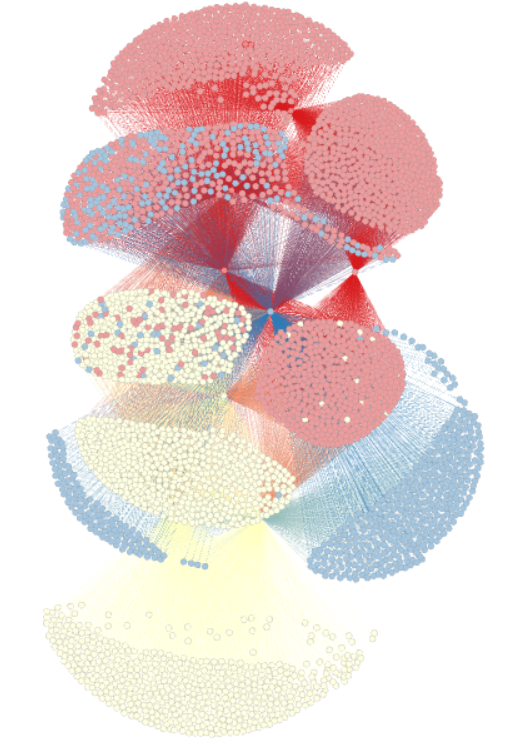}}
    \includegraphics[width=0.4\linewidth]{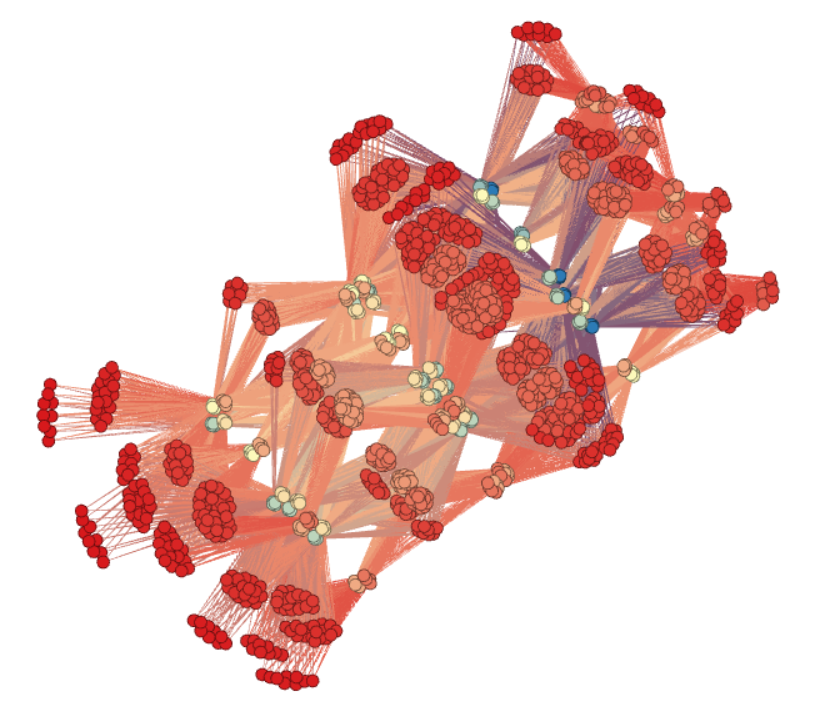}%
    \hspace*{1cm}%
    \includegraphics[width=0.4\linewidth]{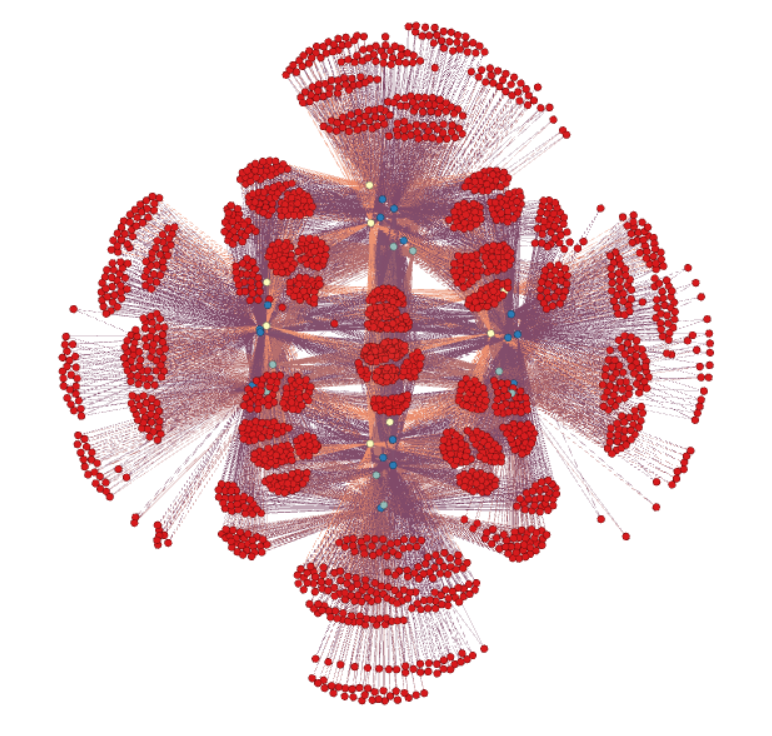}
    \caption{\textbf{Tensor Products with Subgraphs}. \textit{Upper Left}: Graph obtained by identifying neighborhoods of randomly chosen nodes and taking their tensor products with the rest of the graph. Coloring scheme is based on vertex degrees. \textit{Upper Right}: Graph obtained by identifying subgraphs through random walks of length 3 and taking their tensor products with the rest of the graph. Coloring scheme is based on communities. \textit{Bottom Left}: Graph obtained by identifying subgraphs through random walks of length 4 and taking their tensor products with the rest of the graph. Coloring scheme is based on communities. This example makes clear that structures resembling multi-layer neural networks can emerge from extremely simple starting conditions in autodidactic systems. \textit{Bottom Right}: Graph obtained by identifying subgraphs through random walks of length 5 and taking their tensor products with the rest of the graph. Coloring scheme is based on vertex degrees.}
    \label{fig:randomnbhds}
\end{figure}

\subsection{Variety}
\label{sec:variety}

One might wonder whether the statistical learning methods are useful only for illuminating details of an application of existing mathematical methods, or if they can be used to define new \textit{fundamental} quantities, specifically in the context of discrete topological observables. Here, we provide an example of a physically motivated quantity associated to graphs called the 
{\it variety}, first introduced in~\cite{Smolin_Barbour}.  Roughly, the variety is a global measure of dissimilarity of subsystems, i.e., it is a quantitative measurement of the asymmetry present in a discrete structure.

Notions similar to variety date back to Leibniz (\textit{identity of indiscernibles})~\cite{leibniz1989discourse}, and they have recently been applied to quantum foundations in order to reconstruct a version of Bohmian mechanics~\cite{bohm_variety}. Generally speaking, variety has to do with the distinctiveness of \textit{views} that subsets of a system have of each other. As such, it is naturally placed within a \textit{relational} approach to quantum foundations~\cite{Smolin_Barbour,Rovelli,bohm_variety}.

The simplest mathematical model of a fully relational system is a graph, and the definition and study of measures of graph variety will be the main focus of this section. 
We can think of a graph as a representation of a universe of relations: it is a closed system,
defined only by the patterns of links, each defining an elementary relation between the two nodes at its ends.  
The original prescription of variety presented in~\cite{Smolin_Barbour} is that each node of a graph has its {\it view} of the
universe, which is characterized by how it is connected.  One way to describe the view of a node
$i$ is to consider the {\it $n^\mathrm{th}$ neighborhood} of $i$, denoted ${\cal N}^n (i)$.  This is a subgraph consisting of $i$ and all the nodes connected to $i$ by $n$ or fewer hops, plus all the links among them.    One can then say two nodes $i$ and $j$
are $m$-step distinguishable if $m$ is the smallest integer such that the two
neighborhoods,  ${\cal N}^m (i)$ and  ${\cal N}^m (j)$ are not isomorphic.
We define the distinguishability
\f
D(i,j) = \frac{1}{m(i,j)}\,,
\ff

which we use to precisely define the graph variety with the sum over all pairs,
\f
\label{eq:iso_variety}
{ V} = \sum_{i<j}D(i,j)\,.
\ff
While this may seem like the most natural definition of graph variety, it is not the only one, and it is not obvious \textit{a priori} which is most well-suited for a quantum gravity application. Our particular choice may also change if our graphs encode causal structure, e.g. using directed graph edges.

In the theory of causal sets~\cite{surya2019causal}, discrete spacetime is represented by a set of events together with a set of causal relations among them. This causal structure admits a natural representation in terms of a directed acyclic graph. Only a very small fraction of such graphs turn out to be faithfully embeddable in a low-dimensional (for example, 3+1 - dimensional) spacetime~\cite{kleitman1975asymptotic}. While there has been significant progress on understanding discrete Einstein-Hilbert actions, it is not yet clear whether \textit{all} low-dimensional discrete geometries satisfy some unifying optimization principle. It is in this context that we conjecture variety may play a role.
More broadly speaking, the principle that different events should be uniquely determined by their causal pasts has played a role in developing energetic causal 
set models~\cite{energetic_causal}.

A  related implication of Leibniz's principles is that neither classical nor quantum general relativity should admit any global symmetries.   This is strongly suggested  by a theorem of Karel Kuchar that states this is the case in classical general relativity with cosmological boundary conditions~\cite{KK}. 

In the remainder of this section, we explore a handful of definitions of variety and specifically focus on those which are \textit{discriminative} -- maximized for an exponentially small subset of all graphs -- and \textit{complex}, i.e., it is hard to sample graphs which maximize variety. We show that variety is difficult to compute using the usual numerical methods for
discrete topologies, because the computational complexity of candidate functions is at least cubic in the number of nodes, and because the dynamics can be highly non-local. We conjecture that  variety is an example of a kind of quantity that 
may be  best defined and computed through use of an unsupervised learning algorithm which uses \textit{approximate} optimization with bounded error. 
To support this conjecture, 
we report at the end of this section initial results from computational studies of graphs approximately optimizing a new definition of variety based on structural node embeddings.

\subsubsection{Asymmetry, irregularity, and heterogeneity}
In graph theory, similar concepts have appeared under the names asymmetry, irregularity, and heterogeneity. All of these attempt to capture a quantitative measure of graph symmetry. The degree of asymmetry was introduced as a means to classify asymmetric graphs, i.e., graphs which have only a trivial automorphism. This definition of variety, in terms of (the inverse of) the size of the automorphism group, is close to what was originally proposed in~\cite{Smolin_Barbour}.
However, this definition of variety turns out to be trivially satisfied by almost all finite graphs in the large-$N$ limit. It was shown by Erd\"os and R\'enyi~\cite{erdos1963asymmetric} that almost all finite graphs are asymmetric, and that almost all asymmetric graphs have a degree of asymmetry near the maximum $N/2$. Because the ensemble of random graphs is projective, the neighborhood $\mathcal{N}^1(i)$ is an element of the same ensemble defined on $|\mathcal{N}^1(i)|$ nodes and the same fraction of edges; hence, the subgraph induced by the first-nearest neighborhood is also almost surely asymmetric. Further, one can show it is unlikely that any random subgraph $\mathcal{N}^1(i)$ is isomorphic to a different random subgraph $\mathcal{N}^1(j)$, a fact which can intuitively be understood by remembering neighborhoods are a set of $N$ random elements of an ensemble that is \textit{super}-exponential in size.

Graph irregularity comes at the problem from the opposite perspective, attempting to define the degree to which a graph is not regular. There are a number of definitions in literature which we review here. The original definition due to Collatz and Sinogowitz described irregularity as the difference between the adjacency matrix's largest eigenvalue and the average degree~\cite{von1957spektren}. The set of graphs which maximize this notion of irregularity is the set of quasi-complete graphs. Likewise, Bell defined irregularity in terms of the variance of the degree distribution~\cite{bell1992note}, which leads to maximum-irregularity graphs that are either quasi-complete or quasi-stars, depending on the edge density~\cite{abrego2009sum}.  

Two popular measures of graph irregularity are the so-called Albertson index~\cite{albertson1997irregularity}, which measures the imbalance of nearest-neighborhood sizes, and the related definition of total irregularity~\cite{abdo2014total}, which measures the imbalance of degrees of \textit{all} node pairs. The Albertson index is maximized for clique-star graphs, which are stars whose core is a complete subgraph containing at least a third of all nodes~\cite{abdo2014graphs}. The total irregularity is maximized for graphs whose degrees are more heterogeneous, and the resulting ensemble contains $2^{n/2}$ non-isomorphic graphs. Another measure similar to these definitions of irregularity is network heterogeneity~\cite{estrada2010quantifying}. However, this definition is not so interesting for our application as it is defined so as to be extremized for perfect star graphs.

A different definition of irregularity -- specifically, local irregularity -- was introduced in~\cite{chartrand1988define}.  The nodes in a locally irregular graph each have a set of neighbors with distinct degrees, though it is not the case that \textit{all} degrees in the graph must be distinct\footnote{In fact, it is impossible to satisfy the irregularity constraint globally. There are no graphs whose degree set is size $N-1$, and there is only one graph (and its dual) whose degree set is size $N-2$.}. Graphs which satisfy this local definition of irregularity were termed ``highly irregular graphs,'' and it was shown that the size of the associated ensemble has an asymptotic lower bound of $2^{n^2/32}$~\cite{alavi1987highly}.

We consider below a different definition of variety based on structural node embeddings. Several of the above similar definitions fail to capture the properties we desire for a physical variety, specifically because they select graphs which are not topologically interesting.  That is, we do not believe that stars, clique-stars, quasi-stars, etc., have sufficient structure to encode the structure and laws of a spacetime. It may still be the case, however, that metrics such as total or local irregularity are also compelling definitions of variety. We will leave their study to future work.

\subsubsection{Variety of structural node embeddings}
\label{sec:graphwave_variety}

In this section, we propose a \textit{smoother} notion of variety that we conjecture (based on our preliminary results) leads to a novel ensemble of high-variety graphs. To this end, we utilize an algorithm for constructing structural node representations known as \textit{GraphWave}~\cite{diff_wave}.  While we used the published implementation of their method, let us briefly review the algorithm.

\paragraph{A brief description of GraphWave algorithm}

While the previous two definitions of variety make intuitive sense, we will see in the following section that their representative high-variety graphs do not tend to possess enough structure. In order to come up with a notion of variety more aligned with our ideas of what high-variety graphs ought to be, we decided to leverage an unsupervised machine learning algorithm for constructing structural node embeddings called \textit{GraphWave}~\cite{diff_wave}.Let us briefly review this method as explained in their original formulation.

This algorithm starts by considering the diagonalized \textit{Laplacian} matrix $L$, 

\begin{align}
    L = D - A = U \Lambda U^T\,,
\end{align}

where $D$ is the diagonal matrix of node degrees, $A$ is the adjacency matrix, and $\Lambda$ is a diagonal matrix of the Laplacian's eigenvalues. Then, we consider a filter $g_s(\lambda) = e^{- \lambda s}$, which we apply to $\Lambda$, yielding the following result:

\begin{align}
    L \rightarrow L' = U \textrm{Diag}(g_s(\lambda_1),g_s(\lambda_2),\hdots,g_s(\lambda_N)) U^T\,.
\end{align}

If we let $\delta_a$ be the one-hot encoding\footnote{The one-hot encoding of a categorical variable is a binary vector with a single entry `1' and the rest `0'.} corresponding to node $a$, we can construct the associated vector $\Psi_a$, defined as

\begin{align}
    \Psi_a = L' \delta_a\,.
\end{align}

Roughly speaking, we are starting with a signal $\delta_a$ which is peaked at node $a$ and studying how the signal diffuses through the graph. The vector $\Psi_a$ then corresponds to amplitudes of this signal at different nodes. The key idea of the \textit{GraphWave} method is to treat $\Psi_a$ as a probability distribution, and compare such distributions through their \textit{characteristic functions} $\phi_a = E[e^{i \, t \, \psi_a}]$. Values of such characteristic functions evaluated at a set of discrete points $\{t_a\}$ are then put together in vector $\chi_a$ representing each node $a$.

\begin{equation}
    \chi_a = [\textrm{Re}(\phi_a(t_i),\textrm{Im}(\phi_a(t_i))]_{\{t_a\}}
\end{equation}

Additionally, a principal component analysis may be applied to reduce the dimension of the representation.

The idea is that nodes which play the same structural role in the graph will end up with the same representations; nodes from whose perspective the rest of the graph looks slightly different will have nearby, but not identical representations. We identify such node representations as \textit{views}, and need to define the variety as a function of such views.

\paragraph{From node embeddings to variety}

In our application of this method, we worked with 2-dimensional node representations, much like those shown on the example of Fig.~\ref{fig:pentagon}. If such 2-dimensional node representations are denoted as $\tilde{\chi}_a$, the variety is defined as the Coulomb potential 

\begin{equation}
\label{eq:wave_variety}
    V = - \sum_{i,j} \frac{1}{|\tilde{\chi}_i - \tilde{\chi}_j|}
\end{equation}

Thus, variety is defined as the (negative) Coulomb potential energy between node representations, treated as identical positive charges. One piece of inspiration behind this formula is the idea of locality in the space of views being more fundamental than the spacetime locality~\cite{revolution}. Then, we want to introduce variety through a physical interaction between the  views. The $1/r$ potentials are ubiquitous in physics, and would provide the simplest way to push the views apart from each other.

\noindent
\begin{figure}
    \centering
    \includegraphics[width=1.0\linewidth]{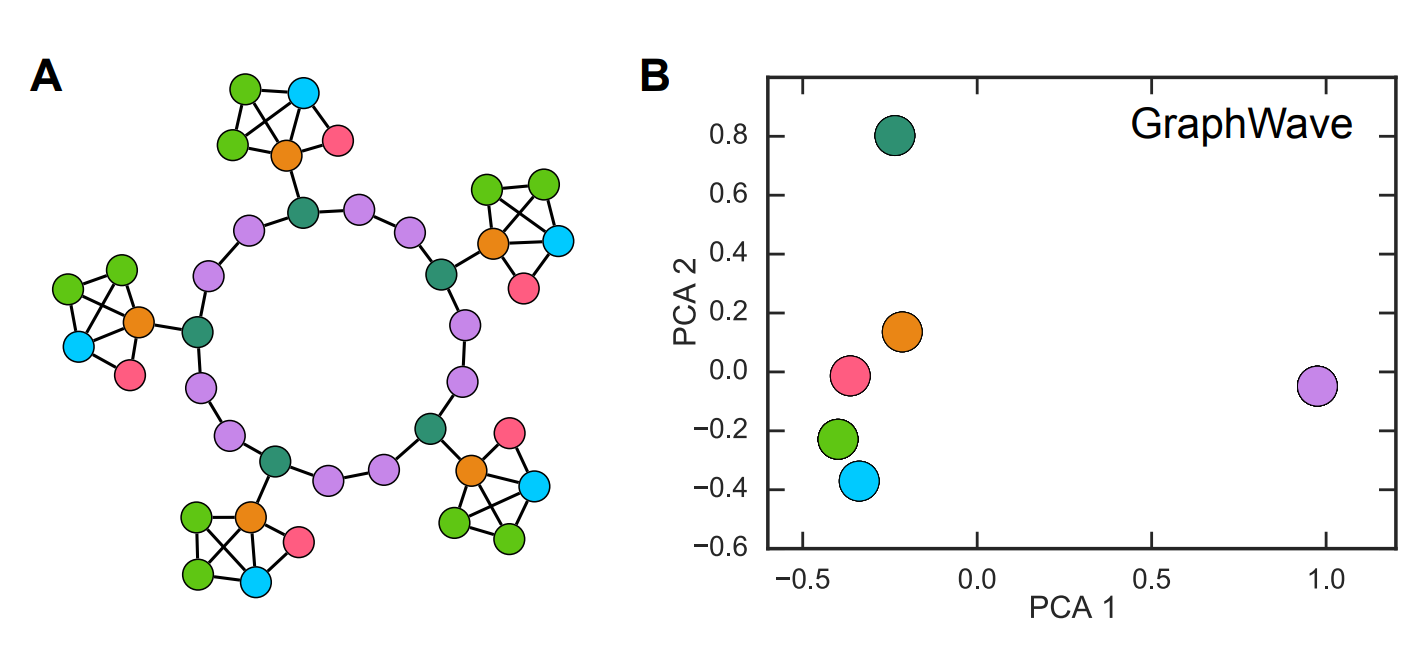}
    \caption{\textbf{GraphWave Embedding Algorithm}. A sample graph (left) is embedded using the two-dimensional PCA projection of the GraphWave embedding algorithm (right). This method provides a way to distinguish between structural roles of different nodes in a graph. This image was first printed in~\cite{diff_wave}.}
    \label{fig:pentagon}
\end{figure}

This measure of variety heavily penalizes graphs whose nodes have completely isomorphic neighborhoods (which would lead to a variety of $- \infty$); in general, node representations try to spread out as much as possible. One can think of this definition of variety as a \textit{smoothening} of notion from earlier in \ref{sec:variety}.

We will see in the following section that this definition of variety leads to some potentially appealing features of high-variety graphs, based on preliminary computations.

\paragraph{Numerical experiments}

We next consider a set of numerical experiments which attempts to characterize high-variety graphs using the above definitions.
After applying the notion of variety from Eq.~(\ref{eq:iso_variety}) based on isomorphisms, we find evidence that for large $N$, a typical element of the Erd\"os-R\'enyi ensemble~\cite{erdos_renyi} with a fraction of connected edges $p = 0.5$ is likely to have all first neighborhoods of nodes non-isomorphic to each other.  This gives the  maximum value of variety that can be achieved with this definition.
\begin{figure}[!t]
    \centering
    \includegraphics[width=0.60\linewidth]{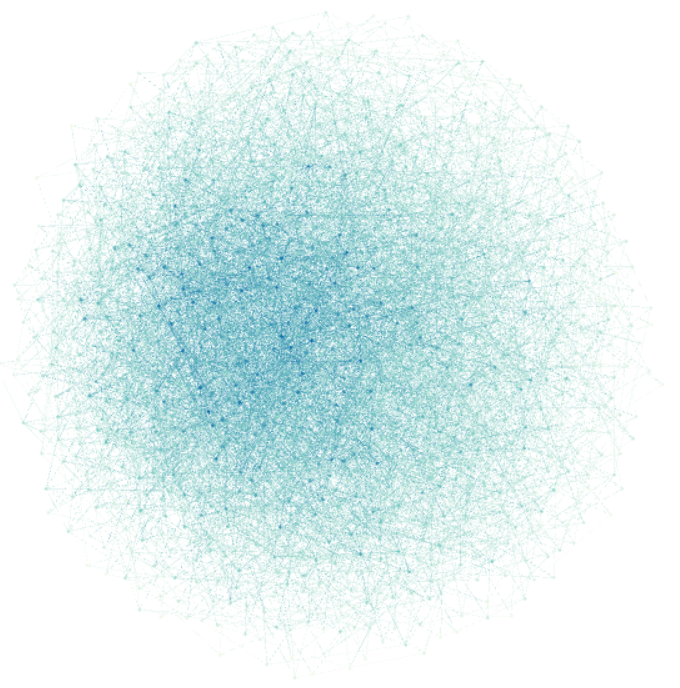}\\
    \includegraphics[width=0.40\linewidth]{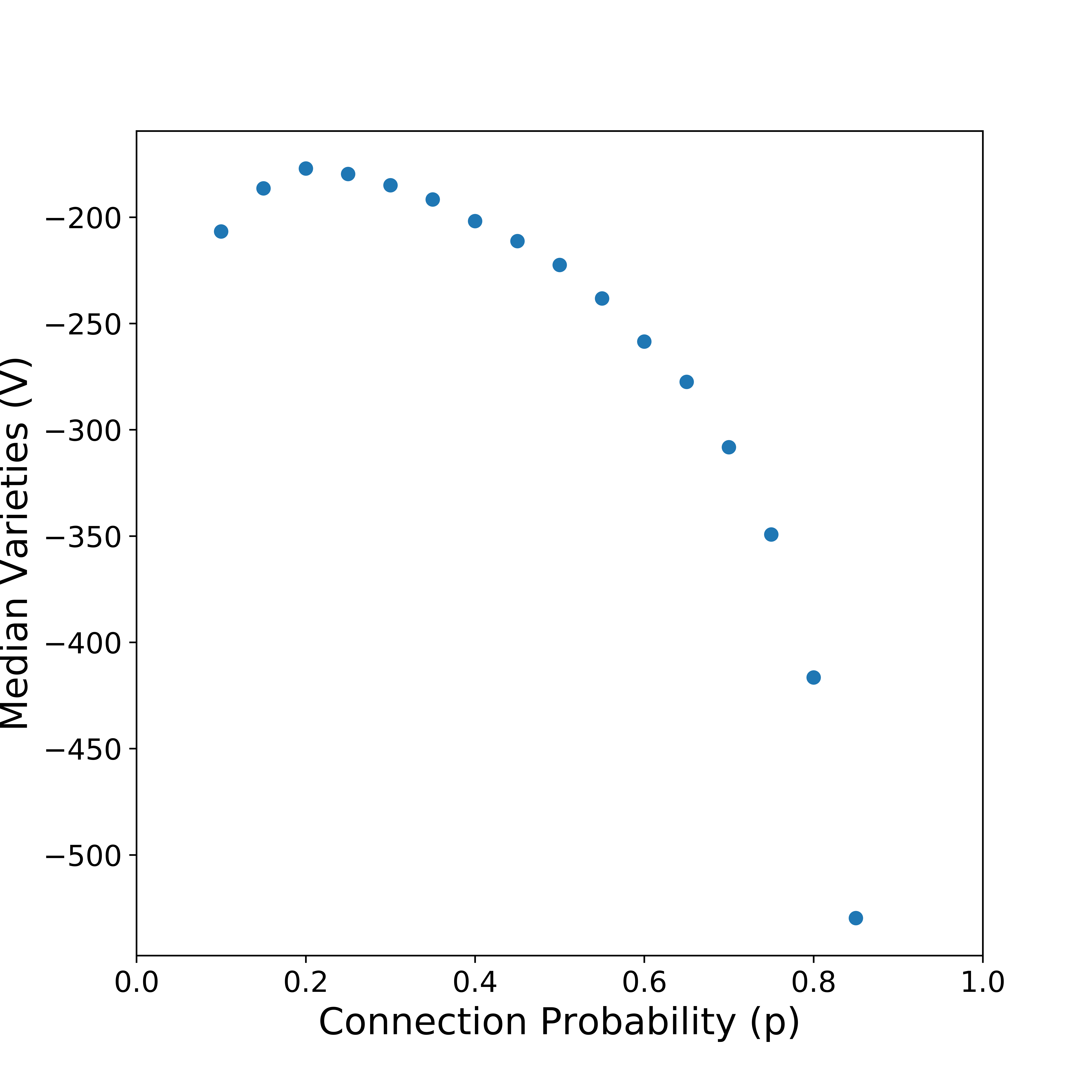}
    \includegraphics[width=0.40\linewidth]{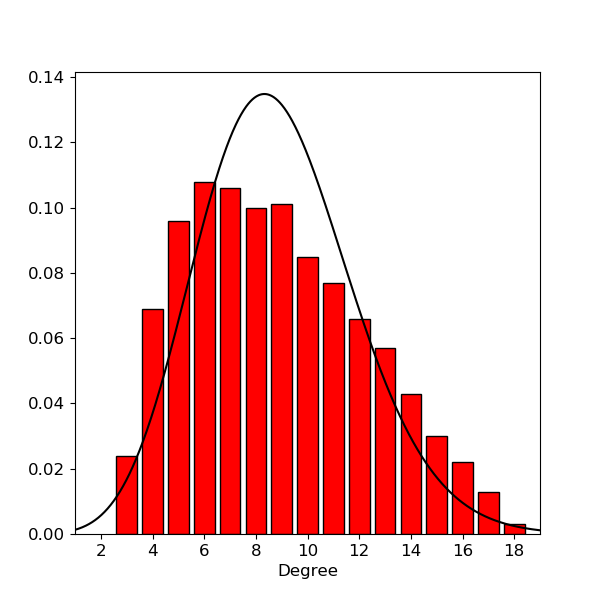}
    \caption{\textbf{High-Variety Graphs}. \textit{Top}: Beginning with a random graph on 1000 nodes and 3920 edges, we construct a high-variety graph with $1000$ nodes using simulated annealing. Node hues reflect their degrees, which lie in the range of [3,18].  \textit{Left}: Erd\"os-R\'enyi random graphs with $N=50$ nodes are studied using the GraphWave-based definition of variety defined in~\eqref{eq:wave_variety}. \textit{Right}: Degree distribution and Poisson distribution with $\mu = 8.8$, i.e. the average degree for nodes in the graph.}
    \label{fig:variety}
\end{figure}

The median values of \textit{GraphWave} - based variety across members of the Erdos-Renyi ensemble with $n=50$ nodes and different values of $p$ are shown in Fig.~\ref{fig:variety}. Note that the \textit{GraphWave} variety of Erdos-Renyi ensembles with $N = 50$ is maximized for $p \sim 0.2$. Even for this value of $p$, we find  that typical members of Erdos-Renyi ensembles do not maximize variety.

Graphs with high values of variety were constructed through the following procedure. First, we estimated the value of $p$ that approximately maximizes the value of variety among Erdos-Renyi graphs, generated 20 different Erdos-Renyi graphs with such values of $p$ and picked the one with the highest variety to serve as the starting point for the optimization procedure. Then, we used a simulated annealing method to make a number of edge insertions and deletions to the graph, with the aim of maximizing the variety. A resulting high-variety 
graph with 1000 nodes is shown in  Fig.~\ref{fig:variety}. 

This graph has the variety that is 22.4 $\%$ greater then its median value with respect to the Erdos-Renyi ensemble with the same connectivity. 
More work remains to be done on the study of high variety graphs. For now, let us content ourselves with the observation that our \textit{GraphWave} based definition of variety led to high-variety graphs that are statistically different from a typical random graph (for example, members of Erdos-Renyi ensembles with $p=0.5$), have significantly higher variety then typical Erdos-Renyi graphs with the same $p$, and have features that agree with our intuitive notions of what variety should mean. Comparing the degree distribution to the Poisson distribution, with mean given by average degree of the graph, we can see that the GraphWave notion of variety does \textit{not} produce a Poisson degree distribution in the large-N limit. It is intriguing that such a definition was obtained not from first principles, but through an application of an unsupervised learning algorithm. 

In future work, it would be interesting to repeat the same simulated annealing procedure a larger number of times and construct enough high-variety graphs so that further statistical conclusions can be made about their properties. We would also like to consider whether a connection may be found between our notion of \textit{variety} and information theory.

Finally, we would like to note that \textit{GraphWave} is certainly not a unique solution to the problem of constructing structural node embeddings, and other algorithms such as \textit{struc2vec}~\cite{struc2vec} are also available. In future work, the effect of different algorithms on the properties of \textit{variety} could be studied.

\subsection{Geometric self-assembly}
\label{sec:geometry}

The problem of geometrogenesis asks how the smooth, continuous geometric structure of space-time may emerge from a sequence of random topological structures~\cite{cheeger1984curvature,wu2015emergent,mulder2018network,kelly2019self,kelly2020emergence,vanderhoorn2020ollivier,boguna2021network,kelly2021convergence}. The various approaches to quantum gravity, such as causal set theory~\cite{surya2019causal}, spin foams~\cite{perez2013spin}, and causal dynamical triangulations~\cite{loll2019quantum}, have considered how a discretized Einstein-Hilbert action may drive the dynamics of discrete space-times into a geometric phase, in which geometry, locality, and possibly transitivity are emergent properties. Yet, no approach has fully succeeded in connecting the discrete and continuum worlds in the form of a complete theory of gravitation. Therefore, it remains an open question which properties are axiomatic and which are emergent, despite many compelling arguments from a variety of perspectives (see also section \ref{sec:protocols} and especially \ref{sec:precedence} on precedence and  consequencers).

We consider in this section two ways in which machine learning may be used to study the emergence of $n$-body local Ising-type interactions in a geometric space. The first method examines how one might learn a near-optimal annealing schedule from random to geometric structures at fixed size using knowledge of the existence of both phases. This scheme represents the moves an omniscient agent would make to generate a geometric space. The second method removes the perfection of the agent in an attempt to understand whether the emergence of geometry is inevitable when the agent is not all-knowing but does have the ability to learn from itself using consequencers in the form of self-attention. We instead use a growing model where we assume exchangeability is inherent (i.e., measures are always label-invariant) but the property of projectivity is learned. The consequence of the latter is that the sequence of states at late times becomes projective, i.e., larger discrete structures encode larger portions of the same geometric space, thereby allowing a continuum limit to exist regardless of early-time behavior. This leads us to conclude the time taken to learn the geometric space is also associated with a fundamental length scale.

In our toy models, we consider a set of Ising spins which form the base set of an abstract simplicial complex used to encode interactions. Abstract simplicial complexes are objects in algebraic topology formed by a collection of subsets of a base set, defined such that every subset is also an abstract simplicial complex. Given a particular realization, the interaction among the elements of the base set, i.e., the Ising spins, are described by the maximal subsets, called facets. A link between a pair of vertices describes a two-spin interaction, a triangular face among a triplet of vertices describes a three-spin interaction, and so forth. Hence, allowing both the Ising spins as well as the connections to change dynamically enables, us to study the optimal interaction patterns among a set of Ising variables.

\subsubsection{Learning an optimal annealing schedule}
We first consider a reinforcement learning protocol to learn the near-optimal trajectories through the discrete phase space which take us between the random and geometric regimes. In the geometric regime, we choose relational structures with well-defined continuum limits, such as random geometric graphs or Delaunay complexes, which are known to converge in geometry and topology to manifolds in a suitable limit~\cite{cheeger1984curvature,vanderhoorn2020ollivier}. The laws are encoded in the undecorated simplicial complexes, i.e., there exist no hyperedges linking three or more vertices, and there do not exist long-range connections between vertices separated by a large graph distance. For a fixed set of laws, the non-geometric counterparts are Erd\"os-R\'enyi random graphs and random piecewise linear manifolds with non-interacting spins\footnote{Note that spins may be non-interacting for two reasons: either the laws dictate they do not interact or the temperature of the system is infinite.}. When we allow the laws to also take a random configuration, the non-geometric phase is described by a random abstract simplicial complex, or possibly a random hypergraph, depending on whether the existence of an $n$-spin interaction also implies the existence of $(n-1)$-spin interactions among the same vertex set. We may also consider other initial configurations depending on our assumptions. The early universe is thought to be highly symmetric \cite{Usym},
so one might postulate the laws may be described very simply. For instance, the initial simplicial complex we use may be either the fully disconnected or fully connected set, which represent the non-interacting and fully-interacting systems, respectively. On the other hand, geometric configurations should be less symmetric but still contain a few relevant symmetry groups, e.g., an SO($d$) rotational symmetry (with respect to a d-dimensional embedding), an SU(2) spin symmetry, and a U(1) symmetry arising from a discrete $\mathbb{Z}_n$ relabeling symmetry.

The objective task now is to generate a near-optimal sequence of updates which moves the system from one regime to the other. We suppose the initial configuration is the fully non-interacting set: the laws are described by a zero-dimensional simplicial complex $K=K_0$, the spins $\{\sigma\}$ take random positions, and the temperature of the system is set to infinity, $\beta=1/T=0$. The energy of the system is given by the generalized Ising Hamiltonian,
\begin{equation}
H(K,\{\sigma\},\vec{a})=a_0\sum\limits_i^{n_0}\sigma_i+a_1\sum\limits_{\delta_{ij}\in K}\sigma_i\sigma_j\delta_{ij}+a_2\sum\limits_{\delta_{ijk}\in K}\sigma_i\sigma_j\sigma_k\delta_{ijk}+\ldots\,.
\end{equation}
The coefficients $\vec{a}=\{a_0,a_1,\ldots\}$ describe the couplings between spins, where spins are denoted by $\sigma$ and existence of facets are denoted by $\delta$s. The sums are over facets in the simplicial complex which encodes the laws, which makes the Hamiltonian a function of all of the variables in the model. When the system transitions to a geometric phase, the Hamiltonian takes an effective form where the couplings scale with the size of the system.

We now form a statistical manifold where each point corresponds to the measure for a particular configuration of the system,
\begin{equation}
\mu(K,\{\sigma\},\vec{a},\beta)=e^{-\beta H(K,\{\sigma\},\vec{a})}\,.
\end{equation}
The optimal annealing schedule minimizes the thermodynamic length~\cite{crooks2007measuring} with respect to the Fisher information metric~\cite{amari2000information}, i.e., it minimizes the change in entropy integrated along the path between the initial and final regimes. Updates which move the system through the manifold occur in several ways. First, we use single-spin flips which change the energy of the system. Second, the laws may change by adding or removing a facet in $K$. Third, the temperature may decrease according to an annealing schedule. We consider a process by which these happen on different timescales, e.g., the spins update much faster than the simplicial complex. We then train an agent using reinforcement learning to select the updates to the laws which move them closer to our target regime. The reward for a single update may be expressed in terms of the (inverse of the) estimated edit distance to the target ensemble, so that the maximum reward follows a geodesic in the statistical manifold. When this distance is expensive to calculate, it may be approximated with a compressed representation using a DNN trained beforehand with supervised learning. As the system cools over the course of the simulation, the agent will have relatively fewer opportunities to update the laws, hence the optimal schedule will be tied to the optimal learning rate. In other words, if the system cools too quickly the system may remain stuck in a quenched non-geometric phase.

\subsubsection{Self-guided assembly using precedence}
We now consider a second process by which an agent may learn from its own actions using the concept of precedence. Here we consider updates which grow the system: at each step the base set is extended by one spin and new facets are added with some probability. Initially, variables are updated purely at random, since there is no prior set of states that a precedence mechanism might exploit. As the walker updates, it begins to learn from its own moves using an attention layer. We first learn a generic precedence function using one or more geometric target ensembles, and then use the precedence function as the kernel of an autodidactic agent.

Let us define an exploration parameter $\xi\in[0,1]$ and a precedence (self-attention) function $f$. At each step, the system size increases, $K\to K'$, and can update one or more spins (generally we want the spins to thermalize between growth steps). One then samples a random variable from the unit interval, and accepts the update to $K$ if the random variable is less than $\xi$. If it is greater than $\xi$, then the update depends on previous states according to $f$. The self-sampling feedback mechanism may be implemented as an attention layer in a deep attention network, which uses a measure on a sequence of prior states to predict the next state. In other words, we either use a uniform measure or a learned measure to decide the next move. Over the course of the walk, the exploration parameter moves from one extreme to the other, so that initial updates are all random and ultimately updates are all determined according to prior updates. The trajectory in the phase space crucially depends on $\xi$ and $f$ which, generally speaking, allow us to formulate a random walk between any pair of endpoints in the phase space.

Now consider the following objective: learn the exploration function $\xi$ and precedence function $f$ which guide the walker from a given source to a given target along an optimal trajectory. The precedence function can be learned in the form of attention coefficients in a DNN using the methods described in~\cite{velivckovic2017graph}. The exploration function changes at a rate compatible with the DNN converging, which makes it similar to a learning rate. The output layer of the DNN is a probability measure on the set of variables one can update, so we may use it as a means for an agent to sample moves, i.e., it encodes the policy. In a reinforcement learning scenario, the agent does not modify the policy directly; rather, it can modify the exploration function, which in turn modifies the measure over moves. In doing this, we assume that we can continuously deform the policy DNN while taking $\xi\to 0$, i.e., that the attention coefficients change adiabatically when the input data is gradually changed. In practice, we presume the learned $\xi$ and $f$ give a near-optimal set of transitions of variables which freeze, in a statistical sense, when the target is reached. The goal, then, is to see if $\xi$ and/or $f$ are universal in some sense. If this is the case, one might argue geometric self-assembly may arise as a generic precedence-based annealing procedure.

\section{Discussion}

This paper is intended to  launch a new
program in theoretical physics and cosmology, whose aim is to investigate a  novel conception of the laws of physics - one in which laws evolve, but evolve by {\it learning}.

In this section we address certain initial philosophical questions that might arise.

Intimations,
from the remarks of Charles Sanders Peirce, to Linde's and Susskinds use of biological language to describe how laws might evolve on a "landscape" of string theories (in the context of eternal inflation),
to works of one of the authors applying the logic of natural selection to cosmological scenarios, are all distinct from what we have
in mind here.  Learning, as we discussed in 1.1, either in living things or in universes
is largely a mechanism to preserve  the good fortune arriving through lucky accidents - what Jacques Monod called {\it chance and necessity}; here we describe autodidactic learning from various perspectives.

Systems with phases that are distinguished in part by being harder- or easier-to-predict are often described using methods from the mathematics of chaos theory, but learning is not the same thing as chaos.  Learning requires an information architecture in which the past is no longer randomly or uniformly influential.  Instead, the past becomes influential in a concentrated way, associated with the consequencer, which will play a  role here as a type of reservoir.

In most physical systems, there is no consequencer because the presence of noise erases any correlations with structures acquired in the past.  But noise-resistant structures can come about.

We've used the concept of a consequencer in order to have a name for state that makes learning causal.  If physical laws emerge from an autodidactic process, then a consequencer is associated with any lack of randomness in which laws come to be.  Emergence guided by a consequencer is more likely to produce "interesting" structures, meaning structures that become optimized for a form of fitness, or support ever higher level structures based on them.  This claim is hard to state or prove formally, but it is clear that machine learning architectures have achieved what other approaches to algorithms have not achieved.    

If laws can evolve, then they can do more: We consider the notion that only a universe that learns its laws can be expected to engender novel phenomena like life and physicists.

\subsection{Traversing and learning in theory space}

We can think of the history of physics as a search through theories that can be expressed in mathematical notation, but that space of theories is not ordered in a way that allows for a machine-learning-like exploration.

The space of cubic matrix models gives rise, under various compactifications, limits, and symmetry breakings, to a space of theories that may be ordered in terms of complexity and other principles.  It is traversable by learning algorithms.

Among other properties, an important feature of learning dynamics on a landscape of theories should be the inclusion of theories we might recognize as low-energy or effective descriptions of other, more fundamental theories. The space of theories described by cubic matrix models is a good example of such a landscape.

In~\cite{universal} it was demonstrated that Chern-Simons theory, BF theory,general
relativity, alone and coupled with Yang-Mills fields, all appear in it. Certain formulations of string theory appear as well. What else might appear?

There are several ways to think about this project.

We might think of this as a search for new theories in a space of theories.  In that case, we might hope to find new formulations for unifying GR and QFT that have more desirable qualities, such as a way to prefer a small number of candidate theories instead of vast numbers of them.  

A similar, but distinct, idea is that we might find an explanatory theory for why nature has settled on GR and QFT  and the standard model out of a vast landscape of other viable theories.

A third idea, which might or might not be taken as distinct from the first two, is that our machine learning model is a model of how nature works, as she finds theories to operate within.

Furthermore, some might wish to speculate about the history and context in which cosmological autodidactic processes might take place.  The readiest idea is to suppose that autodidactic processes are features of the natural world emerging from the big bang, but one might also speculate about a metaphysical context or prefer a more complex idea of histories that could "precede" the big bang.   The latter approaches might be motivated by quests to explain an expanded set of questions about initial conditions, for instance.

Our proposal is that the autodidactic concept need only be operational, at least at this stage in our inquiry,   and can remain  neutral towards or accepting of varied ontological assumptions.  It can be examined independently of such assumptions. There is a useful shade of difference between ontological agnosticism and metaphysics. In particular, this work is neutral regarding hypotheses about the status of reality, as in~\cite{selfsim}.  We  occasionally use a contingent ontology as a story telling device, because it makes the telling easier, but we are not committed to any such frame.  
 
 In this spirit, we have sketched how a search through theory spaces can be undertaken.  We presented a viable approach to specifying a traversable space of theories as well as computational experiments in the iteration of autodidactic systems.

\subsection{The Universe goes to school}

A cosmological consequencer accumulates information based on the history of the universe.  It is causally accessible, which is not typically the case for \textit{all} the information about the past of a system.  

Consequencers can take many forms. The whole state of a system, even the whole Universe, might inform a learning process (such as the evolution of laws, which concerns us here), or it might be that the consequencer is associated with only a subset.  

In the Principle of Precedence, for example, the consequencer for a quantum process might be all the instances of similar processes in the past~\cite{Smolin:2012mk}.

A consequencer is an information reservoir that can correlate and store the consequences of  choices
 made in diverse situations. The
 consequencer within a brain allows an animal to evaluate what is the
 best course of action to take in a given circumstances, for instance, without trying them out in the risky physical world.
 
 Another word for a consequencer is, from a certain perspective, the imagination.  It stores information in a way that reflects patterns, so it also, to a degree,  foresees future circumstances. A learning system effectively recognizes situations similar to ones it may have encountered before, so learning systems also effectively foresee novel situations with similarities to past ones that might arise in the future.  

A consequencer can become recursive, so that it would allow a system to refer to itself, and other versions of itself.  This is an outcome resulting from the general capacity to capture similarities between patterns.   So let us say that a learner has the ability to improve its success in a variety of circumstances, and that it is made possible by consequencers.

Here we consider some further qualities of consequencers.  

\subsubsection{Useless and usable information}

One of the deepest
laws of physics we know is the principle of inertia, discovered by Galileo, which in Newton's formulation tells us that any body undergoing uniform motion continues in that state of motion unless impeded by a force.  Information is exchanged when bodies interact, but usually not in a way that would constitute a consequencer.

Information can be present in the universe and yet be "useless", meaning certain information, such as the spin of a particle, might only be part of a statistical distribution and of no individual consequence.  The information associated with a particle, or even a collection of particles that is smaller than, say, a pebble, needs either an architecture or extraordinary good luck to make it operational as a signal of consequence.   But information architectures with bits as big as pebbles are slow, and have many other problems, making them unlikely to appear either naturally or through engineering.

Therefore, information architectures typically amplify the causal powers of rather small collections of particles\footnote{There have been proposals for computers to be made someday by rearranging stars~\cite{graphstellation}, black holes, or other objects that are much more massive than pebbles.}.

The information in a computer is held in an architecture so that any bit might be of consequence.  Certain bits - which we will define as\textit{ interaction potentials based on measurement thresholds of state within cells in a symmetric, addressable structure} - can matter profoundly more than other bits.  

A bit requires two contextual qualities: 1) a position within a symmetrical structure that can function as an address and 2) operational  consistency in  results of multiple similar measurements of the bit so that it can reliably be known to have a value.  The same applies to qubits, though in that case, multiple measurements require multiple enactments of a calculation.  

The second requirement above is similar to the notion of redundant results in ascertainments of the past, which is described in  \cite{Riedel_2016}.  A coarse grained bundling of similar measurements all giving the same result is needed to explain why bits are different from other distributions.  The notions of computation and architecture build on the idea of bits to allow certain ones to be more causally accessible, or influential, than others.  

A typical portion of the universe - excluding computational architectures - only contains a uniform,  minimalist structure to make information causal,  creating the continuity described by physical law: momentum/position, energy, and so on.

Computers in their rarity are closer to Aristotelian physics: a change in the current of a wire persists until the next instruction to change.  In other words, there is no conservation of momentum, only a preservation of state.  Even computers like quantum or analog computers that run in a non-Aristotelian mode for a time must be set up repeatedly for repeated runs in what is essentially an Aristotelian dynamic at the larger operative scale.

The common interpretation is that the universe initially did not have an information architecture, unless one wishes to interpret conservation as an underlying architecture, but then architecture appeared, initially in the repetitive form of galaxies, black holes, stars, planets, and so on, then in exuberant life, and then later still in our hopeful intellectual designs.  However, we are here considering the possibility that cosmological architecture was more creative (and earlier so) than in the typical story.

Does a creative universe require that Aristotelian systems emerge out of Newtonian ones?  The "latching" of the more abstract Aristotelian sense of memory might be the lowest level quality that allows a system to become self-referential and evolutionary.

   As stated earlier, a cosmic ray that changes a gene is much more likely to be Batesonian, perhaps facilitating a chain of events that leads to flying dinosaurs, for instance.  It is vanishingly unlikely for Shannon information to be Batesonized.  Furthermore, even the  Batesonian information only has a slight chance of being predictably consequential.  DNA only has predictably consequential content when in an egg or a womb that matches it. \footnote{This does not suggest that the context for consequential content cannot ever be duplicated or synthesized, as in ~\cite{artificialuterus}.}

Consequential information can be thought of as  compressed or encrypted information which is only relevant in the presence of matching algorithms.   

Therefore a context, environment, key, or architecture must co-appear to make consequential information effective.  We argued that the presumed history of the DNA/RNA/protein system exemplifies the problem with a consequencer being its own key or compiler:  Complex, cumulative questions cannot be asked, only fleeting ones.  Therefore, the history of a consequencer probably requires alignment of parallel structures.

In the simulations reported earlier, we see parallel structures begin to appear where there was only uniform evolution.  These structures often preclude large sets of futures from being possible.  This reduction of futures is the first step in the appearance of architectures that make information consequential, analogous to a pinhole camera restricting light paths in order to make an image resolvable.

\subsubsection{DNA and machine learning systems compared as consequencers}

The DNA/RNA/protein system and machine learning models in a human technological context can each be thought of as consequencers that are placed in specific slots within larger structurs, but we have already imagined early autodidactic systems in which this was not so.

Coexisting systems distinguished only by the rules followed by their components could overlap and compete in an iterative process to destabilize each other as they evolve, for instance.  In that case there might be learning  without \textit{insulated} consequencers.  It is also conceivable that a learning consequencer would not require a multilevel structure as does the DNA/RNA/protein system.

The DNA system is an example of a consequencer in nature which takes up only a small portion of its universe. The internal states of a machine learning model, which are hidden, or at least unintelligible outside of the model, provide another example. 

The locality and architecture of a consequencer might not be unchanging as a system learns.  As argued earlier, the original consequencer that facilitated the emergence of the DNA/RNA/protein system was probably less insulated than what is operative today.\footnote{ To restate: There is not yet a convincing theory for the origin of it in which it was the first molecular mechanism that could encode information and reproduce, so it is possible that there were earlier  periods when antecedent or alternate chemistries were present. And yet not a trace has been found of these.

This earlier period would have been characterized by multiple interpenetrating molecular information systems.  It would probably have started before molecules evolved membranes to protect themselves to a degree and become persistent enough for experiments to be consequential.  They would have initially interacted with each other with unconstrained promiscuity and violence.  The consequencer in this period would have been more distributed and diffuse than in the later period when the DNA system came to dominate.}

DNA relies on replication to perform new tests in order to gain new lessons.  Machine learning models have been configured in various ways, but do not typically rely on self-replication, though the field of "artificial life" is devoted to ones that do.  A broad framing that includes the humans in the loop might find that all computation is self-replicating, however, in the sense of Marshall McLuhan's quip that humans are the sexual organs of machines.

A mutation of a DNA sequence in a gene - which we would classify as a component of an autodidactic system -
changes both the corresponding protein and its representation, coded into sequences of base pairs.

We can call it an "idea", which is like an "atom" of a consequencer.  It is not always the case that consequencer atoms can be isolated, nor are they likely to mean anything in isolation.  It is going too far to suggest that DNA specifies a protein or the phenotype of a creature. As stated earlier, DNA came to be fine-tuned to interact with molecular machinery in an environment like a womb or an egg to specify a protein or a creature. Consequential information is meaningless without a matching architecture or context, so the chicken and the egg and the DNA must evolve in tandem. An idea isn't an idea without context.

Note how the DNA-RNA-protein system does more than it needs to, if we imagine that the survival of an individual or a species were specified as goals by a cosmic engineer.  The code and mechanisms that enable a creature 
to reproduce are sufficient to encode the construction of a vast number of \textit{other} possible creatures, more than could ever be realized in a Universe.   Biology is not ergodic.  More generally, we conjecture that a learning system that causes novel configurations to come about is also likely to be exploring a space that is too large for the learning system to be ergodic.  

Reproduction and inheritance might be thought of as supervised learning, since a species phenotype is constrained by its ecosystem and environment, but at a larger scale evolution can be thought of as an unsupervised, or indeed an autodidactic system, seemingly propelled to discover endless novelty and revolutionary designs.  This shows that an autodidactic system able to perform in a supervised mode might also perform in an unsupervised one, given the right scale and circumstances.  If laws evolve, they might pass through epochs in which they are more constrained by either the present configuration of their universe (something like being supervised), or past ones, as reflected in a consequencer (more like being unsupervised).

Biology works because  the mechanisms of template  synthesis allow molecular patterns to do their work alongside representations --  ideas -- of themselves\footnote{Learning systems have properties that will outlined in section ?? }.  We have considered how physical laws might evolve in the context a multilevel structure analogous to biology, but the ideas will sometimes apply as well to universes that evolve laws using a self-similar structure, analogous to a hypothetical epoch of pre-biological reproducing molecules on Earth.

Non-trivial learning systems must  come to be able to "learn" from semi-persistent complexes of states-of-self rather than fleeting, disconnected self-states.  This is why species exist in natural evolution.  We stated a canonical example earlier: At some early point in the evolution of life, we presume that early pre-gene-like sequences in self-replicating molecules were maximally promiscuous, with horizontal transfers being the only mechanism of change.  Later on, membranes appeared, and virus-like forms tested more persistent collections of genes against other such collections.

In a rough sense, the internal layers in a machine learning network are similar to species.  They accrue "lessons" based on a complex, but only slowly changing - if changing at all -  measurement architecture. 

Learning systems have some of the properties that make biology possible:  the property of self-reference, i.e. they can "learn" things about themselves and store information in an architecture that makes it influential.  
The same mechanisms make it possible to learn about other learning systems, or variants of themselves;  
the space of alternative versions of themselves is so vast  that a learning process typically cannot achieve ergodicity given the age of the 
universe.

 Furthermore, small changes in genes result in small changes in a phenotype at least to the minimal degree that "learning" can be incremental.  If the relationship of genotype to phenotype was more typically random, then "preadaptation" might not be prevalent enough for evolution to proceed.  The coherence of preadaptations suggests that the process of evolution not only models species, but models it's own change to a degree.

\subsubsection{Signal between a learning system and its environment}

As we argued in 1.1, there are many types of learning.  How would cosmological learning compare to familiar examples?
 
Here are five examples of learning that have drawn great attention:  1) the speciation from which organisms arise, 2) adaptive behaviors found in a wide range of natural organisms (including "simple" ones), 3) intellectual learning in people and occasionally in other big-brained species, 4) social processes such as scientific and legal scholarship, and 5) certain algorithms such as machine learning algorithms.  Other examples have been proposed, such as the Gaia hypothesis, which suggests that the biosphere on Earth displays learning behavior, and many ideas about how human societies at large come to be like intelligent organisms, but these five examples are generally accepted and have clear structures.  What we will note is that in all five cases it would be unimaginable to understand what is happening without treating learning as a fundamental process.  It is not an option but a necessity.  

There are important differences between concepts of learning that have been applied with varying success to the same examples.  For example, Richard Dawkins famously coined the term "meme" initially to suggest that ideas competed and evolved in a Darwinian sense, like genes.  But the term (meme) eventually came to be understood as being somewhat derogatory.  It came to be more often used to tag an item that becomes valuable because of network effect prominence rather than relevance to any other fitness test.  For instance a company's stock or a video clip might be called a meme because it is suddenly popular, but would not otherwise be so valued.  In that case, the machinery of learning is functioning without reference to external circumstances to which it might respond, but only to circumstances within the learning mechanism.   It is no longer adaptation but might at best have a function as untethered,  pre-adaptive churn.

The notion of "learning" as we use it is more than moment-to-moment, brute adaptation.  It is a cumulative process  which  takes on abilities that can be thought of as theorizing, modeling, and predicting the environment in which it is responsive.   For instance, the DNA/RNA/protein system on Earth must have arisen from an adaptive process, and yet it foresees a space of organisms much larger than could be called upon in any given moment of adaptation.  It is doing "more than it needs to" in any one circumstance.  Learning can be said to become "imaginitive" as we put it in 1.1, or perhaps "Maslowian", accruing a tower of ever more refined  criteria enabled by previous criteria.\footnote{This allusion should not be taken to endorse all uses of Maslow's metaphors.}  Fitness for brute survival is  the most fundamental, or "self-bootstrapped", basis for adaptive learning, upon which other bases might arise.  For instance, a scientist might learn out of curiosity, and yet the scientist would not have come to exist were it not for a natural, existential/Darwinian process.

But learning must  also remain moored to reality, and not merely a theater of memes or other artifacts of its own structure.  A learning system must be able to maintain access to signal from the environment even as its own structure becomes noisy.  Another way to put this is that learning must serve the "Kantian whole". 

These remarks are intended to warn against a meme-like concept of cosmological learning.  The formative events in a cosmological autodidactic process might have a meme-like character, but eventually an environment must occur, and then the autodidactic process must be sensitive to the environment, not only to its consequencer.

The example of memes in human social structures show that a learning system that is not constrained by "brute survival" can sometimes become dominated by "economic network effects" in self-reference.  It is interesting to consider cosmological criteria other than brute survival that might give rise to autodidactic structures that are resistent to becoming unconnected to an environment.  The autodidactic principles we present are preliminary candidates for meeting such criteria.  While they concentrate influence from past "lessons", they do not necessarily become dominated by "viral" or "meme-like" events.

\subsection{Might autodidactic ideas become necessary rather than optional?}

We cannot know what unexpected phenomena will be observed in the future, but it is possible that autodidactic ideas will gracefully explain some of them.  But would that be only a matter of preference or might it become  difficult to avoid autodidactic ideas?

The question of irreversibility in learning models was raised earlier, as it is a source of potential tension with reversible theories in physics.  An additional motivation for considering irreversabilty is that if the learning of laws were reversible, then it would be possible to consider only  original laws, the latter laws  being no more than conveniences of interpretation.

\subsubsection{Learning systems and irreversibility}

We are initially motivated to consider change in laws because it is a way to explain why certain laws are in effect now, though a vast landscape of other laws appear to also be possible.  But the laws of physics we observe at present are either unchanging in this epoch or are only changing too slowly for the change to have been observed.  

We are considering natural systems, and natural systems are typically conserved in some  ways.  Therefore we assume a "substantial" process is required for laws to change; that such a process is not entirely arbitrary or free of interactions or constraints.  We treat "consmological learning" as a puzzle solver.

One implication is that if the evolution of laws is real, it is likely to be unidirectional, for otherwise it would be common for laws to revert to previous states, perhaps even more likely than for them to find a new state.  This is because a new state is not random but rather must meet certain constraints, while the immediate past state has already met constraints.  

A reversible but evolving system would randomly explore its immediate past frequently. When we see an evolving system that displays periods of stability, it probably evolves unidirectionally.  

Biology enacts approximate, punctuated stabilities in order for evolutionary experiments to take place.  A species in a successful niche can change little over long periods of time despite phenotypically irrelevant genotypic churn.  Biological evolution can only ever approximately backtrack; it is essentially locked into forward motion only.  (Marine mammals resemble fish but only superficially.)  These two observations are consistent with one another. 

Biology inherits fundamental properties from the physics of its universe.  Physical laws must give rise to irreversible processes for biology to be irreversible.  If we wish to minimize reliance on metaphysical assumptions, we should assume a conscilience between laws and the process that evolves laws.  

If fundamental physics evolves, we should expect laws to  evolve in a way that supports irreversible change, since they are subject to themselves, being in the same universe.  The laws that have evolved support irreversible phenomena, so the evolution of laws is likely to also give rise to operational irreversibility.  This is not the same as being abstractly irreversible, as we will argue in the next section.

It is  natural to ask whether, if fundamental physics is to be modeled by a learning machine, the irreversibility of the learning machine should be part of the explanation for  the irreversibility of nature.

For example, some learning machines incorporate a quench step, which thermalizes the system at temperature $T$ and then takes $T$ lower.  It is  interesting to imagine this playing a role in a model of the early universe.

\subsubsection{Can irreversible learning systems arise from reversible laws?}

We are examining whether the Universe is a learning computer.  How does the reversibility question play out in our investigation?  Can the universe learn in a way that is irreversible while also being based on fundamental laws that are almost entirely conserved, reversible, and deterministic?      

The conventional answer is that it can, because the initial conditions are asymmetric under reversals in time; as such a reversal would turn an expanding universe into a collapsing universe. However one thinks about this consideration, here is sought a different type of explanation for irreversibility: an emergent one.

We know from Landauer's principle that machines that learn are dissipative unless they record their history in order to be reversible.   Such machines become irreversible in time
\cite{CITE3}
at large scales because the memory requirements to retain reversibility become unavailable.  

However, laws of physics are conservative and reversible in time, provided some details related to weak interactions are taken into account.  Therefore, "learning" computation which generates unforeseen states would seem to have a different  character than the universe with which it operates, and might not serve well as a model for that universe.    

This is a general issue, potentially affecting all of the vast number of numerical models of use in physics.  Rounding errors in numerical simulations, for instance, are not merely errors, but potentially deceptive stories.  

A number resulting from a simulation might be what it is for any number of reasons; to know if "a rounding error" is the correct story requires retracing the steps that led to the number.  This is hard to do, but doable, because scientific simulations are designed specifically to support this process.  The nuisance is familiar in astronomy, for instance, when rounding errors become significant.
\cite{CITE4}.

For our purposes, we note that when a result comes out of a computer or some other systemic representation of abstractions, then errors can become better liars -  or if one chooses optimism, can become more likely channels for learning, innovation, or evolution.

We can consider a limited question:  Can a conserved, reversible, deterministic, physical universe without "just so" starting conditions \textit{contain} a learning system that has different qualities? 

 An \textit{operationally} irreversible computation can take place within such a universe.  Here is why:  A reversible process requires not only that the entire information of its past exist in some sense, but that it can specifically be accessed as causal information.  This requires an architecture for accessing it - additional information - that would become very large.  Small reversible computers can be built, but large ones cannot. 

It might be objected that if the universe is fundamentally reversible, then it should not matter if a reverse is only hypothetical.  The laws at the lowest level work - in that case - both forwards and backwards, after all.  

Any classical computer program can hypothetically be reversed, given enough memory to store its history.  If such storage is not enacted, we cannot call an arbitrary program reversible in practice.  The reason is that the additional information disambiguates profoundly vast numbers of possible prior state progressions of the program that would all be otherwise viable as theories about the past. (This is a feature of an \textit{Aristotelian} system.)

Leaving aside the question of whether a previous state is the only one  possible to explain the current state, we must also ask whether the present state presents sufficient opportunity to calculate prior states.

Consider a computer forensics expert asked to reconstruct how a program came to a result.  There are techniques to discern how magnetic marks on a hard drive had been set previously, but those are semi-persistent marks, so it is not surprising that careful study can retrieve traces from previous marks.  

However, consider the previous states of gates in a semiconductor chip.  All the previous gate states are "out there" in the radiation generated by the device, but the only way to retrieve them would require comprehensive measurement of every particle that had a causal relationship with that radiation, and a comparative analysis of those measurements.  Leaving aside measurement problems, and whether one believes the prior states of the gates must somehow be out there, even if one does believe and one can measure, it is clear that the forensics project would require resources of tremendous scope.  The computer needed to reverse a computer that had not planned to be reversed would be have to be profoundly vaster than the original.

If we are talking about physical systems, we must distinguish between what can happen and what cannot.  If a reverse cannot be engineered, even in the most wild speculations, then that reverse should be treated as ontologically different from a reverse that can at least hypothetically take place in reality.  

Another way to put it is that an engineer working with arbitrarily impressive resources from outside of the universe could perhaps take advantage of the reversibility of foundational laws in order to reverse the universe, but no engineer working inside the universe could ever muster the resources - such as a large enough memory - to reverse it.  They would undo their own labor.  

So we can say that computational systems within our universe, including learning systems, can be operationally irreversible, even if the underlying physical laws are reversible. 

This does not imply monotonic changes in any particular property of a learning system.  Some viruses have relatively large genomes and others have relatively small ones, and yet there is little evidence that  phenotype vigor tracks the scale of the genome.  

This framework places limits on how much we can interpret the past of consequencers even when we can recover a detailed history of a consequencer's states.  Old learning only makes sense in the context of the old conditions for learning.  The history of a germ line  can be recovered approximately; we can infer the history of genes.  What we cannot hope to represent is the whole of the "lesson" that influenced the evolution of genes.

\section{Conclusion}

This paper is one of a growing number that attack the question of \textit{why these laws?}  

Why these gauge groups, why these fermion and scalar representations, why the mysteries of chirality,  CP violation, and baryogengesis?  Why the vast hierarchies of scale and why the particular ratios of parameters of the standard model, setting the values of the masses and mixing angles?   It is sobering to contemplate that not one problem of this type has ever been solved, going all the way back to the measurements of the electron's mass and charge.   

Roughly speaking, we are faced with a single stark choice:   

Either: There are no rational reasons for any of  these  choices.   The universe might have been very different, but there will never be a reason why it took the path we observe it on.

Or: There is at least one rational explanation - in which case we are obligated to find it.   To ask the question is to suppose that, for instance, the constants could have been different, but not randomly so.  A scientific explanation would suggest that their values are set as the result of a dynamical process, which means it can be modelled analogously to all the other time dependent processes we are familiar with.

Here we consider a wide class of mechanisms based on the idea of learning.   We ask whether there might be a mechanism woven  into the fabric of the natural world, by means of which the universe could learn its laws.   To investigate this question we had to understand what learning is and how it differs from other ways a dynamical system can evolve.   This led us to the idea of the consequenser, which allows a system to respond to clues in its environment by altering its responses.

Of course, this is just a first step.  Learning is a complex, heterogeneous and broad set of behaviours.  There are many different ways to learn.  This paper reports some of our results from an ongoing search for ways that a system of laws, governing particles and fields, might either naturally or artificially come upon and learn the trick for, well, learning.   These  employ the renormalization group, the  idea that there are no laws except to follow precedence,  self-sampling methods, systems that maximize variety and geometrical self-assembly, and a direct mapping from a mathematical model of a learning system into the equations of motion of a general relativity and gauge fields.  The last-mentioned effort establishes a three way correspondences between learning machines, gauge and gravitational fields, and matrix models.   

There are varied potential spin-offs from our approach: Our models might 
suggest architectures for quantum machine learning systems.  We might find a way to use machine learning  to  simulate quantum or classical  gauge  field theories.  If we add controlled couplings to external  thermal reservoirs, we might be able to simulate the interplay of thermal and quantum effects in a range of models of the early universe.  

Achieving any of this would be a stunning advance. So it is with  trepidation and caution that we mention two more paths these ideas might motivate, each wildly more ambitious than what we have just mentioned.  

Imagine if we could use the correspondences discussed here to construct learning machines out of the degrees of freedom of gauge and gravitational fields.  Perhaps one version of this would be to construct a quark computer
which computes using the spins and isospins of quarks and gluons as qubits.

But beyond that, the correspondences suggest that the effective degree of freedom of the actual vacuum of a quantum gravity or gauge field might naturally evolve to become an autodidactic learning system.   This might be part of the explanation for the choices of gauge fields and representations and values of the coupling constants of the standard model.   

Since the correspondence organizes a landscape of theories, it might lead to a search in such a landscape, which might lead to discoveries of note, or might even serve as a model for what the universe might be doing.

The results here are tiny, baby steps towards these hypotheses, to be further explored in future work.

\section*{Acknowledgments}
 
Research at Perimeter Institute is supported in part by the Government of Canada through the Department of Innovation, Science and Economic Development Canada and by the Province of Ontario through the Ministry of Colleges and Universities.

Microsoft  provided computational, logistical and other general support for this work.  The authors  thank Kevin Scott of Microsoft in particular for support of this project.

Authors Stefan Stanojevic and Michael W. Toomey were supported in this work as research interns at Microsoft Research.  

Authors Jaron Lanier and Dave Wecker were supported as researchers at Microsoft.  Author Jaron Lanier thanks Chetan Nayak for helpful challenges and discussion.

Author Lee Smolin's work in these areas has been
generously supported by NSER, FQXi and the John Temleton Foundation.  He would like to thank Marina Cortes, Andrew
Liddle, Stuart Kauffman, Clelia Verde for collaborations on related projects and Bianca Dittrich, Laurent Friedel, Carlo Rovelli, and Yigit Yargic for encouragement and criticism.

\newpage

\bibliographystyle{unsrt}
\bibliography{bibo}

\begin{thebibliography}{100}

\bibitem{CSP}
Charles~Sanders Peirce.
\newblock Architecture.
\newblock March 1893.

\bibitem{Rees}
B.~J. Carr and M.~J. Rees.
\newblock The significance of numerical coincidences in nature.
\newblock {\em Nature 278}, 1967.

\bibitem{BarrowandTipler}
Barrow and Tipler.
\newblock {\em The Anthropic Cosmological Principle}.
\newblock (Oxford University Press,Oxford), 1986.

\bibitem{LOTC}
Lee Smolin.
\newblock {\em Life of the Cosmos}.
\newblock Oxford University Press, New York, 1997.

\bibitem{Susskind-anthropic}
Leonard Susskind.
\newblock {\em The Athropic landscape of String Theory}.
\newblock 2003.

\bibitem{bousso2004string}
Raphael Bousso and Joseph Polchinski.
\newblock The string theory landscape.
\newblock {\em Scientific American}, 291(3):78--87, 2004.

\bibitem{taylor2015f}
Washington Taylor and Yi-Nan Wang.
\newblock The f-theory geometry with most flux vacua.
\newblock {\em Journal of High Energy Physics}, 2015(12):1--21, 2015.

\bibitem{evolve}
Lee Smolin.
\newblock Did the universe evolve?
\newblock {\em Classical and Quantum Gravity}, pages 173--191, 1992.

\bibitem{cns-review}
Lee Smolin.
\newblock A perspective on the landscape problem, invited contribution for a
  special issue of foundations of physics titled: Forty years of string theory:
  Reflecting on the foundations.
\newblock {\em Foundations In Physics}, feb 2012.

\bibitem{matrix-general}
Washington Taylor.
\newblock Matrix theory: matrix quantum mechanics as a fundamental theory.
\newblock {\em Reviews of Modern Physics}, 73(2):419–461, Jun 2001.

\bibitem{CST}
S.-S. Chern and J.~Simons.
\newblock Characteristic forms and geometric invariants.
\newblock {\em Annals of Mathematics}, 99 (1):48–69, 1974.

\bibitem{CITE-BF}
G.~Horowitz.
\newblock Exactly soluble diffeomorphism invariant theories.
\newblock page 125:417, 1989.

\bibitem{plebanski1977separation}
Jerzy~F Pleba{\'n}ski.
\newblock On the separation of einsteinian substructures.
\newblock {\em Journal of Mathematical Physics}, 18(12):2511--2520, 1977.

\bibitem{CITE1}
Lee Smolin.
\newblock Plebanski action extended to a unification of gravity and yang-mills
  theory.
\newblock {\em Physical Review D}, 80(12), Dec 2009.

\bibitem{CITE2}
A~Garrett Lisi, Lee Smolin, and Simone Speziale.
\newblock Unification of gravity, gauge fields and higgs bosons.
\newblock {\em Journal of Physics A: Mathematical and Theoretical}, 43(44), Oct
  2010.

\bibitem{russell2021artificial}
Stuart Russell and Peter Norvig.
\newblock {\em Artificial intelligence: a modern approach}.
\newblock Pearson, 2021.

\bibitem{cubicm1}
Lee Smolin.
\newblock The cubic matrix model and the duality between strings and loops.
\newblock June 2000.

\bibitem{universal}
Lee Smolin.
\newblock Matrix universality of gauge and gravitational dynamics, 2008.

\bibitem{cubicm2}
Lee Smolin.
\newblock M theory as a matrix extension of chern-simons theory.
\newblock Feb 2000.

\bibitem{cubicm3}
Lee Smolin.
\newblock The exceptional jordan algebra and the matrix string.
\newblock April 2001.

\bibitem{cubicm4}
Lee~Smolin Etera R.~Livine.
\newblock Brst quantization of matrix chern-simons theory.
\newblock Dec 2002.

\bibitem{BFSS}
T.~Banks, W.~Fischler, S.H. Shenker, and L.~Susskind.

\bibitem{HoppePhD}
J~Hoppe.
\newblock {\em Ph.D. thesis (Massachusetts Institute of Technology)}, 1982.

\bibitem{dWHN}
H.~Nicolai B.~de Wit, J.~Hoppe.
\newblock {\em Nuclear Physics B305 (1988) 545}, 1988.

\bibitem{Hoppesolves}
Hoppe J.
\newblock On the construction of zero energy states in supersymmetric matrix
  models i, ii, iii.

\bibitem{moletta2021science}
Chiara Moletta.
\newblock {\em The Science of Can and Can't}.
\newblock Penguin Random House, 2021.

\bibitem{cote2016infinite}
Marc-Alexandre C\^{o}t\'{e} and Hugo Larochelle.
\newblock An infinite restricted boltzmann machine.
\newblock {\em Neural Comput.}, 28(7):1265–1288, July 2016.

\bibitem{Vanchurin_2020}
Vitaly Vanchurin.
\newblock The world as a neural network.
\newblock {\em Entropy}, page 1210, 2020.

\bibitem{cranmer2020lagrangian}
Miles Cranmer, Sam Greydanus, Stephan Hoyer, Peter Battaglia, David Spergel,
  and Shirley Ho.
\newblock Lagrangian neural networks, 2020.

\bibitem{Schmidt81}
Michael Schmidt and Hod Lipson.
\newblock Distilling free-form natural laws from experimental data.
\newblock {\em Science}, 324(5923):81--85, 2009.

\bibitem{greydanus2019hamiltonian}
Sam Greydanus, Misko Dzamba, and Jason Yosinski.
\newblock Hamiltonian neural networks, 2019.

\bibitem{optimal_rg_information}
Patrick~M. Lenggenhager, Doruk~Efe Gökmen, Zohar Ringel, Sebastian~D. Huber,
  and Maciej Koch-Janusz.
\newblock Optimal renormalization group transformation from information theory.
\newblock {\em Physical Review X}, 10(1), Feb 2020.

\bibitem{mutual_information}
Maciej Koch-Janusz and Zohar Ringel.
\newblock Mutual information, neural networks and the renormalization group.
\newblock {\em Nature Physics}, 14(6):578–582, Mar 2018.

\bibitem{Hashimoto_2018}
Koji Hashimoto, Sotaro Sugishita, Akinori Tanaka, and Akio Tomiya.
\newblock Deep learning and the ads/cft correspondence.
\newblock {\em Physical Review D}, 98(4), Aug 2018.

\bibitem{Hashimoto_2019}
Koji Hashimoto.
\newblock Ads/cft correspondence as a deep boltzmann machine.
\newblock {\em Physical Review D}, 99(10), May 2019.

\bibitem{schoenholz2017correspondence}
Samuel~S. Schoenholz, Jeffrey Pennington, and Jascha Sohl-Dickstein.
\newblock A correspondence between random neural networks and statistical field
  theory, 2017.

\bibitem{Lin_2017}
Henry~W. Lin, Max Tegmark, and David Rolnick.
\newblock Why does deep and cheap learning work so well?
\newblock {\em Journal of Statistical Physics}, page 1223–1247, 2017.

\bibitem{Capovilla_1991}
R~Capovilla, J~Dell, T~Jacobson, and L~Mason.
\newblock Self-dual 2-forms and gravity.
\newblock {\em Classical and Quantum Gravity}, 8(1):41--57, jan 1991.

\bibitem{unifysl}
Lee Smolin.
\newblock Unification of the state with the dynamical law, 2012.

\bibitem{lovesknots}
Olafs Vandans, Kaiyuan Yang, Zhongtao Wu, and Liang Dai.
\newblock Identifying knot types of polymer conformations by machine learning.
\newblock {\em Phys. Rev. E}, 101:022502, Feb 2020.

\bibitem{wolfram}
Stephen Wolfram.
\newblock {\em A Project to Find the Fundamental Theory of Physics}.
\newblock Wolfram Media, 2020.

\bibitem{Gordian}
Jaron Lanier.
\newblock Gordian software.
\newblock {\em Edge}, 2003.

\bibitem{cardenas2020process}
J.F. C\'ardenas-Garc\'ia.
\newblock The process of info-autopoiesis -- the source of all information.
\newblock {\em Biosemiotics}, 13:199--221, 2020.

\bibitem{Kochchen}
Robert de Mello Koch Ling~Cheng Ellen~de Mello~Koch.
\newblock Is deep learning a renormalization group flow.
\newblock 12 Jun 2019.

\bibitem{kadanoff_rg}
Leo~P Kadanoff.
\newblock Scaling laws for ising models near t c.
\newblock {\em Physics Physique Fizika}, 2(6):263, 1966.

\bibitem{original}
Cédric Bény.
\newblock Deep learning and the renormalization group, 2013.

\bibitem{mera}
G.~Vidal.
\newblock Class of quantum many-body states that can be efficiently simulated.
\newblock {\em Physical Review Letters}, 101(11), Sep 2008.

\bibitem{neural_rg}
Shuo-Hui Li and Lei Wang.
\newblock Neural network renormalization group.
\newblock {\em Physical Review Letters}, 121(26), Dec 2018.

\bibitem{nvp}
Laurent Dinh, Jascha Sohl-Dickstein, and Samy Bengio.
\newblock Density estimation using real nvp, 2016.

\bibitem{variational_rg}
Pankaj Mehta and David~J. Schwab.
\newblock An exact mapping between the variational renormalization group and
  deep learning, 2014.

\bibitem{super_resolving}
Stavros Efthymiou, Matthew J.~S. Beach, and Roger~G. Melko.
\newblock Super-resolving the ising model with convolutional neural networks.
\newblock {\em Physical Review B}, 99(7), Feb 2019.

\bibitem{Smolin:2012mk}
Lee Smolin.
\newblock {Precedence and freedom in quantum physics}.
\newblock 5 2012.

\bibitem{hardy2016operational}
Lucien Hardy.
\newblock Operational general relativity: possibilistic, probabilistic, and
  quantum.
\newblock {\em arXiv preprint arXiv:1608.06940}, 2016.

\bibitem{Masanes:2010tt}
Lluis Masanes and Markus~P. Muller.
\newblock {A Derivation of quantum theory from physical requirements}.
\newblock {\em New J. Phys.}, 13:063001, 2011.

\bibitem{weinstein2017learning}
Steven Weinstein.
\newblock Learning the einstein-podolsky-rosen correlations on a restricted
  boltzmann machine, 2017.

\bibitem{bateson2000steps}
Gregory Bateson.
\newblock {\em Steps to an ecology of mind: Collected essays in anthropology,
  psychiatry, evolution, and epistemology}.
\newblock University of Chicago Press, 2000.

\bibitem{Game_of_Life}
Martin Gardner.
\newblock Mathematical games: The fantastic combinations of john conway's new
  solitaire game "life".
\newblock pages 120--123, 1970.

\bibitem{silver2018general}
David Silver, Thomas Hubert, Julian Schrittwieser, Ioannis Antonoglou, Matthew
  Lai, Arthur Guez, Marc Lanctot, Laurent Sifre, Dharshan Kumaran, Thore
  Graepel, et~al.
\newblock A general reinforcement learning algorithm that masters chess, shogi,
  and go through self-play.
\newblock {\em Science}, 362(6419):1140--1144, 2018.

\bibitem{Smolin_Barbour}
Julian Barbour and Lee Smolin.
\newblock Extremal variety as the foundation of a cosmological quantum theory,
  1992.

\bibitem{leibniz1989discourse}
Gottfried~Wilhelm Leibniz.
\newblock Discourse on metaphysics.
\newblock In {\em Philosophical papers and letters}, pages 303--330. Springer,
  1989.

\bibitem{bohm_variety}
Lee Smolin.
\newblock Quantum mechanics and the principle of maximal variety.
\newblock {\em Foundations of Physics}, 46(6):736–758, Mar 2016.

\bibitem{Rovelli}
Carlo Rovelli.
\newblock Relational quantum mechanics.
\newblock {\em International Journal of Theoretical Physics},
  35(8):1637–1678, Aug 1996.

\bibitem{surya2019causal}
Sumati Surya.
\newblock The causal set approach to quantum gravity.
\newblock {\em Living Reviews in Relativity}, 22(1):1--75, 2019.

\bibitem{kleitman1975asymptotic}
Daniel~J Kleitman and Bruce~L Rothschild.
\newblock Asymptotic enumeration of partial orders on a finite set.
\newblock {\em Transactions of the American Mathematical Society},
  205:205--220, 1975.

\bibitem{energetic_causal}
Marina Cortês and Lee Smolin.
\newblock Quantum energetic causal sets.
\newblock {\em Physical Review D}, 90(4), Aug 2014.

\bibitem{KK}
"Kuchar and Karel.
\newblock 1971.

\bibitem{erdos1963asymmetric}
P.~Erd\"os and A.~R\'enyi.
\newblock Asymmetric graphs.
\newblock {\em Acta Mathematica Hungarica}, 14:295, 1963.

\bibitem{von1957spektren}
Lothar Von~Collatz and Ulrich Sinogowitz.
\newblock Spektren endlicher grafen.
\newblock In {\em Abhandlungen aus dem Mathematischen Seminar der
  Universit{\"a}t Hamburg}, volume~21, pages 63--77. Springer, 1957.

\bibitem{bell1992note}
Francis~K Bell.
\newblock A note on the irregularity of graphs.
\newblock {\em Linear Algebra and its Applications}, 161:45--54, 1992.

\bibitem{abrego2009sum}
B.M. \'Abrego, S.~Fern\'andez-Merchant, M.G. Neubauer, and W.~Watkins.
\newblock Sum of squares of degrees in a graph.
\newblock {\em J. Inequal. Pure Appl. Math.}, 10:1, 2009.

\bibitem{albertson1997irregularity}
M.O. Albertson.
\newblock The irregularity of a graph.
\newblock {\em Ars. Comb.}, 46:219, 1997.

\bibitem{abdo2014total}
H.~Abdo, S.~Brandt, and D.~Dimitrov.
\newblock The total irregularity of a graph.
\newblock {\em DMTCS}, 16:201, 2014.

\bibitem{abdo2014graphs}
Hosam Abdo, Nathann Cohen, and Darko Dimitrov.
\newblock Graphs with maximal irregularity.
\newblock {\em Filomat}, 28(7):1315--1322, 2014.

\bibitem{estrada2010quantifying}
Ernesto Estrada.
\newblock Quantifying network heterogeneity.
\newblock {\em Physical Review E}, 82(6):066102, 2010.

\bibitem{chartrand1988define}
Gary Chartrand, Paul Erd{\"o}s, and Ortrud~R Oellermann.
\newblock How to define an irregular graph.
\newblock {\em The College Mathematics Journal}, 19(1):36--42, 1988.

\bibitem{alavi1987highly}
Yousef Alavi, Gary Chartrand, Fan~RK Chung, Paul Erd{\"o}s, Ronald~L Graham,
  and Ortrud~R Oellermann.
\newblock Highly irregular graphs.
\newblock {\em Journal of Graph Theory}, 11(2):235--249, 1987.

\bibitem{diff_wave}
Claire Donnat, Marinka Zitnik, David Hallac, and Jure Leskovec.
\newblock Learning structural node embeddings via diffusion wavelets.
\newblock {\em Proceedings of the 24th ACM SIGKDD International Conference on
  Knowledge Discovery \& Data Mining}, Jul 2018.

\bibitem{revolution}
Lee Smolin.
\newblock {\em Einstein's unfinished revolution: The search for what lies
  beyond the quantum}.
\newblock Penguin Books, 2020.

\bibitem{erdos_renyi}
P~Erdos and A~Renyi.
\newblock On random graphs i.
\newblock {\em Publ. math. debrecen}, 6(290-297):18, 1959.

\bibitem{struc2vec}
Leonardo~F.R. Ribeiro, Pedro~H.P. Saverese, and Daniel~R. Figueiredo.
\newblock struc2vec.
\newblock {\em Proceedings of the 23rd ACM SIGKDD International Conference on
  Knowledge Discovery and Data Mining}, Aug 2017.

\bibitem{cheeger1984curvature}
Jeff Cheeger, Werner M{\"u}ller, and Robert Schrader.
\newblock On the curvature of piecewise flat spaces.
\newblock {\em Communications in mathematical Physics}, 92(3):405--454, 1984.

\bibitem{wu2015emergent}
Zhihao Wu, Giulia Menichetti, Christoph Rahmede, and Ginestra Bianconi.
\newblock Emergent complex network geometry.
\newblock {\em Scientific reports}, 5(1):1--12, 2015.

\bibitem{mulder2018network}
Daan Mulder and Ginestra Bianconi.
\newblock Network geometry and complexity.
\newblock {\em Journal of Statistical Physics}, 173(3):783--805, 2018.

\bibitem{kelly2019self}
Christy Kelly, Carlo~A Trugenberger, and Fabio Biancalana.
\newblock Self-assembly of geometric space from random graphs.
\newblock {\em Classical and Quantum Gravity}, 36(12):125012, 2019.

\bibitem{kelly2020emergence}
Christy Kelly, Carlo Trugenberger, and Fabio Biancalana.
\newblock Emergence of the circle in a statistical model of random cubic
  graphs, 2020.

\bibitem{vanderhoorn2020ollivier}
Pim van~der Hoorn, William~J. Cunningham, Gabor Lippner, Carlo Trugenberger,
  and Dmitri Krioukov.
\newblock Ollivier-ricci curvature convergence in random geometric graphs,
  2020.

\bibitem{boguna2021network}
Mari{\'a}n Bogu{\~n}{\'a}, Ivan Bonamassa, Manlio De~Domenico, Shlomo Havlin,
  Dmitri Krioukov, and M~{\'A}ngeles Serrano.
\newblock Network geometry.
\newblock {\em Nature Reviews Physics}, pages 1--22, 2021.

\bibitem{kelly2021convergence}
Christy Kelly, Carlo Trugenberger, and Fabio Biancalana.
\newblock Convergence of combinatorial gravity, 2021.

\bibitem{perez2013spin}
Alejandro Perez.
\newblock The spin-foam approach to quantum gravity.
\newblock {\em Living Reviews in Relativity}, 16(1):1--128, 2013.

\bibitem{loll2019quantum}
Renate Loll.
\newblock Quantum gravity from causal dynamical triangulations: a review.
\newblock {\em Classical and Quantum Gravity}, 37(1):013002, 2019.

\bibitem{Usym}
Neil Turk.
\newblock The universe within.
\newblock {\em Massay Lectures}, page 312, September 2012.

\bibitem{crooks2007measuring}
Gavin~E Crooks.
\newblock Measuring thermodynamic length.
\newblock {\em Physical Review Letters}, 99(10):100602, 2007.

\bibitem{amari2000information}
Shun-ichi Amari.
\newblock Information geometry.
\newblock {\em Japanese Journal of Mathematics}, pages 1--48, 2000.

\bibitem{velivckovic2017graph}
Petar Veli{\v{c}}kovi{\'c}, Guillem Cucurull, Arantxa Casanova, Adriana Romero,
  Pietro Lio, and Yoshua Bengio.
\newblock Graph attention networks.
\newblock {\em arXiv preprint arXiv:1710.10903}, 2017.

\bibitem{selfsim}
Klee Irwin, Marcelo Amaral, and David Chester.
\newblock The self-simulation hypothesis interpretation of quantum mechanics.
\newblock {\em Entropy}, 22(2), 2020.

\bibitem{graphstellation}
Jaron Lanier.
\newblock Rearranging stars to communicate with aliens.
\newblock {\em Discover}, February 2008.

\bibitem{Riedel_2016}
C.~Jess Riedel, Wojciech~H. Zurek, and Michael Zwolak.
\newblock Objective past of a quantum universe: Redundant records of consistent
  histories.
\newblock {\em Physical Review A}, 93(3), Mar 2016.

\bibitem{artificialuterus}
Shani Aguilera-Castrejon, Oldak.
\newblock Ex utero mouse embryogenesis from pre-gastrulation to late
  organogenesis.
\newblock {\em Nature}, 2021.

\bibitem{CITE3}
Simon~Portegeise Zwart.
\newblock The future.
\newblock {\em Computational astrophysics}, 361(6406 979), September 2018.

\bibitem{CITE4}
Simon F.~Portegies Zwart and Tjarda C.~N. Boekholt.
\newblock Numerical verification of the microscopic time reversibility of
  newton’s equations of motion: Fighting exponential divergence.
\newblock {\em Communications in Nonlinear Science and Numerical Simulation},
  1-10(00), 2018.

\bibitem{kaufmann1}
S.~Kauffman.
\newblock Reinventing the Sacred.
\newblock {\em Basic Books}, N.~Y. (2008).

\bibitem{kaufmann2}
S.~Kauffman.
\newblock A World Beyond Physics: The Emergence and Evolution of Life?
\newblock {\em Oxford University Press}, Oxford (2019).

\bibitem{kaufmann3}
S.~Kauffman
\newblock Answering Schrodinger's ``What is Life?''
\newblock {\em Entropy} 22, 815 (2020).

\end{thebibliography}

\end{document}